\documentclass[%
 reprint,
superscriptaddress,
 amsmath,amssymb,
]{revtex4-1}

\usepackage{graphicx}
\usepackage{dcolumn}
\usepackage{bm}
\usepackage{svg}
\usepackage{placeins}

\usepackage[english]{babel}
\usepackage{hyperref}
\usepackage{cleveref}

\usepackage{tikz}
\usetikzlibrary{bayesnet}

\usepackage{bbold}

\usepackage{listings}
\usepackage{upquote}

\usepackage{beramono} 
\usepackage[T1]{fontenc}

\colorlet{punct}{red!60!black}
\definecolor{background}{HTML}{EEEEEE}
\definecolor{delim}{RGB}{20,105,176}
\colorlet{numb}{magenta!60!black}

\lstdefinelanguage{json}{
    basicstyle=\footnotesize\ttfamily,
    numbers=left,
    numberstyle=\scriptsize,
    stepnumber=1,
    numbersep=8pt,
    showstringspaces=false,
    breaklines=true,
    frame=lines,
    backgroundcolor=\color{background},
    literate=
     *{0}{{{\color{numb}0}}}{1}
      {1}{{{\color{numb}1}}}{1}
      {2}{{{\color{numb}2}}}{1}
      {3}{{{\color{numb}3}}}{1}
      {4}{{{\color{numb}4}}}{1}
      {5}{{{\color{numb}5}}}{1}
      {6}{{{\color{numb}6}}}{1}
      {7}{{{\color{numb}7}}}{1}
      {8}{{{\color{numb}8}}}{1}
      {9}{{{\color{numb}9}}}{1}
      {:}{{{\color{punct}{:}}}}{1}
      {,}{{{\color{punct}{,}}}}{1}
      {\{}{{{\color{delim}{\{}}}}{1}
      {\}}{{{\color{delim}{\}}}}}{1}
      {[}{{{\color{delim}{[}}}}{1}
      {]}{{{\color{delim}{]}}}}{1},
}

\lstdefinelanguage{R}{
  basicstyle=\footnotesize\ttfamily, 
  numbers=left,
  numberstyle=\scriptsize,
  stepnumber=1,
  numbersep=5pt,
  backgroundcolor=\color{background},
  showspaces=false,
  showstringspaces=false,
  showtabs=false,
  frame=single,                   
  rulecolor=\color{black},        
  tabsize=2,
  captionpos=b,
  breaklines=true,
  breakatwhitespace=false,
  title=\lstname,                 
  keywordstyle=\color{blue},
  commentstyle=\color{green},
  stringstyle=\color{violet},
  escapeinside={\%*}{*)}
} 

\crefname{lstlisting}{listing}{listings}
\Crefname{lstlisting}{Listing}{Listings}

\newcommand{\mat}[1]{\mathbf{#1}}

\newcommand{\Cov}[2]{\operatorname{Cov}\left[#1, #2\right]}

\newcommand{\Var}[1]{\operatorname{Var}\left[ #1 \right]}

\begin{document}

\preprint{APS/123-QED}

\title{Nuclear data evaluation with Bayesian networks}

\author{Georg Schnabel}
\email{g.schnabel@iaea.org}
\affiliation{%
    NAPC–Nuclear Data Section, International Atomic Energy Agency, Vienna, Austria
}%

\author{Roberto Capote}
\affiliation{%
    NAPC–Nuclear Data Section, International Atomic Energy Agency, Vienna, Austria
}%

\author{Arjan Koning}
\affiliation{%
    NAPC–Nuclear Data Section, International Atomic Energy Agency, Vienna, Austria
}%

\author{David Brown}
\affiliation{%
    National Nuclear Data Center, Brookhaven National Laboratory, Upton, NY 11973-5000, USA
}%

\date{\today}

\begin{abstract}
    Bayesian networks are graphical models to represent the deterministic and probabilistic relationships between variables within the Bayesian framework.
    The knowledge of all variables can be updated using new information about some of the variables.
    The Bayesian Generalized Linear Least Squares method can be regarded as an inference method for Bayesian networks of variables with multivariate normal priors and linear relationships between them. 
    We show that relying explicitly on the Bayesian network interpretation enables large scale inference and gives more flexibility in incorporating prior assumptions and constraints into the nuclear data evaluation process, such as the constraints that some cross sections equal linear combinations of other cross sections and that all cross sections must be non-negative.
    The latter constraint is accounted for by a non-linear transformation and therefore we also discuss inference in Bayesian networks with non-linear relationships between variables.
    Using Bayesian networks, the evaluation process yields more detailed information, such as posterior estimates and uncertainties of all statistical and systematic errors associated with the experiments.
    We further elaborate on a sparse Gaussian process construction that can be well integrated into the Bayesian network framework and applied to, e.g., the modeling of energy-dependent model parameters, model deficiencies of the physics model or energy-dependent systematic errors of experiments.
    We present three proof-of-concept examples that emerged in the context of the neutron data standards project and in the ongoing international evaluation efforts of $^{56}$Fe.
    In the first example we demonstrate the modelization and explicit estimation of relative energy-dependent error components associated with experimental datasets.
    Then we show that Bayesian networks in combination with the outlined Gaussian process construction may be applied to an evaluation of $^{56}$Fe in the energy range between one and two MeV, where it is difficult to obtain satisfactory evaluations by R-Matrix and nuclear model fits.
    Finally, we present a model-based evaluation of $^{56}$Fe between 5\,MeV and 30\,MeV with a consistent and statistically sound treatment of model deficiencies.
    The R scripts to reproduce the Bayesian network examples and the \textit{nucdataBaynet} package for Bayesian network modeling and inference have been made publicly available.
\end{abstract}

\pacs{Valid PACS appear here}
\maketitle

\tableofcontents

\section{\label{sec:introduction}Introduction}
An accurate knowledge of nuclear quantities, such as cross sections, branching ratios, and half-lifes,  of the nuclei is important in various branches of nuclear physics and engineering. The risk and safety assessment of nuclear power plants, the modeling of astrophysical processes and dose calculations in nuclear medicine all depend on nuclear data.
The aim of nuclear data evaluation is to provide accurate estimates of these quantities along with reliable uncertainties quantifying their accuracy.

Evaluated nuclear data are the result of an evaluation process that begins with the identification and collection of suitable experimental data.
Selected data are scrutinized regarding potential errors and biases.
It is common practice to characterize the knowledge about the experimental features in the following way.
The determined possible ranges of the errors are expressed as uncertainties.  
Biases, i.e., errors that affect distinct measured values in the same way, are accounted for by introducing correlations.
Uncertainties and correlations are usually summarized as a covariance matrix, referred to as experimental covariance matrix.
Finally, a statistical method is applied to fuse the information of the indvidual experiments, i.e., the measured values and the associated covariance matrix, in order to obtain estimates and an associated evaluated covariance matrix of quantities of interest.

An established method in the field of nuclear data evaluation to achieve this fusion is the Bayesian version of the  Generalized Least Squares (GLS) method, e.g.,~\cite{muir_evaluation_1989,smithProbabilityStatisticsData1991,capote_nuclear_2010}, which enables the inclusion of prior knowledge.
It relies on the assumption that relationships between variables are linear and the knowledge of the variables is expressed in terms of a multivariate normal distribution, which is uniquely characterized by a mean vector and a covariance matrix.
Therefore, the experimental covariance matrix can be directly used in the GLS method.
Especially in the field of nuclear data, the name Generalized Linear Least Squares (GLLS) is often preferred to emphasize the linearity assumption.

From now on, we always refer to the Bayesian version when we use the term GLS method, as is common practice in the field of nuclear data evaluation. 
The GLS method is often presented as a method to obtain estimates of parameters, e.g., parameters of a physics model or cross sections on a mesh, and an associated evaluated covariance matrix based on measured data and the associated experimental covariance matrix.
This viewpoint is also reflected in the mathematical equations with the appearance of the experimental covariance matrix.
Unfortunately, the experimental covariance matrix is dense if systematic experimental uncertainties are taken into account,  posing an obstacle to the application of the GLS method to a large number of data points.
One solution to deal with large covariance matrices in the GLS method is to update the prior covariance matrix block-wise as implemented in the GANDR evaluation code~\cite{muir_global_2007} and another one is to exploit the structure of the prior covariance matrix when it is derived from a sample of model predictions~\cite{schnabel_modified_2017}.
However, these approaches only address the issue of a large prior covariance matrix and not a large experimental covariance matrix.

Regarding the flexibility of the GLS method in terms of possible modeling assumptions, Gaussian processes (GPs), e.g.,~\cite{rasmussen_gaussian_2006,steinInterpolationSpatialData1999,tarantolaInverseProblemTheory2005}, have been included in the nuclear data evaluation process in various ways, e.g., to account for model deficiencies~\cite{pigni_uncertainty_2003,leeb_consistent_2008,schnabelDifferentialCrossSections2016,helgessonFittingDefectNonlinear2017b}, as priors on energy-dependent model parameters~\cite{helgessonTreatingModelDefects2018,schnabelConceptionSoftwareImplementation2021}, and as a flexible fitting function instead of a physics model~\cite{iwamotoGenerationNuclearData2020a}.
Even though these approaches may be regarded as new methods, as soon as we discretize the function for which the GP serves as prior, these approaches merely represent specific constructions of the prior covariance matrix which is then used in the GLS method.
There is still a vast space of modeling possibilities that has not been thoroughly exploited in nuclear data evaluations yet.
For example, to our knowledge, the modeling of model deficiencies with Gaussian processes obeying simultaneously the constraints that some cross sections are defined as sums of other cross sections and that cross sections must be non-negative has not been demonstrated so far.
The developments in this direction are again impeded by the resulting large and potentially dense covariance matrix and its negative impact on the computation time of the GLS method.
Furthermore, the GLS method can only be applied to variables with linear relationships, which may be a serious limitation in certain nuclear data evaluation scenarios, e.g.,~\cite{smith_large_2002,capote_investigation_2008}.

The limitations of the GLS method can be pushed back by changing the perspective on the inference process.
Model parameters, model predictions, model deficiencies, measurements, statistical and systematic errors are linked to each other by functional relationships.
These quantities with the exception of the measured values are uncertain and can therefore be represented as random variables.
These random variables and their probablistic functional relationships can be modeled as a Bayesian network, e.g.,~\cite{koller_probabilistic_2009,pearlProbabilisticReasoningIntelligent2014}.
Inference procedures for Bayesian networks exploit the connection structure of the variables to speed up computations and thus enable inference at a larger scale. 

Bayesian networks have a long history and a sound mathematical foundation. 
Early and seminal contributions to their theory were made by Judea Pearl in the eighties, e.g.,~\cite{Pearl_CSS85}.
Importantly, a Bayesian network is a specific \textit{representation} of the information required to perform Bayesian inference, which is both intuitive for humans and computationally efficient.
As an aside, also hierarchical models, e.g.,~\cite{bussHierarchicalMontecarloApproach2010,schnabelFittingAnalysisTechnique2018} in the context of nuclear data, can be interpreted in terms of Bayesian networks.   

Relying on Bayesian networks as a mental abstraction helps to find good and valid Bayesian models and makes the modeling assumptions more explicit.
Besides increased computational efficiency, one benefit for nuclear data evaluation is that consistent and joint evaluations of several isotopes coupled together by cross section measurements of materials in natural composition and the consistent evaluation of exclusive cross sections using measurements of inclusive ones, such as residual production cross sections, become straight-forward from the modeler point of view.
Another possible application is the consistent updating of evaluations in library projects, such as JEFF~\cite{plompenJointEvaluatedFission2020}, JENDL~\cite{shibataJENDL4NewLibrary2011a}, ENDF/B~\cite{brownENDFBVIIITh2018a}.
Many past evaluations represent valuable knowledge but are unfortunately not reproducible anymore.
The Bayesian network framework enables the consistent update of such evaluations using new measurements.
In the future, the unresolved resonance range may also be evaluated using Bayesian networks.
For instance, a computational mesh of several hundred thousand points would be required to properly represent point-wise the resonances in the unresolved resonance range of $^{239}$Pu~\cite{levittProbabilityTableMethod1972}.
This mesh size is nowadays intractable within the conventional formulation of the GLS method but may be tractable in the Bayesian network framework. 
Please note that the usual treatment of the unresolved resonance range does assume a non-Gaussian probability distribution and the hypothetical treatment using Gaussian processes within the GLS method would be conceptually different.

In this paper, we first focus on Bayesian networks with linear relationships between variables and prior knowledge represented by multivariate normal distributions. 
These are the same assumptions as in the GLS method, and inference in such Bayesian networks therefore yields the same results.
Because the GLS method is well-known in the nuclear data field, we anchor the exposition of the Bayesian network interpretation at the GLS method and tailor it to the specifics of nuclear data evaluation.
The aim of this paper is therefore to demonstrate potential advantages of the Bayesian network interpretation for nuclear data evaluation and not to give a comprehensive account of the general theory of Bayesian networks.
A comprehensive account of their theory can be found in~\cite{pearlProbabilisticReasoningIntelligent2014}.

As non-linear relationships between variables appear frequently in nuclear data evaluation, it is important to take them into account properly in the inference procedure.
Consequently we also discuss how inference in Bayesian networks with non-linear relationship between variables can be performed, keeping the assumption of a multivariate normal prior on the variables.
To this end, we use a customized Levenberg-Marquardt algorithm~\cite{helgessonFittingDefectNonlinear2017b} to find the most likely values of the variables according to the exact posterior distribution.
Assumptions, such as an uncertainty being relative to the underlying measured quantity or a quantitity being non-negative, can be rigorously taken into account.
The exact treatment of non-linear relationships also enables the use of other probability distributions related to the (multivariate) normal distribution, such as the log-normal distribution for strictly positive variables, e.g.,~\cite{smith_large_2002}, and the application of transformations to obtain approximately normally distributed variables in cases where the original variables are not normally distributed, e.g.,~\cite{hoeferAssessingImprovingModel2021}.  

Approximate posterior covariances based on the linearization of the non-linear relationships between variables at the posterior maximum can be quickly computed for a subset of the variables despite a possible large total number of variables, e.g., several hundred thousand. 
Furthermore, sample vectors from the approximate posterior distribution including all variables can be obtained using the full approximate posterior covariance matrix, even though it is never explicitly computed.  

Regarding the flexibility of modeling assumptions, we elaborate on a sparse Gaussian process construction that can be well integrated into the Bayesian network framework and scales well with the number of data points.
A similar construction has already been employed in a prototype of a nuclear data evaluation pipeline~\cite{schnabelConceptionSoftwareImplementation2021} in the step concerned with the automatic correction of uncertainties of inconsistent experimental datasets. 
This GP construction allows the inclusion of prior knowledge about the expected range, slope, and smoothness of the function for which the Gaussian process serves as prior.
The prior on these properties can be different at different (usually energy) locations of the function.
This GP construction may be used instead of the originally proposed ones in, e.g.,~\cite{leeb_consistent_2008,schnabelDifferentialCrossSections2016,helgessonFittingDefectNonlinear2017b,schnabelEstimatingModelBias2018a} to take into account model deficiencies, unknown parameter functions in a nuclear model~\cite{helgessonTreatingModelDefects2018,schnabelConceptionSoftwareImplementation2021} or be used in place of a nuclear model to fit cross sections~\cite{iwamotoGenerationNuclearData2020a}.
It is however also perfectly possible to use other GP constructions in the Bayesian network framework, although inference may not be possible for a large number of variables then.

Finally, we present three example applications to demonstrate the flexibility of the Bayesian network framework and its usefulness for nuclear data evaluation.
The first example application emerged in the context of the neutron data standards project~\cite{carlsonInternationalEvaluationNeutron2009,carlsonEvaluationNeutronData2018} and showcases the inclusion of energy-dependent unrecognized sources of uncertainty~\cite{capoteUnrecognizedSourcesUncertainties2020} and the rigorous treatment of uncertainties given relative to the underlying true cross sections.
The two other examples emerged in the context of the international evaluation efforts of neutron-induced reaction of $^{56}$Fe within the International Nuclear Data Evaluation Network (INDEN).
As R-matrix codes, e.g.,~\cite{larsonUpdatedUserGuide1998,moxonREFIT2009LeastSquareFitting2010,desaintjeanCONRADCodeNuclear2021}, and nuclear physics model codes, e.g.,~\cite{youngGNASHPreequilibriumStatistical1977,hermanEMPIRENuclearReaction2007,koning_modern_2012,iwamotoCCONECodeSystem2016,kawanoCoH3CoupledChannelsHauserFeshbach2021}, struggle to provide good fits in the incident energy range from several hundred keV to about 5\,MeV, we show how an evaluation may be performed using only Gaussian processes in the Bayesian network framework in the energy range between one and two MeV, while preserving consistency between the various channels and ensuring that all evaluated cross sections are non-negative.
In the final example, we incorporate model defects into a model-based evaluation with TALYS~\cite{koning_modern_2012,koningTENDLCompleteNuclear2019}, as already suggested and demonstrated in schematic examples~\cite{schnabelDifferentialCrossSections2016,helgessonFittingDefectNonlinear2017b}, but using the GP construction proposed in this paper. 
In all the examples, comprehensive uncertainty information of all evaluated quantities is available, such as for the evaluated cross sections and systematic error components. 
The R scripts to create these Bayesian networks and do inference in them are provided as examples as a part of the publicly available \textit{nucdataBaynet} R package~\cite{georgNucdataBaynetNuclearData}.

The description of the Bayesian networks given as examples in~\cref{sec:examples} is focused on the modeler point of view.
A modeler can think in terms of nodes and mappings between them, such as linear interpolations, convolutions and non-linear transformations, and does not need to be concerned about the details of the inference algorithms described in section~\ref{sec:methodology}.
Therefore a reader may start with~\cref{subsec:baynet-primer} and then jump to the examples in~\cref{sec:examples} to get a feel for Bayesian network modeling before possibly delving into the backing math in~\cref{subsec:glsrecap} and subsequent sections.

Finally, we stress that nuclear model code and evaluation systems exist that already offer advanced Bayesian inference capabilities with a significant methodological overlap with the methods presented in this paper, such as CONRAD~\cite{desaintjeanCONRADCodeNuclear2021}, EMPIRE~\cite{hermanEMPIRENuclearReaction2007,hermanEMPIRE3MaltaModularSystem2013}, GANDR~\cite{muir_global_2007}, GMAP~\cite{poenitzDataInteractionObjective1981,poenitzSimultaneousEvaluationStandards1997}, KALMAN~\cite{kawanoCovarianceEvaluationSystem1997}, SAMMY~\cite{larsonUpdatedUserGuide1998}, SOK~\cite{kawanoSimultaneousEvaluationFission2000} and the T6 code system~\cite{koningTENDLCompleteNuclear2019,koning_bayesian_2015}.
A detailed account of Bayesian statistics in the context of the analysis of nuclear resonance data is given in~\cite{frohner_evaluation_2000} and a comprehensive general introduction with an eye to physics in~\cite{lindenBayesianProbabilityTheory2014}.
However, to the best of our knowledge, the employed Bayesian methods have never been discussed in terms of the Bayesian network interpretation in the nuclear data field.
Bayesian neural networks have already been employed for nuclear mass prediction~\cite{niuNuclearMassPredictions2018} but a neural network is conceptually different from a Bayesian network even though Bayesian inference was applied to adjust the weights of the network.
Regarding the application of Bayesian networks in other nuclear-related fields, they have been discussed for example in the context of power plant safety analysis~\cite{chenApplyingBayesianNetworks2010,mohanApplicationBayesianNetworks2020,chenUseBayesianNetworks2021}, software reliability quantification of system critical software~\cite{kangDevelopmentBayesianBelief2018}, reactor control~\cite{ramosDynamicBayesianNetworks2021}, and analysis of nuclear acquisitions~\cite{Freeman2009BayesianNA}. 

\section{\label{sec:methodology}Methodology}

\subsection{\label{subsec:baynet-primer}Primer on Bayesian networks}

Bayesian inference is a framework for inference under uncertainty.
The central formula in Bayesian statistics is given by Bayes theorem, e.g.,~\cite{siviaDataAnalysisBayesian1996},
\begin{equation}
    P(H\,|\,E) = \frac{P(E\,|\,H) P(H)}{P(E)} \,,
\end{equation}
where $H$ is an hypothesis and $E$ represents an observation.
For instance, $H$ could be an hypothesis from the set $\{H_1: \textrm{`It rained'}, H_2: \textrm{`It did not rain'}\}$ and the evidence $E$ an observation from a set of possible observations $\{\textrm{`Floor is wet'}, \textrm{`Floor is dry'}\}$.
The notation $P(.)$ denotes a probability, a number between zero and one, which is interpreted as degree of belief in Bayesian statistics.
The probability $P(H)$ is referred to as prior probability and represents the degree of belief that hypothesis $H$ is true without taking into account the evidence $E$.
The probability $P(E \,|\, H)$ denotes the probability for making the observation $E$ given that $H$ is true.
For instance, $P(\textrm{Floor is wet}\,|\,\textrm{It rained}) = 1$ if we know that rain always causes the floor to be wet. 
The probability $P(E)$ is referred to as marginal likelihood or model evidence and is the probability of making a specific observation $P(E)$ based on the modeling assumptions.
The probability $P(H\,|\,E)$ is referred to as posterior probability and represents the degree of belief that $H$ is true after the observation $E$ has been made. 

The sets we have introduced above are discrete and we can exhaustively enumerate their elements.
However, for continuous quantities, such as temperatures or cross sections in nuclear data, the notion of probability needs to be replaced by the concept of probability density function.
Given a probability density function $\rho(x)$, the probability that a value is within the range enclosed by $a$ and $b$ is given by
\begin{equation}
    P(a \le x \le b) = \int_{a}^b \rho(x) dx \,.
\end{equation}
Due to this relationship, probability density functions must be non-negative everywhere and their integral over the full range one.
We do not emphasize the distinction between probabilities and probability density functions in the following because this aspect is not crucial for the concept of Bayesian networks.

Even though the Bayesian theorem in its basic form is at the core of Bayesian inference as it formalizes the mechanics of learning from experience, it only deals with a very simple case.
In the following we generalize the perspective on the inference problem to prepare the discussion of Bayesian networks.
The distinction between hypotheses and evidence is to some extent arbitrary.
For instance, the possible observation `Floor is wet' we mentioned as an example is also a hypothesis, but we named it evidence because it is an hypothesis that we observed to be true. 
In contrast, which hypothesis $H$ among the possible ones is true was not directly observed and Bayes theorem allows us to indirectly improve our knowledge about their likelihood by using the evidence.

Therefore, at a higher level of abstraction, we can equally say that we have a system of hypotheses and the Bayesian inference framework provides a mechanism to update our knowledge about their likelihood by observing some of the hypotheses to be true or false.
Assume that we have a number of variables $H_1, H_2, H_3, \dots$ with each one being an hypothesis from a certain class of hypotheses.
For instance, $H_1$ could be the value of the elastic cross section, $H_2$ the value of the non-elastic cross section and $H_3$ the value of the total cross section.
Their joint prior probability density function is denoted by $P(H_1, H_2, H_3)$.
To give an example of the form of the Bayesian update formula in the case of three variables, assume that we know the elastic cross section $H_1$.
The Bayesian update formula takes then the form
\begin{equation}
    P(H_2, H_3 \,|\, H_1) = \frac{P(H_1 \,|\, H_2, H_3) P(H_2, H_3)}{P(H_1)} \,.
\end{equation}

Even though the Bayesian update formula can be written down for an arbitrary number of variables in this way, typical inference tasks with many variables pose computational challenges.
A relevant inference query is to find an assignment of values to $H_2$ and $H_3$ that maximizes the value of the posterior probability density function $P(H_2, H_3\,|\,H_1)$.
This assignment is called a maximum a posteriori probability (MAP) estimate.

The idea of Bayesian networks is to exploit the structure of the relationships between variables.
Every joint probability distribution can be decomposed into a product of conditional probability distributions.
For instance, for three variables, we can write the joint probability density function as
\begin{equation}
    \label{eq:jointprobdecomp}
    P(H_1, H_2, H_3) = 
    P(H_1) P(H_2 \,|\, H_1) P(H_3 \,|\, H_1, H_2) \,.
\end{equation}
In this product, going from left to right, each conditional probability distribution is conditioned on all variables that appeared before.
These conditional dependencies can be visualized as a Bayesian network:
\begin{center}
\tikz{
    \node[latent](H3){$H_3$};
    \node[latent, left=of H3, yshift=-0.5cm](H2){$H_2$};
    \node[latent, left=of H2, yshift=0.5cm](H1){$H_1$};
    \edge{H1,H2}{H3}
    \edge{H1}{H2}
}
\end{center}
Variables (or a collection of variables of the same type) are associated with nodes and directed arrows go from the variables being conditioned on to the dependent variables.
Bayesian networks are not allowed to have cycles, hence it is impossible to arrive at the same node twice by following the arrows along their pointing direction.

In the product in \cref{eq:jointprobdecomp} conditional dependencies can be dropped if variables are \textit{conditionally independent}.
Two random variables $X$ and $Y$ are conditional independent given a third random variable $Z$ if
\begin{equation}
    P(X \,|\, Y, Z) = P(X \,|\, Z)
    \;\;\textrm{and}\;\;
    P(Y \,|\, X, Z) = P(Y \,|\, Z) \,.
\end{equation}
In words, knowledge about the value of $Y$ does not improve the knowledge about the value of $X$ if the value of $Z$ is already known.

In the example of the three hypotheses: If $H_2$ and $H_3$ are conditionally independent given $H_1$, \cref{eq:jointprobdecomp} simplifies to
\begin{equation}
    P(H_1, H_2, H_3) = P(H_1) P(H_2 \,|\, H_1) P(H_3 \,|\, H_1) \,.
\end{equation}
and the corresponding Bayesian network loses one arrow:
\begin{center}
\tikz{
    \node[latent](H1){$H_1$};
    \node[latent, right=of H1, yshift=0.5cm](H2){$H_2$};
    \node[latent, right=of H1, yshift=-0.5cm](H3){$H_3$};
    \edge{H1}{H2,H3}
}
\end{center}

Please note that there is not a unique way to factorize a joint probability distribution.
For instance, for two variables we can choose $P(H_1, H_2) = P(H_1) P(H_2 \,|\, H_1)$ or $P(H_1, H_2) = P(H_2) P(H_1 \,|\, H_2)$.
Consequently, there are several Bayesian network topologies to describe the same joint probability distribution which usually differ in both the number of arrows and their orientations.
An objective in Bayesian network modeling is therefore to find a topology which is simple, e.g., possesses a smaller number of connections or a topology with arrows reflecting the causal relationships between variables.
The specific topology of Bayesian network can be exploited in Bayesian inference.

Bayesian networks may be classified according to whether variables are continuous or discrete in the network or the type of prior distribution imposed on variables, e.g., Gaussian distributions.
Three general inference tasks can be identified, e.g.,~\cite{koller_probabilistic_2009}:
\begin{enumerate}
    \item Determinig the links between variables and their orientation,
    \item Estimating the parameters of the conditional prior distributions, and,
    \item Inferring the values of unobserved variables.
\end{enumerate}

In this paper, we restrict ourselves to Bayesian networks with all prior conditional distributions being multivariate normal and possibly non-linear relationships between variables.
We deal only with the inference of unobserved variables (3), which is the essential inference task in nuclear data evaluation.
To develop the concepts and notation, we first review the Generalized Least Squares method, already linking it to Bayesian networks, and afterwards extend the discussion to nested and non-linear relationships.

\subsection{\label{subsec:glsrecap}Generalized Least Squares recapitulated}

The Bayesian version of the Generalized Least Squares (GLS) method, e.g.,~\cite{muir_evaluation_1989}, is usually regarded as a method to update the parameters of a model based on data from experiments affected by statistical and systematic errors.
In the nuclear data context, parameters may be the cross sections associated with the points of an energy mesh or the parameters of a nuclear physics model.
This view suggests that model parameters and experimental errors are quantities of different nature both from the point of physics and Bayesian statistics.
However, from the point of Bayesian statistics, all these quantities are variables with uncertain values and can be treated in the same way.
In order to introduce the Bayesian network interpretation, we present the GLS formulas without regard to the particular meaning of the variables to emphasize this symmetry.

The GLS method is based on the multivariate normal distribution, which is of the form
\begin{multline}
    \label{eq:mvndef}
    \mathcal{N}(\vec{x} \,|\, \vec{\mu}, \mat{\Sigma}) =
    \frac{1}{\sqrt{(2\pi)^N \det \mat{\Sigma}}}
    \\
    \exp\left(
    -\frac{1}{2}
    (\vec{x}-\vec{\mu})^T
    \mat{\Sigma}^{-1}
    (\vec{x}-\vec{\mu})
    \right) \,,
\end{multline}
and characterized by a mean vector $\vec{\mu}$ and a covariance matrix $\mat{\Sigma}$.
We use the notation $\vec{x} \sim \mathcal{N}(\vec{\mu}, \mat{\Sigma})$ to indicate that the random vector $\vec{x}$ is distributed according a multivariate normal distribution with mean vector $\vec{\mu}$ and covariance matrix~$\mat{\Sigma}$, hence its probability density function given by~\cref{eq:mvndef}.

Assume we have a set of variables $\{y_i\}_{i=1..N}$ whose values are uncertain and we assemble these variables to a vector $\vec{y}$ . 
Some variables are either completely or to a certain degree determined by other variables.
For instance, the values of the parameters of a deterministic nuclear model determine completely the resulting predictions.
In contrast, for a stochastic nuclear model, the model parameters certainly influence the predictions but do not completely determine them due to the stochastic nature of the simulation.
In a nuclear experiment, the measurement is influenced by the underlying true values of nuclear properties but not completely determined due to various measurement errors.

To account for these dependencies, we partition the variables in $\vec{y}$ into two subvectors, $\vec{y}_I$ and $\vec{y}_J$.
We refer to variables whose indices are in the set $I$ as independent variables and assume them to be governed by a multivariate normal distribution, i.e., $\vec{y}_I \sim \mathcal{N}(\vec{u}_I, \mat{U}_{I,I})$.
The variables with their indices being in $J$ are referred to as dependent variables and they are assumed to be functions of the independent variables,
\begin{equation}
    \vec{y}_J =
    f(\vec{y}_I, \vec{\tau}) =
    g(\vec{y}_I) + \vec{\tau}
    \label{eq:stochastic_link} \,.
\end{equation}
We introduced the random variable $\vec{\tau}$ to allow for stochastic perturbation of the otherwise deterministic link.
The vectors $\vec{\tau}$ and $\vec{y}_J$ are of the same size.
Please note that the variables in $\vec{\tau}$ have also to be regarded as independent variables, even though their indices are not included in $I$ for notational reasons.
In nuclear data evaluation, $\vec{\tau}$ typically contains the statistical errors.
We assume $\vec{\tau}$ to be governed by a multivariate normal distribution, i.e., $\vec{\tau}\sim\mathcal{N}(\vec{u}_J, \mat{U}_{J,J})$, and the random variable $\vec{y}_I$ to be independent of $\vec{\tau}$.
With these specifications, the joint probability distribution $P(\vec{y}_J, \vec{y}_I, \vec{\tau})$ factorizes therefore into
\begin{equation}
    P(\vec{y}_J, \vec{y}_I, \vec{\tau}) = P(\vec{y}_I) P(\vec{\tau}) P(\vec{y}_J \,|\, \vec{y}_I, \vec{\tau}) 
\end{equation}
with the following corresponding Bayesian network:
\begin{center}
\tikz{
    \node[latent](H3){$\vec{y}_J$};
    \node[latent, below=of H3, yshift=0.7cm, xshift=-0.5cm](H2){$\vec{\tau}$};
    \node[latent, left=of H3, xshift=0.0cm](H1){$\vec{y}_I$};
    \edge{H1,H2}{H3}
}
\end{center}


A non-linear relationships $g(\vec{y}_I,\vec{\tau})$ in \cref{eq:stochastic_link} can be cast into the GLS framework by constructing a linear Taylor approximation.
A linear Taylor approximation of \cref{eq:stochastic_link} is of the form
\begin{equation}
    \label{eq:stochastic_linear_link}
    \vec{y}_J = \vec{y}_{\textrm{ref},J} +
    \mat{T} \, 
    \left(
      \vec{y}_I - \vec{y}_{\textrm{ref},{I}}
    \right) +
    (\vec{\tau} - \vec{\tau}_{\textrm{ref}})
\end{equation}
where $\vec{y}_{\textrm{ref},J} = f(\vec{y}_{\textrm{ref},I},\vec{\tau}_{\textrm{ref}})$ and $\mat{T}$ is the Jacobian matrix of $f$ evaluated at $\vec{y}_{\textrm{ref},I}$.
The Jacobian matrix is somestimes also referred to as sensitivity matrix.
At this point it may be regarded as unnecessary to introduce a potentially non-zero vector $\vec{\tau}_\textrm{ref}$ because $\vec{\tau}$ is an additive (i.e., also linear) contribution to the result. 
However, this term will become important later in the treatment of nested relationships.

Depending on the specific construction of $\mat{T}$, this equation may implement interpolation using splines, a Legendre polynomial, a Fourier polynomial or any other type of polynomial.
For instance, for spline interpolation $\mat{T}$ may map the values at the knot points given in $\vec{y}_I$ to the locations of the data points $\vec{y}_J$.
In the case of a Legendre polynomial, the Jacobian $\mat{T}$ maps the values of the Legendre coefficients given in $\vec{y}_I$ to the function values at the positions of the data points whose values are given in $\vec{y}_J$.
\Cref{eq:stochastic_linear_link} is therefore general and can accomodate many different types of interpolation schemes.
Importantly, this equation as a building block does not only cover interpolation but, for instance, is also used to distribute normalization errors given in $\vec{y}_I$ to the data points in $\vec{y}_J$.

It will be helpful to have a compact formula to express $\vec{y}$, i.e., both $\vec{y}_I$ and $\vec{y}_J$, as a function of the independent variables $\vec{y}_I$ and $\vec{\tau}$.
In other words, we want to extend \cref{eq:stochastic_linear_link} so that its result is the full vector $\vec{y}$.
To this end, we introduce a vector $\vec{z}$ of the same size as $\vec{y}$ whose components are given by
\begin{equation}
    \label{eq:zdef}
    \vec{z}_{I} = \vec{y}_I \;\;\;\;\textrm{and}\;\;\;\; \vec{z}_{J} = \vec{\tau} \,.
\end{equation}
This vector $\vec{z}$ contains all independent variables, which completely determine the values in $\vec{y}_J$. 

Analogously, we also introduce a vector $\vec{z}_\textrm{ref}$ with elements given by
\begin{equation}
    \vec{z}_{\textrm{ref},I} = \vec{y}_{\textrm{ref},I} \;\;\;\;\textrm{and}\;\;\;\; \vec{z}_{\textrm{ref},J} = \vec{\tau}_\textrm{ref} \,.
\end{equation}
Finally, we regard $\mat{T}$ as the respective block in a larger matrix $\mat{S}$ whose blocks are defined as
\begin{equation}
    \label{eq:matS_definition}
    \mat{S}_{I,I} = \mathbb{1}, \;\;\;\;
    \mat{S}_{I,J} = \mat{0} \;\;\;\;
    \mat{S}_{J,I} = \mat{T}, \;\;\;\;
    \mat{S}_{J,J} = \mathbb{1} \,.
\end{equation}
Please note that even though we use the term block, the ordering of the rows and columns does not matter, as long as the index sets $I$ and $J$ do not contain common indices and all indices from $1$ to $N$ are present in $A = I \cup J$.
The same is true for the vectors $\vec{z}$ and $\vec{z}_{\textrm{ref}}$.

The introduction of these quantities allows us to rewrite \cref{eq:stochastic_linear_link} to cover both dependent and independent variables,
\begin{equation}
    \label{eq:baynet_mapping}
    \vec{y} = \vec{y}_{\textrm{ref}} +
    \mat{S} \, 
    \left(
      \vec{z} - \vec{z}_{\textrm{ref}}
    \right) \,.
\end{equation}
Regarding the Bayesian network interpretation, this equation propagates the values of the variables without parent nodes, i.e., $\vec{y}_I$ and $\vec{\tau}$ to all the other variables.
Now with all variables and their mutual relationships defined, we can deal with the question of inference in the Bayesian framework.

Bayesian statistics can be seen as a framework to update the knowledge about the possible values of variables based on observation of some variables.
Using our notation, the Bayesian inference formula to achieve this update is given by
\begin{equation}
    P(\vec{y}_I \,|\, \vec{y}_J) = \frac{1}{P(\vec{y}_I)} P(\vec{y}_J \,|\, \vec{y}_I) P(\vec{y}_I)
\end{equation}
where we marginalized over $\vec{\tau}$ to treat it as a nuisance variable, i.e., $P(\vec{y}_J \,|\, \vec{y}_I) = \int P(\vec{\tau}) P(\vec{y}_J \,|\, \vec{y}_I, \vec{\tau}) d \vec{\tau}$.

The GLS method is a special case of the general Bayesian update formula to update the knowledge about $\vec{y}_I$ based on the observed values of $\vec{y}_J$ in \cref{eq:stochastic_link} if the following assumptions are met:
\begin{itemize}
    \item The function $f(.)$ in \cref{eq:stochastic_link} is linear, i.e., it can be exactly represented by \cref{eq:stochastic_linear_link}
    \item The prior distribution of $\vec{y}_I$ and $\vec{\tau}$ are multivariate normal and no prior correlations exist between elements in $\vec{y}_I$ and elements in $\vec{\tau}$
\end{itemize}
The resulting posterior distribution $P(\vec{y}_I\,|\, \vec{y}_J)$ is then also multivariate normal and the posterior mean vector and covariance matrix can be obtained by analytic matrix formulas.

To state the GLS formulas, we introduce the prior mean vector $\vec{u}$ and prior covariance matrix $\mat{U}$ which characterize the prior on $\vec{y}_I$ and $\vec{\tau}$ simultaneously.
The priors on these vectors have been already introduced along with~\cref{eq:stochastic_link}, which we repeat here for the convenience of the reader: $\vec{y}_I \sim \mathcal{N}(\vec{u}_I, \mat{U}_{I,I})$ and $\vec{\tau} \sim \mathcal{N}(\vec{u}_J, \mat{U}_{J,J})$.
The matrices $\mat{U}_{I,I}$ and $\mat{U}_{J,J}$ are now considered to be the blocks of the compound covariance matrix $\mat{U}$ at the positions defined by index sets $I$ and $J$, respectively.
Due to the independence assumption between $\vec{y}_I$ and $\vec{\tau}$, we have $\mat{U}_{I,J}=\mat{U}_{J,I}^T=\mat{0}$.
Let us assume that we have observed the values of the vector $\vec{y}_{J}$, i.e., $\vec{y}_{J}=\vec{r}$.
Observed nodes in the Bayesian network are colored grey:
\begin{center}
\tikz{
    \node[obs](H3){$\vec{y}_J$};
    \node[latent, below=of H3, yshift=0.7cm, xshift=-0.5cm](H2){$\vec{\tau}$};
    \node[latent, left=of H3, xshift=0.0cm](H1){$\vec{y}_I$};
    \edge{H1,H2}{H3}
}
\end{center}
The update formula to obtain the posterior covariance matrix $\mat{U}_{I,I}'$ is given by
\begin{equation}
    \label{eq:gls_upd_cov}
    \mat{U}_{I,I}' = \left(
        \mat{S}_{J,I}^T \mat{U}_{J,J}^{-1} \mat{S}_{J,I} + \mat{U}_{I,I}^{-1}
    \right)^{-1} \,.
\end{equation}
and the update formula for the posterior mean vector by
\begin{multline}
    \label{eq:gls_upd_mean}
    \vec{u}_I' = \vec{y}_{\textrm{ref},I} + 
    \mat{U}_{I,I}'
    \Big(
         \mat{S}_{J,I}^T \mat{U}_{J,J}^{-1}  (\vec{r} - \vec{y}_{\textrm{ref},J}) \\
         + \mat{U}_{I,I}^{-1} (\vec{u}_I - \vec{y}_{\textrm{ref},I})
    \Big) \,.
\end{multline}
A derivation of~\cref{eq:gls_upd_cov,eq:gls_upd_mean} can be found in~\cref{app:GLS-LM-derivation}.

These equations can be written in more compact form.
The assumption of zero cross-covariance elements between $\vec{y}_I$ and $\vec{\tau}$, i.e., $\mat{U}_{I,J}=\mat{U}_{J,I}^T=\mat{0}$, allows us to write
\begin{equation}
    \mat{U}_{I,I}' = \left(
        \mat{S}_{A,I}^T \mat{U}_{A,A}^{-1} \mat{S}_{A,I}
    \right)^{-1} \,.
\end{equation}
If in addition taking into account the structure of $\mat{S}$ in \cref{eq:matS_definition} and defining the vector $\vec{v}$ with elements 
\begin{equation}
    \vec{v}_I = \vec{u}_I ,\, \vec{v}_J = \vec{r} \,,
\end{equation}
we can rewrite \cref{eq:gls_upd_mean} as
\begin{equation}
    \vec{u}_I' = \vec{y}_{\textrm{ref},I} + 
    \mat{U}_{I,I}'
         \mat{S}_{A,I}^T \mat{U}_{A,A}^{-1}  (\vec{v} - \vec{y}_{\textrm{ref}})
    \,.
\end{equation}
The vector $\vec{u}_I'$ and the covariance matrix $\mat{U}_{I,I}'$ are the updated distribution parameters of the multivariate normal distribution associated with $\vec{y}_I$ due to observing the values in $\vec{y}_J$, i.e., $\vec{y}_I \sim \mathcal{N}(\vec{u}_I', \mat{U}_{I,I}')$

To discuss the application of the GLS method for a Bayesian network with more nodes, we consider the following scenario for the purpose of illustration.
So far we did not impose further constraints on the prior of $\vec{\tau}$ except being a multivariate normal distribution.
If, however, $\vec{\tau}$ can be decomposed into two a priori mutually independent blocks $\vec{\tau}_K, \vec{\tau}_L$, i.e., $\mat{U}_{K,L} = (\mat{U}_{L,K})^T = \mat{0}$, we may split up \cref{eq:stochastic_linear_link} into two equations:
\begin{align}
    \vec{y}_K &= \vec{y}_{\textrm{ref},K} +
    \mat{T}_{K,I} \, 
    \left(
      \vec{y}_I - \vec{y}_{\textrm{ref},{I}}
    \right) +
    (\vec{\tau}_K - \vec{\tau}_{\textrm{ref},K}) \,,
\\
    \vec{y}_L &= \vec{y}_{\textrm{ref},L} +
    \mat{T}_{L,I} \, 
    \left(
      \vec{y}_I - \vec{y}_{\textrm{ref},{I}}
    \right) +
    (\vec{\tau}_L - \vec{\tau}_{\textrm{ref},L}) \,.
\end{align}
If we regard the values in $\vec{y}_I$ as fixed, the remaining contribution to $\vec{y}_K$ and $\vec{y}_L$ given by $\vec{\tau}_K$ and $\vec{\tau}_L$, respectively, are independent of each other due to our independence assumption for the blocks in $\vec{\tau}$ above.
Due to this conditional independence of $\vec{y}_K$ and $\vec{y}_L$ there is no arrow between $\vec{y}_K$ and $\vec{y}_L$.
Considering this conditional independence and the values in $\vec{y}_K$ and $\vec{y}_L$ being given as a function of $\vec{y}_I$, we have the following Bayesian network:
\begin{center}
\tikz{
    \node[obs](H3){$\vec{y}_K$};
    \node[latent, below=of H3, yshift=0.7cm](H4){$\vec{y}_L$};
    \node[latent, left=of H3, xshift=0.5cm, yshift=0.3cm](H2){$\vec{\tau}_K$};
    \node[latent, left=of H4, xshift=0.5cm, yshift=-0.3cm](H5){$\vec{\tau}_L$};
    \node[latent, left=of H3, xshift=0.0cm, yshift=-0.5cm](H1){$\vec{y}_I$};
    \edge{H1,H2}{H3}
    \edge{H1,H5}{H4}
}
\end{center}
In this scenario, one may assume that only the variables in $\vec{y}_K$ have been observed, i.e., the variables in the vector $\vec{y}$ referred to by index set K.

Keeping this example in mind, we want to discuss the general form of the GLS equations for a Bayesian network without intermediate nodes in which some of the terminal nodes have been observed, which amounts to knowing the values of some variables in the vector $\vec{y}_J$ in~\cref{eq:stochastic_linear_link}.
The treatment of nested relationship will be discussed later in~\cref{subsec:nested_relationships}.

We denote by $D$ the index set referring to the observed values in $\vec{y}$, i.e., $\vec{y}_D$ is the subvector with observed values.
For instance, in the example above we can identify $D = K$.
Let us denote by $\neg{D}$ the index set with all indices that are not in $D$, i.e., $\neg{D} = A \setminus D$ with $A = \{1,\dots,N\}$.
In this case, we can consider \cref{eq:baynet_mapping} to realize that the variables in the vector $\vec{z}_{\neg D}$, which includes all variables in $\vec{y}_I$ and a subset of the variables in $\vec{\tau}$, can be regarded as the independent variables.
The dependent and measured variables are now those in $\vec{y}_D$.
For this more general case, the structure of the update formulas remains the same.

For the posterior covariance matrix we have
\begin{equation}
    \label{eq:postcov-notDnotD}
    \mat{U}_{\neg D,\neg D}' = \left(
        \mat{S}_{D,\neg D}^T \mat{U}_{D,D}^{-1} \mat{S}_{\neg D,D} + \mat{U}_{\neg D,\neg D}^{-1}
    \right)^{-1} \,,
\end{equation}
and for the posterior mean vector
\begin{multline}
    \vec{u}_{\neg D}' = \vec{z}_{\textrm{ref},\neg D} + 
    \mat{U}_{\neg D,\neg D}'
    \Big(
         \mat{S}_{D,\neg D}^T \mat{U}_{D,D}^{-1}  (\vec{r} - \vec{y}_{\textrm{ref},D}) \\
         + \mat{U}_{\neg D,\neg D}^{-1} (\vec{u}_{\neg D} - \vec{z}_{\textrm{ref},\neg D})
    \Big) \,.
\end{multline}
If we assume that the random variables in $\vec{z}_{D}$ are independent of those in $\vec{z}_{\neg D}$, we can again simplify the equations. 
Later when dealing with Bayesian networks with more nodes, this independence assumption will be fulfilled because variables of the form $\vec{\tau}$ associated with different nodes are a priori independent and all variables associated with each single node are either observed or unobserved at the same time.
We do not consider situations in which only a part of the variables associated with a node is observed.

If the observed values are $\vec{y}_{D} = \vec{r}$,
the vector $\vec{v}$ is now defined by
\begin{equation}
    \label{eq:vvec_general}
    \vec{v}_{\neg D} = \vec{u}_{\neg D} ,\; \vec{v}_D = \vec{r} \,.
\end{equation}
In addition we define a vector $\vec{v}_{\textrm{ref}}$ as
\begin{equation}
    \vec{v}_{\textrm{ref},\neg D} = \vec{z}_{\textrm{ref},\neg D} ,\;
    \vec{v}_{\textrm{ref},D} = \vec{y}_{\textrm{ref},D} \,.
\end{equation}
With these specifications, the posterior covariance matrix of the independent variables $\vec{z}_{\neg D}$ can be written as
\begin{equation}
    \label{eq:baynet_upd_cov}
    \mat{U}_{\neg D,\neg D}' = \left(
        \mat{S}_{A,\neg D}^T \mat{U}_{A,A}^{-1} \mat{S}_{A,\neg D}
    \right)^{-1} \,,
\end{equation}
and the posterior mean vector as
\begin{multline}
    \label{eq:baynet_upd_mean}
    \vec{u}_{\neg D}' = \vec{z}_{\textrm{ref},\neg D} + 
    \mat{U}_{\neg D,\neg D}'
         \mat{S}_{D,A}^T \mat{U}_{A,A}^{-1}  (\vec{v} - \vec{v}_{\textrm{ref}}) \,.
\end{multline}
The outer inversion in \cref{eq:baynet_upd_cov} has never to be evaluated explicitly.
In~\cref{subsec:postcov-approx}, we discuss how blocks of the posterior covariance matrix of interest can be computed and how to draw samples from the posterior distribution.
The multiplication with $\mat{U}_{\neg D, \neg D}'$ in \cref{eq:baynet_upd_mean} can be reformulated as the task to solve a system of linear equations,
\begin{align*}
    \mat{A} \vec{x} = \vec{b}
    \;\;\;\textrm{with}\;\;\;
    \mat{A} &= \mat{S}_{A,\neg D}^T \mat{U}_{A,A}^{-1} \mat{S}_{A,\neg D} \\
    \vec{b} &= \mat{S}_{D,A}^T \mat{U}_{A,A}^{-1}  (\vec{v} - \vec{v}_{\textrm{ref}}) \,,
\end{align*}
with the solution vector $\vec{x}$ representing the result of the matrix product in \cref{eq:baynet_upd_mean}.
As all matrices $\mat{U}^{-1}$, $\mat{S}$, and $\mat{S}_{A,\neg D}^T \mat{U}_{A,A}^{-1} \mat{S}_{A,\neg D}$ are typically sparse, we employ sparse matrix algorithms to greatly speed up the computation of the posterior expectation vector $\vec{u}_{\neg D}'$.

Please note that so far we have only recapitulated one version of the formulas of the Generalized Least Squares method and introduced specific notation.
These formulas will be the basis for inference in Bayesian networks with linear relationships and with a slight modification explained in~\cref{subsec:nonlinearfuns} also in the case of non-linear relationships.
Using these formulas implies that observed nodes are always attached to a vector $\vec{\tau}$ with non-zero prior uncertainties.
We feel that this assumption is not a serious limitation in the context of nuclear data evaluation because measurements are always associated with a statistical uncertainty component.
If this is not the case, the prior uncertainty associated with the elements of $\vec{\tau}$ can be made very small which effectively eliminates the impact of these noise node on the results although it may be detrimental to numerical stability. 

Deterministic relationships, i.e., those associated with a zero diagonal element in the covariance matrix $\mat{U}$, must be treated as a special case because $\mat{U}^{-1}$ is not defined anymore.
Assume that $Z$ is the subset of $\neg D$ associated with zero prior uncertainties and $\neg Z$ is the subset of $\neg D$ with non-zero prior uncertainties.
The approach to deal with deterministic relationships is to use $\neg Z$ instead of $\neg D$ in~\cref{eq:baynet_upd_mean} and~\cref{eq:baynet_upd_cov}.
Afterward the full posterior mean vector $\vec{u}_{\neg D}$ and covariance matrix $\mat{U}_{\neg D,\neg D}$ can be obtained by imputing the prior mean vector and covariance matrix to the missing blocks,
\begin{equation}
    \label{eq:baynet_upd_deterministic}
    \vec{u}_{Z}' = \vec{u}_Z \;\;\;\textrm{and}\;\;\;
    \mat{U}_{Z,Z}' = \mat{0},
    \mat{U}_{Z,\neg{Z}} = (\mat{U}_{\neg Z,Z})^T = \mat{0} \,.
\end{equation}

In order to obtain the posterior expectation $\vec{u}_{D}$ of $\vec{z}_D$, we note that the known vector $\vec{r}$ of the observed node is given by the sum of the posterior expectation $\vec{u}_D$ and the posterior expectations in $\vec{u}_{\neg D}$ propagated to the observed node by means of~\cref{eq:baynet_mapping}.
With the result of the propagation given by
\begin{equation}
    \label{eq:baynet_propagate_neg_d}
    g_{D}(\vec{u}_{\neg D}) := 
    \vec{y}_{\textrm{ref},\neg D} +
    \mat{S}_{\neg D,D}
    (\vec{u}_{\neg D}' - \vec{z}_{\textrm{ref},\neg D})
\end{equation}
we can therefore compute the posterior expectation $\vec{u}_{D}'$ by
\begin{equation}
    \label{eq:compute_statistical}
    \vec{u}_{D}' = \vec{r} - g_{D}(\vec{u}_{\neg D}) \,.
\end{equation}

To compute the posterior covariance matrix blocks $\mat{U}_{D,D}$ and $\mat{U}_{\neg D, D}$, we remark that the following relationships for the covariance matrix of random vectors $\vec{x}, \vec{y}, \vec{z}$ hold: 
\begin{align*}
    \Cov{\vec{x}}{\vec{y}} &= \left( \Cov{\vec{y}}{\vec{x}} \right)^T, \\
    \Cov{\vec{x}+\vec{y}}{\vec{z}} &= \Cov{\vec{x}}{\vec{z}} + \Cov{\vec{y}}{\vec{z}}, \\
    \Cov{\mat{S}\vec{x}}{\vec{y}} &= \mat{S} \Cov{\vec{x}}{\vec{y}} \,,
\end{align*}
where the notation $\Cov{\vec{x}}{\vec{y}}$ denotes the covariance matrix between random vectors $\vec{x}$ and $\vec{y}$.

Introducing the abbreviation $\Var{\vec{x}}:=\Cov{\vec{x}}{\vec{x}}$ and calculating the variance of both sides of the equation $\vec{z}_D  = \vec{r} - g_D(\vec{z}_{\neg D})$, we find that $\Var{\vec{z}_D} = \Var{g_D(\vec{z}_{\neg D})}$ and in terms of the corresponding posterior covariance matrix blocks:
\begin{equation}
    \label{eq:baynet_upd_U_D_D}
    \mat{U}_{D,D}' = \mat{S}_{D,\neg D} \mat{U}_{\neg D, \neg D}' \left(\mat{S}_{D,\neg D}\right)^T
\end{equation}
The application of the variance operator on both sides of the rearranged equation $\vec{z}_D + g_D(\vec{z}_{\neg D}) = \vec{r}$ yields
\begin{equation}
    \label{eq:baynet_upd_U_D_negD}
    \mat{U}_{D,\neg D}' = \left( \mat{U}_{\neg D, D}' \right)^T = \mat{S}_{D,\neg D} \mat{U}_{\neg D, \neg D}' \,.
\end{equation}
In summary, with \cref{eq:baynet_upd_cov,eq:baynet_upd_mean,eq:baynet_upd_deterministic,eq:baynet_propagate_neg_d,eq:compute_statistical,eq:baynet_upd_U_D_D,eq:baynet_upd_U_D_negD} we obtain the posterior expectation $\vec{u}'$ and covariance matrix $\mat{U}'$ for the full vector of independent variables $\vec{z}$ by conditioning on the known vector $\vec{r}$ of one (or several) observed nodes for both stochastic and deterministic linear functional relationship between variables.
The posterior expectation $\vec{y}_\textrm{post}$ associated with the dependent nodes in $\vec{y}$ can be obtained by using \cref{eq:baynet_mapping} with the posterior expectation of independent variables:
\begin{equation}
    \vec{y}_\textrm{post} = \vec{y}_\textrm{ref} + \mat{S} \left( \vec{u}' - \vec{z}_\textrm{ref} \right)
\end{equation}
The posterior covariance matrix $\mat{Y}_\textrm{post}$ of dependent nodes is given by
\begin{equation}
    \mat{Y}_\textrm{post} = \mat{S} \mat{U}' \mat{S}^T \,.
\end{equation}

The important feature of these formulas is that all matrices involved are typically sparse in nuclear data evaluation if we do not absorb functional relationships, such as those for normalization errors into the covariance matrix but explicitly keep them in $\mat{S}$.
The matrix $\mat{S}$ contains the information about the connection structure of the Bayesian network.
In this paper, we do not need specialized algorithms for Bayesian networks to exploit the connection structure but can rely on general libraries for sparse matrix operations instead.
In the next section, we elaborate on ways to deal with the large posterior covariance matrix by exploiting sparsity.
Afterwards, we focus on the Bayesian network interpretation and discuss the proper handling of nested and non-linear relationships.

\subsection{\label{subsec:postcov-approx}Evaluating the posterior covariance matrix}

When working with ten or even hundred thousands of variables, it is often not practical to compute or store the full posterior covariance matrix explicitly.
Even if we just need to evaluate the posterior covariance block $\mat{U}_{I,I}'$ of independent variables, it may be still too large in practice.

In nuclear data evaluation, the ability to compute at least the posterior covariance matrix block associated with the evaluated cross sections is very important because this information is needed for linear error propagation through application codes.
However, sometimes linear error propagation based on the covariance matirx is not feasible with a reasonable amount of computational resources.
For instance, if we have to use a very fine energy mesh, e.g., several tens of thousands of mesh points, in energy regions with quickly oscillating cross sections, even the handling of the posterior covariance matrix block associated with cross sections becomes impractical.
In such cases, we may prefer to draw samples from the posterior distribution and to subsequently propagate these samples through application codes according to the \textit{Total Monte Carlo} (TMC) method~\cite{koning_towards_2008,rochmanUncertaintiesCriticalitySafetyBenchmarks2009,rochmanEvaluation23NaNeutroninduced2010}.

Therefore, feasible approaches to evaluate blocks of the posterior covariance matrix and to sample from the posterior distribution using the full posterior covariance matrix are both important in the field of nuclear data. 

Following we explain the computation of blocks from the posterior covariance matrix and the sampling using the full posterior covariance matrix.
For both requirements, we exploit the key insight that the inverse of the posterior covariance matrix in \cref{eq:baynet_upd_cov},
\begin{equation}
    \left( \mat{U}_{\neg D,\neg D}' \right)^{-1} =
        \mat{S}_{A,\neg D}^T \mat{U}_{A,A}^{-1} \mat{S}_{A,\neg D}
    \,,
\end{equation}
can be very sparse even though the posterior covariance matrix is not.
The reason is that off-diagonal elements in the posterior covariance matrix are non-zero if the associated variables are correlated, irrespective of whether this correlation is mediated by a third variable.
In contrast to that, elements in the inverse posterior covariance matrix are zero if the associated variables are conditionally independent given all other variables, e.g.,~\cite[Chap. 7.1.3]{koller_probabilistic_2009}.
In other words, if two variables are only correlated because they are both influenced by one or a set of other variables, they are conditionally independent. 
The approaches described in the following to obtain random samples from the posterior distribution and to calculate blocks of the full posterior covariance matrix rely on the property that the inverse posterior covariance matrix is sparse. 
This sparseness property is present in all the Bayesian network examples we present in \cref{subsec:example-usu,subsec:example-ironslow,subsec:example-ironfast} thanks to the sparse Gaussian process construction explained in~\cref{subsec:sparse-gp-construction} and the specific handling of normalization errors.

First we discuss one way to obtain samples from the posterior distribution that is amenable to optimization by exploiting sparsity.
To sample from a multivariate normal distribution $\vec{x} \sim \mathcal{N}(\vec{\mu}, \mat{\Sigma})$, we can first compute a Cholesky decomposition of the inverse covariance matrix $\mat{\Sigma}^{-1} = \mat{L}^T\mat{L}$ with $\mat{L}$ being an upper triagonal matrix.
Making use of the fact that the elements in $\vec{z} = \mat{L} \vec{x}$ are distributed according to a standard normal distribution, i.e., $z_i \sim \mathcal{N}(0,1)$, we can generate a vector $\vec{z}$ with elements from a standard normal distribution and solve the set of linear equations $\mat{L} \vec{y} = \vec{z}$.
Because $\mat{L}$ is an upper triagonal matrix, this system of linear equations can be efficiently solved using backward substitution, e.g.,~\cite[Chap. 2.2]{pressNumericalRecipesArt2007}.
A sample from the multivariate normal distribution is then given by $\vec{s} = \vec{\mu} + \vec{y}$. 

Therefore, we can make a sparse Cholesky decomposition of $\left( \mat{U}_{\neg D,\neg D}' \right)^{-1}$ in order to apply the approach described in the previous paragraph to obtain samples from the posterior distribution.
The essential idea of a sparse Cholesky decomposition of a sparse matrix $\mat{A}$ is to find a permutation matrix $\mat{P}$ so that the Cholesky decomposition of the permuted matrix $\mat{P}\mat{A}\mat{P}^T$ is sparse.
Several algorithms exist to find such a so-called fill-reducing permutation, e.g.,~\cite{davisDirectMethodsSparse2006a}, and a popular one is the minimum-degree algorithm, e.g.,~\cite{georgeEvolutionMinimumDegree1989}.
Without a permutation applied before the Cholesky decomposition, the resulting Cholesky factor $\mat{L}$ is in general not sparse.
We employed the \textit{Matrix} package~\cite{batesMatrixSparseDense2021} available in the programming language R, which itself relies on the \textit{CHOLMOD} routines~\cite{chenAlgorithm887CHOLMOD2008} written in C to perform the sparse Cholesky decomposition.
The sparsity of the Cholesky factor $\mat{L}$ does not only help to keep the storage requirement low but also speeds up the solution of the system of linear equations $\mat{L}\vec{x}=\vec{z}$.
Finally, because we only have obtained a sample of the variables associated with indices $\neg D$, we need to compute the missing variables associated with indices $D$ using \cref{eq:compute_statistical}.
Please note that the values in the resulting vector are realizations of the independent variables (those in $\vec{z}$ in the previous section).
To obtain a sample of the values at dependent nodes, we have to propagate these sampled values associated with independent nodes by means of \cref{eq:baynet_mapping}.

Although the computation of the full posterior covariance matrix may be intractable, it is still valuable to be able to compute posterior covariances between quantities of interest.
The sparse Cholesky decomposition of the inverse posterior covariance matrix also helps to solve this task.
We know that $\left( \mat{U}_{\neg D,\neg D}' \right)^{-1} = \mat{P}^T \mat{L}^T\mat{L}\mat{P}$ with $\mat{P}$ the permutation matrix and $\mat{L}$ an upper triagonal matrix.
We denote by $\vec{e}_i$ a vector with all elements being zero except the element at the i$^\textrm{th}$ position being one.
The element $c_{ij}$ in the $i^\textrm{th}$ row and $j^\textrm{th}$ column of the posterior covariance matrix is given by
\begin{multline}
    \vec{e}_i^T \left( \mat{U}_{\neg D,\neg D}' \right) \vec{e}_j =
    \vec{e}_i^T \left( (\mat{U}_{\neg D,\neg D}')^{-1} \right)^{-1} \vec{e}_j = \\
    \vec{e}_i^T \left( \mat{P}^T \mat{L}^T \mat{L} \mat{P} \right)^{-1} \vec{e}_j =
    \vec{e}_i^T \mat{P}^{-1} \mat{L}^{-1} \left( \mat{L}^T \right)^{-1} \left(\mat{P}^T\right)^{-1} \vec{e}_j = \\
    \left( (\mat{L}^T)^{-1} \left(\mat{P}^T\right)^{-1} \vec{e}_i \right)^T \left( \mat{L}^T \right)^{-1} \left(\mat{P}^T\right)^{-1} \vec{e}_j = \\
    \left( (\mat{L}^T)^{-1} \mat{P} \vec{e}_i \right)^T \left( \mat{L}^T \right)^{-1} \mat{P} \vec{e}_j
    \,,
\end{multline}
where we made use of the fact that for permutation matrices $\mat{P}^T=\mat{P}^{-1}$.
The multiplications $\left(\mat{L}^T\right)^{-1} \mat{P}\vec{e}_i$ and $\left(\mat{L}^T\right)^{-1}\mat{P}\vec{e}_j$ can be restated as the solution of the set of linear equations
\begin{equation}
    \mat{L}^T \vec{r}_i = \mat{P} \vec{e}_i \;\;\;\textrm{and}\;\;\;
    \mat{L}^T \vec{r}_j = \mat{P} \vec{e}_j
\end{equation}
which allows the computation of the covariance element as $c_{ij} = \vec{r}_i^T \vec{r}_j$.
Blocks of the posterior covariance matrix can be evaluated by assemblig the (column) vectors $\vec{e}_i$ with indices of interest to matrices $\mat{E}_I = (\vec{e}_{i_1}, \vec{e}_{i_2}, \dots)$ and $\mat{E}_J = (\vec{e}_{j_1},\vec{e}_{j_2}, \dots)$, and then solve the set of linear equations
\begin{equation}
    \mat{L}^T \mat{R}_I = \mat{P} \mat{E}_I \;\;\;\textrm{and}\;\;\;
    \mat{L}^T \mat{R}_J = \mat{P} \mat{E}_J
\end{equation}
to finally compute $\mat{C}_{I,J} = (\mat{R}_I)^T \mat{R}_J$.
This covariance block is associated with a subset of the independent variables in $\vec{z}_{\neg D}$.
To calculate the covariance matrix block associated with the independent variables in $\vec{z}_{D}$, we can use \cref{eq:baynet_upd_U_D_D}.
The product in this equation can be computed by solving the set of linear equations
\begin{equation}
    \left( \mat{U}_{\neg D, \neg D}' \right)^{-1} \mat{M} = \mat{S}_{D, \neg D}
\end{equation}
and then computing $\mat{S}_{D,\neg D} \mat{M}$.
Instead of using the index set $D$, it is perfectly possible to use only a subset of D comprising only the indices associated with variables of interest.

\subsection{A simple Bayesian network}

Besides computational advantages thanks to exploiting sparsity, the strength of Bayesian networks is their interpretability.
Instead of formalizing an estimation problem in nuclear data evaluation in terms of an experimental covariance matrix and parameter prior covariance matrix, hence converting all functional relationships between variables into elements of a covariance matrix, we maintain and work explicity with these relationships.
Consequently, we are also able to obtain updated posterior distributions of quantities, which are often marginalized out, such as systematic errors of experiments. This offers a convenient way to spot inconsistencies in the posterior distribution.
For instance, a posterior expectation of a normalization error not consistent with its prior specification is a clear warning that some modeling assumptions are wrong.

To emphasize the advantages of the Bayesian network interpretation and also to prepare the discussion of nested relationships in the next section, we give a simple example of how the fitting of model parameters to an experimental dataset with $M$ measurement points affected by statistical and normalization errors may be modeled as a Bayesian network.
We then show how to compute the quantities that are required to evaluate the GLS formuals in \cref{eq:baynet_upd_cov,eq:baynet_upd_mean}.

The relationship of measured values, associated systematic and statistical errors, and the model parameters can be written as
\begin{equation}
    \label{eq:simple_baynet_model_obsvec}
    \vec{\sigma}_\textrm{exp} =
    g(\vec{p})
    + \mat{J}_\textrm{norm} \eta 
    + \mat{J}_\textrm{stat} \vec{\tau} \,,
\end{equation}
with $g(\vec{p})$ assumed to be a linear model, e.g., the linearized version of a nuclear physics model,
\begin{equation}
    \label{eq:linmod}
    g(\vec{p}) = \vec{y}_\textrm{ref}^\textrm{mod} + \mat{J}_\textrm{mod} (\vec{p} - \vec{p}_\textrm{ref}) \,.
\end{equation}
The matrix $\mat{J}_\textrm{mod}$ denotes the Jacobian matrix of the model.
The mapping matrix $\mat{J}_\textrm{norm}$ of dimension $M \times 1$ maps the normalization error $\eta$ systematicly to each data point.
We did not use the same symbol $\mat{T}$ as in~\cref{eq:stochastic_linear_link} to denote the Jacobian matrices because the Jacobian matrices here have to be regarded as submatrices of $\mat{T}$ and therefore also as submatrices of $\mat{S}$ defined in~\cref{eq:matS_definition} as we will discuss in a moment.

For an absolute normalization error, all elements of $\mat{J}_{\textrm{norm}}$ are one.
Finally, $\mat{J}_\textrm{stat}$ is given by the identity matrix to map the statistical errors to all the data points. 
To be fully consistent with the notation introduced in~\cref{subsec:glsrecap}, we need to introduce the reference vectors associated with the Taylor approximation,
\begin{multline}
    \label{eq:simple_baynet_model_obsvec2}
    \vec{\sigma}_\textrm{exp} = \vec{\sigma}_{\textrm{ref}}^{\textrm{exp}}
    + \mat{J}_\textrm{mod} (\vec{p} - \vec{p}_\textrm{ref}) \\
    + \mat{J}_\textrm{norm} (\eta - \eta_\textrm{ref}) 
    + \mat{J}_\textrm{stat} (\vec{\tau} - \vec{\tau}_\textrm{ref}) \,,
\end{multline}
with $\vec{\sigma}_{\textrm{ref}}^\textrm{exp} = \vec{y}_{\textrm{ref}}^\textrm{mod} + \mat{J}_\textrm{norm} \eta_\textrm{ref} + \mat{J}_\textrm{stat} \vec{\tau}_\textrm{ref}$.
The corresponding Bayesian network is given by:
\begin{center}
\tikz{
    \node[obs](sigmaexp){$\vec{\sigma}_{\textrm{exp}}$};
    \node[latent, left=of sigmaexp, yshift=-1cm, xshift=0.5cm](eta){$\eta$};
    \node[latent, left=of sigmaexp, xshift=0.0cm](pvec){$\vec{p}$};
    \node[latent, below=of sigmaexp, yshift=0.5cm](staterr){$\vec{\tau}$};
    \edge{pvec,eta,staterr}{sigmaexp}
}
\end{center}
The vectors $\vec{y}$ and $\vec{z}$ are
\begin{equation}
    \vec{y} =
    \begin{pmatrix}
        \vec{\sigma}_{\textrm{exp}} \\
        \vec{p} \\
        \eta
    \end{pmatrix}
    \;\;\;\textrm{and}\;\;\;\;
    \vec{z} =
    \begin{pmatrix}
        \vec{\tau} \\
        \vec{p} \\
        \eta
    \end{pmatrix} \,.
\end{equation}
The respective reference vectors for a Taylor expansion are given by
\begin{equation}
    \vec{y}_\textrm{ref} =
    \begin{pmatrix}
        \vec{\sigma}_{\textrm{ref}}^\textrm{exp} \\
        \vec{p}_\textrm{ref} \\
        \eta_\textrm{ref} \\
    \end{pmatrix}
    \;\;\;\;\textrm{and}\;\;\;\;
    \vec{z}_\textrm{ref} =
    \begin{pmatrix}
        \vec{\tau}_\textrm{ref} \\
        \vec{p}_\textrm{ref} \\
        \eta_\textrm{ref} \\
    \end{pmatrix}
\end{equation}
and the matrix $\mat{S}$ as required for \cref{eq:baynet_mapping} is of the form
\begin{equation}
    \mat{S} = 
    \begin{pmatrix}
        \mat{J}_\textrm{stat} & \mat{J}_\textrm{mod} & \mat{J}_\textrm{norm} \\
        \mat{0} & \mathbb{1} & \mat{0} \\
        \mat{0} & \mat{0} & \mathbb{1}
    \end{pmatrix} \,.
\end{equation}
Please note that $\mat{J}_\textrm{stat} = \mathbb{1}$ and the vector $\vec{\tau}$ has the same purpose as in the last section, i.e., to introduce stochasticity into the otherwise deterministic link which connects the model parameters and the normalization error with the experimental measurement vector.

As the values of the experimental measurement vector $\vec{\sigma}_{\textrm{exp}}$ are observed, we can use the GLS update formula in \cref{eq:baynet_upd_cov,eq:baynet_upd_mean} to obtain the joint posterior distributions of $\vec{p}$ and $\eta$.
To that end, the index set $D$ refers to the indices of the elements associated with the block $\vec{\sigma}_\textrm{exp}$ in $\vec{y}$.

If we assume the prior covariance matrix for the model parameter vector $\vec{p}$ to be diagonal, the full prior covariance matrix $\mat{U}$ summarizing at the same time our knowledge on model parameters, the normalization error, and the statistical errors is diagonal as well.
The mapping matrix $\mat{S}$ is very sparse, except possibly for the block $\mat{J}_\textrm{mod}$ associated with the mapping of model parameters to the corresponding predictions. 
However, the matrix $\mat{J}_\textrm{mod}$ is very rectangular for a nuclear model because there are usually much fewer model parameters than predicted values.
If using a mathematical fitting function, we can opt for piecewise linear interpolation or similar local interpolation schemes that yield a sparse Jacobian.
Prior knowledge on possible shapes of the function, such as its smoothness, can be introduced via the prior specification.
\Cref{subsec:sparse-gp-construction} elaborates on a sparse Gaussian process construction for this purpose. 

\subsection{\label{subsec:nested_relationships}Nested relationships}

In the previous section, we discussed a very simple Bayesian network that directly connected the independent variables to the dependent variables that are measured.
However, the full potential of Bayesian networks can only be exploited if we permit nested relationships.
Before we discuss inference in the presence of nested relationships in general, we give a motivating example with a nested relationship to develop intuition. 

To this end, we extend the discussion of the previous section to incorporate a positivity constraint.
We keep the linear model specification in \cref{eq:linmod}, which may represent a piecewise linear model, spline or any other sort of polynomial depending on the form of $\mat{J}_\textrm{mod}$.
Because a linear model may produce negative predictions, we apply a transformation to the model output $\vec{y}_\textrm{pred} = f(\vec{y}_\textrm{mod})$ with the effect of $f$ being element-wise exponentiation to enforce positivity.
This transformation represents a non-linearity in the model description.
In the current discussion, we linearize the non-linear relationship and focus on the nested property.
In the next section, we show how to deal exactly with the non-linearity.

In vicinity of a vector $\vec{y}_{\textrm{pred}}^{\textrm{ref}}$, the linear Taylor approximation of $f(\vec{y}_\textrm{mod})$ can be written as
\begin{equation}
    \label{eq:posmap_explink}
    \vec{y}_\textrm{pred} = \vec{y}_\textrm{pred}^\textrm{ref} + \mat{J}_\textrm{pos} (\vec{y}_\textrm{mod} - \vec{y}_\textrm{mod}^\textrm{ref}) \,,
\end{equation}
with $\mat{J}_\textrm{pos}$ being a diagonal matrix with the diagonal elements $J_{\textrm{mod},ii} = \exp(z_i)$ with $z_i$ being the i$^\textrm{th}$ element of $\vec{y}_\textrm{mod}^\textrm{ref}$.  
Please note that we did not add a random variable $\vec{\tau}$ to \cref{eq:posmap_explink}  because we deal with a deterministic relationship.
For the application of \cref{eq:zdef} and later equations, such a node must be formally present.
Formally, we therefore set all elements in the respective blocks in the reference vector $\vec{z}_\textrm{ref}$, the prior estimate $\vec{u}$ and covariance matrix $\mat{U}$ to zero, which is equivalent to removing the node altogether. 

The Taylor approximation for the compound function can be obtained by plugging the functional form of $g(p)$ into the Taylor approximation of $f$:
\begin{multline}
    \vec{y}_\textrm{pred} = \vec{y}_\textrm{pred}^\textrm{ref} + \mat{J}_\textrm{pos} \big(\vec{y}_\textrm{mod}^\textrm{ref} + \mat{J}_\textrm{mod}(\vec{p} - \vec{p}_\textrm{ref}) - \vec{y}_\textrm{mod}^\textrm{ref} \big) = \\
    = \vec{y}_\textrm{pred}^\textrm{ref} + \mat{J}_\textrm{pos} \mat{J}_\textrm{mod} (\vec{p} - \vec{p}_\textrm{ref}) \,.
\end{multline}
Finally, the Taylor approximation of this compound function can be combined with the normalization and statistical error, as has been already done in \cref{eq:simple_baynet_model_obsvec2},
\begin{multline}
    \vec{\sigma}_\textrm{exp} = \vec{\sigma}_{\textrm{exp}}^{\textrm{ref}}
    + \mat{J}_\textrm{pos} \mat{J}_\textrm{mod} (\vec{p} - \vec{p}_\textrm{ref}) \\
    + \mat{J}_\textrm{norm} (\eta - \eta_\textrm{ref}) 
    + \mat{J}_\textrm{stat} (\vec{\tau} - \vec{\tau}_\textrm{ref}) \,,
\end{multline}
with $\vec{\sigma}_{\textrm{exp}}^{\textrm{ref}} = \vec{y}_{\textrm{pred}}^{\textrm{ref}} + \mat{J}_\textrm{norm} \eta_\textrm{ref} + \mat{J}_\textrm{stat} \vec{\tau}_{\textrm{ref}}$.
The Bayesian network representing the relationships between the variables in this case is illustrated here:
\begin{center}
\tikz{
    \node[obs](sigmaexp){$\vec{\sigma}_{\textrm{exp}}$};
    \node[latent, left=of sigmaexp, yshift=-1cm, xshift=0.5cm](eta){$\eta$};
    \node[latent, left=of sigmaexp, xshift=0.0cm](predvec){$\vec{y}_{\textrm{mod}}$};
    \node[latent, left=of predvec, xshift=0.0cm](pvec){$\vec{p}$};
    \node[latent, below=of sigmaexp, yshift=0.5cm](staterr){$\vec{\tau}$};
    \edge{predvec,eta,staterr}{sigmaexp}
    \edge{pvec}{predvec}
}
\end{center}

The quantities to evaluate \cref{eq:baynet_upd_mean,eq:baynet_upd_cov} are therefore given by
\begin{equation}
    \vec{y} =
    \begin{pmatrix}
        \vec{\sigma}_{\textrm{exp}} \\
        \vec{y}_\textrm{mod} \\
        \vec{p} \\
        \eta 
    \end{pmatrix}
    \;\;\;\textrm{and}\;\;\;
    \vec{z} =
    \begin{pmatrix}
        \vec{\tau} \\
        \vec{0} \\
        \vec{p} \\
        \eta 
    \end{pmatrix} \,.
\end{equation}
The second element in $\vec{z}$ is zero because we regard the link from the node $\vec{p}$ to $\vec{y}_\textrm{mod}$ as deterministic, which is formally solved as stating that the associated parentless node is always zero and also its reference point, prior expectation and associated prior covariance matrix element in $\mat{U}$ are zero as well.   

The reference vectors of the Taylor expansion are given by
\begin{equation}
    \vec{y}_\textrm{ref} =
    \begin{pmatrix}
        \vec{\sigma}_\textrm{exp}^\textrm{ref} \\
        \vec{y}_\textrm{mod}^\textrm{ref} \\
        \vec{p}_\textrm{ref} \\
        \eta_\textrm{ref} \\
    \end{pmatrix} 
    \;\;\;\textrm{and}\;\;\;
    \vec{z}_\textrm{ref} = \begin{pmatrix}
        \vec{\tau}_\textrm{ref} \\
        \vec{0} \\
        \vec{p}_\textrm{ref} \\
        \eta_\textrm{ref}
    \end{pmatrix}
\end{equation}
and the sensitivity matrix is of the form
\begin{equation}
    \label{eq:compmat_S_example}
    \mat{S} = 
    \begin{pmatrix}
        \mathbb{1} &  \mat{0} & \mat{J}_\textrm{pos}\mat{J}_\textrm{mod} & \mat{J}_\textrm{norm} \\
        \mat{0} & \mathbb{1} & \mat{J}_\textrm{mod} & \mat{0} \\
        \mat{0} & \mat{0} & \mathbb{1} & \mat{0} \\
        \mat{0} & \mat{0} & \mat{0} & \mathbb{1} \\
    \end{pmatrix} \,.
\end{equation}
The matrix $\mat{S}$ is again very sparse, with the potential exception of the blocks containing $\mat{J}_\textrm{mod}$.

Because the values of the experimental measurement vector $\vec{\sigma}_{\textrm{exp}}$ are known, the indices in the index set $D$ introduced in the paragraph above \cref{eq:postcov-notDnotD} refer to the positions of elements of $\vec{\sigma}_{\textrm{exp}}$ in the vector $\vec{y}$.
The respective overall Jacobian matrix $\mat{S}$ was constructed by hand as an example of how it is done in principle.
For more complex Bayesian networks with more deeply nested functions, the manual approach is impractical.
In such scenarios, a programmatic recipe to construct $\mat{S}$ based on the Jacobian matrices of the individual functional relationships is pertinent.

To this end, we remind ourselves of the notation established in~\cref{subsec:glsrecap}.
There we introduced a vector $\vec{y}$ that comprised the subvector of independent variables $\vec{y}_I$ and the dependent variables $\vec{y}_J$ whose values are partially determined as a function of the independent variables, $\vec{y}_J = f(\vec{y}_I) + \vec{\tau}$. 
In the Bayesian network picture, the vector $\vec{\tau}$ contains all the parentless nodes directly attached to the dependent nodes.
We also combined the vectors $\vec{y}_I$ and $\vec{\tau}$ to the vector $\vec{z}$ that contained the variables associated with all independent nodes.
The mental distinction between the parentless nodes in $\vec{y}_I$ and those associated with variables in $\vec{\tau}$ was to structurally ensure that we can apply the GLS method by using \cref{eq:baynet_upd_cov} and \cref{eq:baynet_upd_mean}.

The GLS formula for the posterior expectation \cref{eq:baynet_upd_mean} relies on the availability of the vector $\vec{y}_\textrm{ref}$ containing the propagated values of the reference values $\vec{z}_\textrm{ref}$ through the network and the overall Jacobian matrix $\mat{S}$.
An example of the overall Jacobian matrix for a nested relationship was given in \cref{eq:compmat_S_example}.

To describe the necessary propagation and computation of the overall Jacobian, we first introduce the notion of what we call a \textit{mapping}.
Each individual functional relationship between two blocks of variables can be characterized by a set of source indices $\mathcal{S}$ and a set of target indices $\mathcal{T}$ and a function $f$.
The two sets must not have indices in common, i.e., $\mathcal{S} \cap \mathcal{T} = \emptyset$.

Each individual mapping takes the elements in $\vec{y}$ indexed by $\mathcal{S}$, applies a function $f(.)$ to it in order to obtain the contribution to the values of the elements in $\vec{y}$ indexed by $\mathcal{T}$.
In other words, the application of a mapping on a vector $\vec{y}$ produces a new vector $\vec{y}'$ of the same size given by
\begin{equation}
    \label{eq:def_mapping}
    \vec{y}_\mathcal{T}' = \vec{y}_\mathcal{T} + f(\vec{y}_{\mathcal{S}})
    \;\;\;\;\textrm{and}\;\;\;\;
    \vec{y}_\mathcal{\neg T}' = \vec{y}_{\neg T}
\end{equation}

In the presented example, for instance, the function $f$ to enforce positivity together with the set of source indices referring to the block $\vec{y}_\textrm{mod}$ in $\vec{y}$ and the set of target indices referring to the block $\vec{\sigma}_\textrm{exp}$ in $\vec{y}$ is a mapping, which we denote by $\mathcal{M}_\textrm{pos}$.
The other mappings are associated with the linear model $g(\vec{p})$ and the distribution of normalization errors and statistical errors to the experimental data points, and are denoted by $\mathcal{M}_\textrm{mod}$, $\mathcal{M}_\textrm{norm}$ and $\mathcal{M}_\textrm{stat}$, respectively.
The mappings $\mathcal{M}_\textrm{pos}$, $\mathcal{M}_\textrm{norm}$ and $\mathcal{M}_\textrm{stat}$ contribute additively to the experimental measurement $\vec{\sigma}_\textrm{exp}$, hence their sets of target indices are identical.

Illustrated on the current example, the propagation of the independent values in $\vec{z}$ to obtain the values associated with $\vec{y}_J$ works in the following way.
Given that the subset $\vec{y}_I$ is specified, we can compute the elements of the dependent variables $\vec{y}_J$ by applying the mappings in the correct order.
In the example with the positivity constraint, we first have to apply $\mathcal{M}_\textrm{mod}$ to obtain $\vec{y}_\textrm{mod}$ and then apply $\mathcal{M}_\textrm{pos}$, $\mathcal{M}_\textrm{norm}$ and $\mathcal{M}_\textrm{stat}$ to obtain $\vec{\sigma}_\textrm{exp}$.
One mapping that depends on the output of another mapping must not be applied before that other mapping.
Importantly, after all mappings that refer to the same target index set $\mathcal{T}$ have been applied and before the application of any subsequent mapping, we need to also add the vector $\vec{z}_\mathcal{T}$ to $\vec{y}_\mathcal{T}$ to account for the additive contributions represented by $\vec{z}_{\mathcal{T}}$.

As a general recipe, the order of the mappings $\mathcal{M}_1, \mathcal{M}_2, \dots$ is established by the following requirement:
For two mappings $\mathcal{M}_i$ and $\mathcal{M}_j$ with $i < j$, the set of source indices $\mathcal{S}_i$ must not contain elements of the set of target indices $\mathcal{T}_j$, i.e., $\mathcal{T}_j \cap \mathcal{S}_i = \emptyset$. 
Please note that this criterion may leave some ambiguity in the order, which is not a problem.
For instance, in our running example, it does not matter whether $\mathcal{M}_\textrm{stat}$ is applied before or after $\mathcal{M}_\textrm{norm}$.

Having the mappings in the correct order, the propagation of values in a general situation can be achieved as follows.
Initially, we set $\vec{y} = \vec{z}$.
The maps are then applied one by one in an order consistent with the criterion above.
The resulting vector $\vec{y}$ of one mapping serves as input to the next mapping.
After the application of the final mapping, the vector $\vec{y}$ contains the correctly propagated values of the independent variables in $\vec{z}$.

The construction of the overall Jacobian matrix $\mat{S}$ can be achieved similarly.
We start with an initial mapping matrix $\mat{S}$ given by
\begin{equation}
    \mat{S}_{I,I} = \mathbb{1}, \;\;\;\;
    \mat{S}_{I,J} = \mat{0}, \;\;\;\;
    \mat{S}_{J,I} = \mat{0}, \;\;\;\;
    \mat{S}_{J,J} = \mathbb{1} \,.
\end{equation}
We go through the mappings in a valid order.
For each mapping $\mathcal{M}_i$ in this sequence, we multiply the Jacobian matrix $\mat{W}$ of the current mapping by the resulting mapping matrix $\mat{S}$ of the previous mapping to obtain the updated mapping matrix $\mat{S}'$,
\begin{equation}
    \mat{S}' = \mat{W} \mat{S} \,.
\end{equation}

The $N \times N$ Jacobian matrix $\mat{W}$ of a mapping $\mathcal{M}$ given by a function $f$ with source index set $\mathcal{S}$ and target index set $\mathcal{T}$ is given by
\begin{equation}
    \label{eq:mapping_jacobian}
    \mat{W}_{\mathcal{T},\mathcal{S}} = \mat{J} ,\;\;
    \mat{W}_{\mathcal{S},\mathcal{S}} = \mathbb{1} ,\;\;
    \mat{W}_{\mathcal{T},\mathcal{T}} = \mathbb{1} ,\;\;
    \mat{W}_{\mathcal{S},\mathcal{T}} = \mat{0} . 
\end{equation}
where $\mat{J}$ is the Jacobian matrix of the function $f(.)$.
Please note that in the case of a non-linear relationship $f$, the Jacobian of a mapping depends on the actual values in $\vec{y}_{\mathcal{S}}$, see \cref{eq:def_mapping}.

If we deal with observations of intermediate nodes, we cannot use the overall Jacobian matrix $\mat{S}$ in the GLS method but need to use a slightly modified matrix $\mat{S}'$. 
This Bayesian network demonstrates a case with an observation of an intermediate node:
\begin{center}
\tikz{
    \node[obs](sigmaexp){$\vec{y}_{2}$};
    \node[latent, left=of predvec, yshift=-1cm, xshift=0.5cm](eta){$\vec{\tau}_1$};
    \node[obs, left=of sigmaexp, xshift=0.0cm](predvec){$\vec{y}_{1}$};
    \node[latent, left=of predvec, xshift=0.0cm](pvec){$\vec{p}$};
    \node[latent, left=of sigmaexp, yshift=-1cm, xshift=0.5cm](staterr){$\vec{\tau}_2$};
    \edge{predvec,staterr}{sigmaexp};
    \edge{eta}{predvec};
    \edge{pvec}{predvec}
}
\end{center}
To calculate the matrix $\mat{S}'$, we need to make the following alteration in the prescription we outlined for the computation of $\mat{S}$.
In the sequential application of the mappings, whenever the set of source indices $\mathcal{S}$ of the current mapping makes reference to variables of an observed node, we need to use instead of \cref{eq:mapping_jacobian} the specification
\begin{equation}
    \mat{W}_{\mathcal{T},\mathcal{S}} = \mat{0} ,\;\;
    \mat{W}_{\mathcal{S},\mathcal{S}} = \mathbb{1} ,\;\;
    \mat{W}_{\mathcal{T},\mathcal{T}} = \mathbb{1} ,\;\;
    \mat{W}_{\mathcal{S},\mathcal{T}} = \mat{0} \,.
\end{equation}
This change means that we remove the functional dependence of the variables associated with the target indices in $\mathcal{T}$ from those associated with the source indices $\mathcal{S}$.

We also need to make a similar adjustment to the propagation of the values in $\vec{z}_\textrm{ref}$ to obtain the values in $\vec{y}_\textrm{ref}$.  
We start with the initial assignment $\vec{y} = \vec{z}_\textrm{ref}$.
In the sequential application of the mappings, i.e., $\vec{y}' = \vec{y} + f_i(\vec{y}_{\mathcal{S}})$, whenever $\mathcal{S}$ makes reference to an observed node, we need to propagate the observed values $\vec{r}_{\mathcal{S}}$ instead, $\vec{y}' = \vec{y} + f_i(\vec{r}_{\mathcal{S}})$.
The resulting vector $\vec{y}_\textrm{ref}'$ together with $\mat{S}'$ must be used instead of their un-primed counterparts to evaluate the GLS formulas in \cref{eq:baynet_upd_cov,eq:baynet_upd_mean}.

In the graphical representation of the schematic Bayesian network, it amounts to removing the respective arrow:
\begin{center}
\tikz{
    \node[obs](sigmaexp){$\vec{y}_{2}$};
    \node[latent, left=of predvec, yshift=-1cm, xshift=0.5cm](eta){$\vec{\tau}_1$};
    \node[obs, left=of sigmaexp, xshift=0.0cm](predvec){$\vec{y}_{1}$};
    \node[latent, left=of predvec, xshift=0.0cm](pvec){$\vec{p}$};
    \node[latent, left=of sigmaexp, yshift=-1cm, xshift=0.5cm](staterr){$\vec{\tau}_2$};
    \edge{staterr}{sigmaexp};
    \edge{eta}{predvec};
    \edge{pvec}{predvec}
}
\end{center}
Thereby the flow of information from observations downstream of $\vec{y}_1$, e.g., the observation of $\vec{y}_2$, does not impact anymore the posterior distribution of independent variables upstream of $\vec{y}_1$.
However, the observed values of $\vec{y}_1$ have still to be propagated downstream, as they contribute with certain proportion to the values observed in $\vec{y}_2$, which does not need to be explained by independent variables anymore connected to $\vec{y}_2$, such as $\vec{\tau}_2$.
The special treatment of observed nodes in the computation of $\vec{y}_\textrm{ref}$ ensures that the contribution of the observed nodes is properly accounted for in downstream nodes in the application of the GLS method in \cref{eq:baynet_upd_mean}.
Please note that the functionality to incorporate observations of intermediate nodes has not been implemented in the \textit{nucdataBaynet} package yet.

Finally, we want to address the possibly counter-intuitive aspect that an observed intermediate node blocks any information flow from downstream nodes to upstream nodes if one considers the adjustment of differential cross sections using both differential and integral experimental data.
The Bayesian network used as illustration to deal with an intermediate node does not represent the situation of integral data assimilation.
If $\vec{y}_1$ represented the differential measurements, this vector would not be propagated through the application code, e.g., a neutron transport calculation, but the inferred \textit{true} cross section would be used instead.
The following Bayesian network correctly captures the estimation of differential cross sections by using both differential and integral experimental data:
\begin{center}
\tikz{
    \node[latent](truediff){$\vec{\sigma}_\textrm{true}^\textrm{diff}$};
    \node[obs, right=of truediff, yshift=0.7cm](diffexp){$\vec{\sigma}_\textrm{exp}^\textrm{diff}$};
    \node[obs, right=of truediff, yshift=-0.7cm](intexp){$\vec{\sigma}_\textrm{exp}^\textrm{int}$};
    \node[latent, left=of intexp, xshift=0.5cm, yshift=-0.7cm](intexperr){$\vec{\tau}_\textrm{exp}^\textrm{int}$};
    \node[latent, left=of diffexp, xshift=0.5cm, yshift=0.7cm](diffexperr){$\vec{\tau}_\textrm{exp}^\textrm{diff}$};
    \edge{intexperr}{intexp};
    \edge{truediff}{diffexp,intexp};
    \edge{diffexperr}{diffexp}
}
\end{center}
The true differential cross section $\vec{\sigma}_\textrm{true}^\textrm{diff}$ explains the differential measurement $\vec{\sigma}_\textrm{exp}$ up to the experimental error $\vec{\tau}_\textrm{exp}^\textrm{diff}$.
The integral measurement $\vec{\sigma}_\textrm{exp}^\textrm{int}$ is explained by the true differential cross section propagated through the transport code, and again up to a measurement error $\vec{\sigma}_\textrm{exp}^{int}$.
In this scenario, there are not any observed intermediate nodes.

\subsection{\label{subsec:nonlinearfuns}Non-linear relationships}

So far we assumed all relationships to be linear.
However, as non-linear relationships appear frequently in nuclear data evaluation, their adequate treatment is important.
Some examples are:
\begin{itemize}
    \item The relationship between parameters of nuclear models and the associated predictions are usually non-linear.
    \item Ratios of cross sections are non-linear functions of the cross sections.
    \item Measured cross sections are non-linear functions of the true cross section and the relative normalization error
    \item To enforce positivity of a cross section $y$, a non-linear variable transformation $y=\exp(x)$ can be introduced and the auxiliary variable $x$ treated as the independent variable.
\end{itemize}

It is therefore tremendously useful to have an inference procedure that is capable to deal with these non-linearities.
At this point of the discussion, it is pertinent to recall that the GLS formula in \cref{eq:baynet_upd_mean} locates the maximum of the posterior distribution given by
\begin{equation}
    \label{eq:bayesupdate_nonlinear}
    P(\vec{z}_{\neg D} \,|\, \vec{y}_D) = \frac{1}{P(\vec{y}_{D})} P(\vec{y}_D \,|\, \vec{z}_{\neg D}) P(\vec{z}_{\neg D})
\end{equation}
if likelihood and prior are given by the multivariate normal distributions $P(\vec{y}_D \,|\, \vec{z}_{\neg D}) = \mathcal{N}(g(\vec{z}_{\neg D}) + \vec{u}_D, \mat{U}_{D,D})$ and $P(\vec{z}_{\neg D}) = \mathcal{N}(\vec{u}_{\neg D}, \mat{U}_{\neg D,\neg D})$, respectively, and $g$ is a linear function.
The distribution parameters of the likelihood are due to the prior associated with the parentless node $\vec{\tau}_D \sim \mathcal{N}(\vec{u}_D, \mat{U}_{D,D})$ attached to the observed node.

The Levenberg-Marquardt (LM) algorithm~\cite{levenbergMethodSolutionCertain1944,marquardtAlgorithmLeastSquaresEstimation1963} is an iterative algorithm to solve the non-linear least squares problem.
A customized LM algorithm has been presented in~\cite{helgessonFittingDefectNonlinear2017b} to take into account the same prior information as the GLS method.
More precisely, it can locate exactly the posterior maximum of \cref{eq:bayesupdate_nonlinear} also for a non-linear relationship $g$.
The specific values of $\vec{z}_{\neg D}$ associated with the posterior maximum represent a so-called maximum a posteriori probability (MAP) estimate.
In the following we discuss the adjustments to the GLS method to obtain the customized LM algorithm.
A detailed derivation of the update formula employed in the LM algorithm is provided in~\cref{app:GLS-LM-derivation}.

The LM algorithm proceeds by applying in each iteration the GLS method with the inverse of the prior covariance matrix augmented by a damping term.
This mechanism enables the algorithm to dynamically transition between the gradient ascent method and the GLS method depending on the degree of the non-linearities present.
As \cref{eq:baynet_upd_cov,eq:baynet_upd_mean} implement the GLS method, the application of the LM algorithm in our situation is straight-forward.
In each iteration, we have to use \cref{eq:baynet_upd_cov} augmented by a damping term,
\begin{equation}
    \mat{U}_{I,I}' =
    \left( \mat{S}_{A,I}^T \mat{U}^{-1} \mat{S}_{A,I} + \lambda \mat{D} \right)^{-1} \,,
\end{equation}
with $\lambda$ being a real positive scalar, $\mat{D}$ a diagonal matrix and the Jacobian matrix $\mat{S}$ evaluated at the current reference vector $\vec{z}_\textrm{ref}$, which is associated with the largest value of the posterior distribution found so far.
The index set $A$ contains all $1..N$ indices with $N$ being the number of rows (or columns) of $\mat{U}$.

In each iteration, a proposal vector $\vec{z}_{\textrm{prop},\neg D}$ is computed using \cref{eq:baynet_upd_mean}.
If the proposed vector is associated with a larger value of the posterior distribution, it becomes the new reference vector $\vec{z}_{\textrm{ref},\neg D}$ for the next iteration, which also represents the new best guess for the maximum of the posterior distribution.
Importantly, also $\vec{y}_\textrm{ref}$ has to be updated accordingly by propagating $\vec{z}_{\textrm{ref},\neg D}$ using the non-linear mappings as described in the previous section.
If the proposed vector leads to a lower value of the posterior distribution than the current reference vector, it is rejected.

The value of the control parameter $\lambda$ to adjust the step size is changed in each iteration depending on the gain defined by
\begin{equation}
    \rho = \frac{
        f_\textrm{ex}(\vec{z}_{\textrm{prop},\neg D}) - f_\textrm{ex}(\vec{z}_{\textrm{ref},\neg D})
    }{
        f_\textrm{lin}(\vec{z}_{\textrm{prop},\neg D}) - f_\textrm{lin}(\vec{z}_{\textrm{ref},\neg D})
    } \,,
\end{equation}
where $f_\textrm{ex}$ is the exact logarithmized posterior density function using the non-linear relationship $g$ and $f_\textrm{lin}$ the expected value of the logarithmized posterior distribution using a linear approximation to $g$ constructed at the vector $\vec{z}_{\textrm{ref},\neg D}$.
The following prescription was suggested by~\cite{marquardtAlgorithmLeastSquaresEstimation1963} to update the parameter $\lambda$ from one iteration to the next one,
\begin{equation}
    \lambda' =
    \begin{cases}
    2 \lambda & \textrm{if } \rho < 0.25 \\
    \lambda & \textrm{if } 0.25 \leq \rho < 0.75 \\
    \lambda/3 & \textrm{if } 0.75 \leq \rho
    \end{cases} \,.
\end{equation}
In words, if the expected improvement using the linear approximation is similar to the real improvement, the value of $\lambda$ is decreased which leads to a larger step size and an update more similar to the GLS update.
On the other hand, if the expected improvement and real improvement are very different, $\lambda$ is increased which reduces the step size and proposes vectors more according to the gradient ascent method.
Please note that the iterative application of the GLS update without the adaptive damping term is not guaranteed to converge, e.g., demonstrated in~\cite[p. 117]{schnabel_large_2015}.

Overall, the customized LM algorithm is very efficient.
It exploits the fact that the result of the non-linear mappings appears as the mean vector in the multivariate normal likelihood.
This structure in the equations enables to analytically obtain and use second-order information about the posterior pdf even though only first-order information (i.e., the Jacobian matrix) of the non-linear relationships is available.

Its suitability and efficiency for model-based nuclear data evaluation has already been demonstrated in a full scale evaluation within a nuclear data evaluation pipeline prototype~\cite{schnabelConceptionSoftwareImplementation2021} to fit energy-dependent model parameters of TALYS~\cite{koning_modern_2012,koningTENDLCompleteNuclear2019}.

If several local maxima are present due to non-linear mappings, such as induced by relative normalization errors, a stage-wise approach is helpful to ensure that a good local and ideally the global maximum is found.
Only a subset of the variables may be adjusted in each stage and the other adjustable variables are kept fixed.
Mathematically, this is achieved by setting the prior estimates of fixed variables in $\vec{u}$ equal to the associated elements in the reference vector $\vec{z}_\textrm{ref}$ and setting the respective elements in the prior covariance matrix $\mat{U}$ to zero.
With these specifications, the result of \cref{eq:baynet_upd_mean} is conditioned on the values of the fixed variables.
Note that due to the zero prior uncertainty of the fixed elements, also the prescription to deal with deterministic relationships described in the paragraph below~\cref{eq:baynet_upd_mean} needs to be applied.

As a final remark, sometimes it is suggested to treat relative normalization uncertainties by updating the elements of the covariance matrix from one step to another in an iterative optimization scheme.
The convergence of the LM algorithm, however, is only guaranteed if the prior covariance matrix $\mat{U}$ is assumed to be constant throughout the iteratitive scheme.
The correct modeling of relative errors (and thereby induced relative uncertainties) to ensure convergence is explained in~\cref{subsec:imp-nonlinear-maps}.

In the next section, we elaborate on an efficient and flexible approach to define priors on functions. 
This construction may be used to model the prior knowledge about an excitation function, which is attempted to be measured in an experiment.
Afterwards we elaborate on two important experimental aspects, relative normalization errors and energy calibration errors, that can be properly taken into account in the presented Bayesian network framework but not with the GLS method.

\subsection{\label{subsec:sparse-gp-construction}Sparse Gaussian process construction}

Gaussian processes, e.g.,~\cite{rasmussen_gaussian_2006}, have been applied in various ways in the nuclear data evaluation field.
They have been suggested to account for model deficiencies of the nuclear model in the fast energy region, e.g.,~\cite{leeb_consistent_2008,schnabel_large_2015,schnabelDifferentialCrossSections2016},
as prior on energy-dependent model parameters of nuclear models~\cite{helgessonTreatingModelDefects2018,schnabelConceptionSoftwareImplementation2021},
to model energy-dependent systematic errors of experiments, e.g.,~\cite{schnabelFittingAnalysisTechnique2018},
and to fit cross section curves to experimental data~\cite{iwamotoGenerationNuclearData2020a}.
This list demonstrates the versatility of GPs.

A disadvantage of GPs is that the time needed to fit them to data scales with $N^3$ with $N$ being the number of data points.  
This scaling behavior arises due to the occurence of the inverse of a covariance matrix of size $N\times N$ in the regression formula.
Approaches to extend the applicability of GP regression to larger datasets are sparse approximations, e.g.,~\cite{quinonero-candela_unifying_2005,snelson_sparse_2006} and references therein, and numerical methods relying on iterative solvers, e.g.,~\cite{NEURIPS2019_01ce8496,2020arXiv200611267P}.
Another approach, which is application dependent, is to exploit a specific structure of the covariance matrix.
For instance, the equations for GP inference can be more efficiently evaluated if a covariance matrix is of toeplitz structure or can be decomposed into kernels acting independently on different dimensions, e.g.,~\cite{saatchiScalableInferenceStructured2011}.

In this section, we elaborate on a construction of a sparse Gaussian process, which enmeshes well with the outlined Bayesian network interpretation, scales to large datasets, and allows for the flexible incorporation of prior knowledge on the features of the unknown function, such as its expected range, slope and smoothness.
This construction has already been discussed to some extent in the context of a nuclear data evaluation pipeline~\cite{schnabelConceptionSoftwareImplementation2021}.
Also in~\cite[Chap.~13.2]{lindenBayesianProbabilityTheory2014} essentially the same construction is described in good detail and derived using the MaxEnt principle~\cite{jaynesJaynesPapersProbability1983,shoreAxiomaticDerivationPrinciple1980} and variational calculus, but relying on a spline basis instead of a piecewise linear function as done here.

Given two mesh points associated with energies $E_{i}$ and $E_{i+1}$, where $E_{i} < E_{i+1}$ and function values $y_i$ and $y_{i+1}$, respectively, values at intermediate energies can be determined by linear interpolation:
\begin{multline}
    g(E) = 
    \left(
        \frac{E_{i+1}-E}{E_{i+1}-E_i}
    \right) y_i
    +
    \left(
        \frac{E-E_i}{E_{i+1}-E_i} 
    \right) y_{i+1} \\
    \textrm{if}\;\; E_i \leq E < E_{i+1}
    \label{eq:basic_linearinterpol}
\end{multline}
To state the formulas in a general way for a complete mesh of energies, we introduce the abbreviation
\begin{equation}
    c_i(E) = \begin{cases}
        \frac{E_{i+1} - E}{E_{i+1}-E_i} & \textrm{if}\;\; E_i \leq E < E_{i+1} \\
        0 & \textrm{otherwise}
    \end{cases}
\end{equation}
and
\begin{equation}
    d_i(E) = \begin{cases}
        \frac{E - E_{i-1}}{E_{i}-E_{i-1}} & \textrm{if}\;\; E_{i-1} \leq E < E_{i} \\
        0 & \textrm{otherwise}
    \end{cases}
\end{equation}
as well as their sum,
\begin{equation}
    f_i(E) = c_i(E) + d_i(E) \,.
\end{equation}
Piecewise linear interpolation, i.e., locating for an energy of interest $E$ the enclosing energies on the mesh and then performing linear interpolation according to \cref{eq:basic_linearinterpol}, can now be concisely written as
\begin{equation}
    g(E) =
    \sum_{i=1}^{M} f_i(E) y_{i} \,.
    \label{eq:pwlinint}
\end{equation}
Thanks to the bilinearity of the covariance operator, the covariance between function values at arbitrary energies can be computed by
\begin{equation}
    \Cov{g(E)}{g(E')} = \\
    \sum_{i=1}^{M} \sum_{j=1}^{M}
    f_i(E) f_j(E') \Cov{y_i}{y_j}.
    \label{eq:covpwlinint}
\end{equation}
This formula shows that the covariance matrix $\tilde{\mat{K}}$ with $\tilde{K}_{ij}=\Cov{y_i}{y_j}$ associated with a \textit{finite} number of variables $M$, in combination with linear interpolation enables the computation of covariances for arbitrary pairs of energies $E, E'$.
In other words, it is a valid specification of a covariance function $\kappa(E,E')$ and together with a mean function $m(E)$ completely characterizes a Gaussian process.

To make the link to the Bayesian network interpretation, we assume that observations of some energy-dependent quantity have been made at experimental energies $\{E_i'\}_{i=1..N}$ and the computational mesh used for the sparse Gaussian process construction in \cref{eq:covpwlinint} is given by $\{E_j\}_{j=1..M}$.
The vector of function values on the computational mesh is denoted by $\vec{y}_\textrm{GP}$ and the vector of function values at the experimental mesh by $\vec{\sigma}_\textrm{exp}$.
The Jacobian matrix $\mat{J}_\textrm{GP}$ that performs the mapping from the computational to the experimental mesh can be computed by
\begin{equation}
    J_{\textrm{GP},ij} = \frac{\partial g(E_i')}{\partial y_j} = f_j(E_i') \,.
\end{equation}

The mapping of the function values of the computational to the experimental mesh can now be written as
\begin{equation}
    \vec{\sigma}_\textrm{exp} =
    \vec{\sigma}_{\textrm{exp}}^{\textrm{ref}} 
    + \mat{J}_\textrm{GP} (\vec{y}_\textrm{GP} - \vec{y}_{\textrm{GP}}^{\textrm{ref}}) 
    + (\vec{\tau} - \vec{\tau}^\textrm{ref}) \,,
\end{equation}
with $\vec{\sigma}_{\textrm{exp}}^{\textrm{ref}} = \mat{J}_\textrm{GP} \vec{y}_{\textrm{GP}}^{\textrm{ref}} + \vec{\tau}_\textrm{ref}$.
The corresponding Bayesian network is of the form
\begin{center}
\tikz{
    \node[obs](H3){$\vec{\sigma}_\textrm{exp}$};
    \node[latent, below=of H3, yshift=0.7cm, xshift=-0.5cm](H2){$\vec{\tau}$};
    \node[latent, left=of H3, xshift=0.0cm](H1){$\vec{y}_\textrm{GP}$};
    \edge{H1,H2}{H3}
}
\end{center}
The vectors to evaluate the GLS equations in \cref{eq:baynet_upd_cov,eq:baynet_upd_mean} are given by
\begin{equation}
    \vec{y} = \begin{pmatrix}
        \vec{\sigma}_\textrm{exp} \\
        \vec{y}_\textrm{GP}
    \end{pmatrix}
    \;\;\;\textrm{and}\;\;\;
    \vec{z} = \begin{pmatrix}
        \vec{\tau} \\
        \vec{y}_\textrm{GP}
    \end{pmatrix} \,.
\end{equation}
Regarding the reference vectors of the Taylor expansion, we have
\begin{equation}
    \vec{y}_\textrm{ref} = \begin{pmatrix}
        \vec{\sigma}_{\textrm{exp}}^\textrm{ref} \\
        \vec{y}_{\textrm{GP}}^\textrm{ref}
    \end{pmatrix}
    \;\;\;\textrm{and}\;\;\;
    \vec{z}_\textrm{ref} = \begin{pmatrix}
        \vec{\tau}_\textrm{ref} \\
        \vec{y}_\textrm{GP}^\textrm{ref}
    \end{pmatrix} \,,
\end{equation}
and for the overall mapping matrix
\begin{equation}
    \mat{S} = \begin{pmatrix}
        \mathbb{1} & \mat{J}_\textrm{GP} \\
        \mat{0} & \mathbb{1}
    \end{pmatrix} \,.
\end{equation}
The matrix $\mat{J}_\textrm{GP}$ contains only two non-zero elements per row and the whole matrix $\mat{S}$ is therefore very sparse.
The compound prior covariance matrix is given by
\begin{equation}
    \mat{U} = \begin{pmatrix}
        \mat{U}_{\vec{\tau}} & \mat{0} \\
        \mat{0} & \mat{U}_\textrm{GP} 
    \end{pmatrix} \,.
\end{equation}
We usually assume that $\mat{U}_{\vec{\tau}}$ is a diagonal matrix because it contains the uncertainties about the statistical errors of the experimental data points.
Regarding the prior covariance matrix $\mat{U}_\textrm{GP}$ of the discretized Gaussian process, we have to make a choice which covariance function to use.
A common choice is the so-called squared exponential,
\begin{equation}
    \label{eq:sqrexpGP_example}
    \Cov{y_i}{y_j} = \delta^2 \exp\left(-\frac{1}{2\lambda^2} (E_i - E_j)^2 \right) \,.
\end{equation}
The length scale $\lambda$ determines how rapidly the function is a priori expected to change or oscillate and the amplitude $\delta$ defines the range the unknown function values are expected to cover.

However, there are several potential issues with this covariance function.
One issue being that the same length scale is applied in all energy ranges, which is clearly an invalid assumption for an energy range covering both the resonance region and the fast energy range.
Depending on the application, hand-tailored composite covariance functions can be used, e.g.,~\cite{helgessonFittingDefectNonlinear2017b,schnabelEstimatingModelBias2018}, or an energy-dependent length scale employed, e.g.,~\cite{schnabelFirstSketchConstruction2018}.
Another issue is that the prior induced by this covariance function incorporates the assumption that all potential solution functions have derivatives of all orders, e.g.,~\cite[Chap. 4.2.1]{rasmussen_gaussian_2006}, which may be regarded as an unreasonably strong assumption.
Using a covariance function of the Matern class instead is a possible solution, which only possesses derivatives up to a certian order, e.g.,~\cite[Chap. 4.2.1]{rasmussen_gaussian_2006}.
Finally, the resulting covariance matrix associated with the values at the mesh points is dense which is detrimental to scale up the inference in Bayesian networks to a large number of variables, e.g., several hundred thousand.
Sparse approximations to the full Gaussian process are one approach to deal with this issue, e.g.,~\cite{quinonero-candela_unifying_2005}.
The discretization introduced in \cref{eq:covpwlinint} in combination with linear interpolation is already a sparse approximation to the full Gaussian process.
However, if using the squared exponential covariance function, the block $\mat{U}_\textrm{GP}$ is still dense.
In the following we present a GP construction that renders $\mat{U}_\textrm{GP}$ also very sparse.

A diagonal covariance matrix $\mat{U}_\textrm{GP}$ is very sparse but does not impose any constraint on the smoothness of possible solution functions.
The idea is therefore to introduce pseudo-observations of the second derivatives of this function to enforce a certian degree of smoothness.
To this end, we introduce the discretized version of the first derivative at location $E_i$,
\begin{equation}
    \Delta_i = \frac{y_{i+1} - y_i}{E_{i+1}-E_i} \,,
\end{equation}
and use this definition recursively to get a discretized version of the second derivative at location $E_i$,
\begin{multline}
    \label{eq:finitediff2nd}
    \Delta_i^2 = \frac{\Delta_{i+1} - \Delta_{i}}{E_{i+1} - E_{i}} 
    = 
    \\
    \frac{1}{(E_{i+1} - E_{i})^2} \, y_i +
    \frac{1}{(E_{i+2} - E_{i+1})(E_{i+1}-E_i)} y_{i+2}
    \\ 
    - \left(
    \frac{1}{(E_{i+1}-E_i)^2} + \frac{1}{(E_{i+2}-E_{i+1})(E_{i+1}-E_i)} 
    \right) y_{i+1}
    \,.
\end{multline}
This discretized version of the second derivative can be cast into a matrix equation with a matrix $\mat{J}_{2nd}$ that maps the values in the vector $\vec{y}_\textrm{GP}$ associated with the energies $E_1, \dots, E_M$ to the second derivatives $\vec{y}_\textrm{2nd}$ at energies $E_1, \dots, E_{M-2}$,
\begin{equation}
    \vec{y}_\textrm{2nd} = \vec{y}_\textrm{2nd}^\textrm{ref} 
    + \mat{J}_\textrm{2nd} \, (\vec{y}_\textrm{GP} - \vec{y}_\textrm{GP}^\textrm{ref})
    + (\vec{\tau}_\textrm{2nd} - \vec{\tau}_\textrm{2nd}^\textrm{ref}) \,,
\end{equation}
with $\vec{y}_\textrm{2nd}^\textrm{ref} = \mat{J}_\textrm{2nd} \vec{y}_\textrm{GP}^\textrm{ref} + \vec{\tau}_\textrm{2nd}^\textrm{ref}$.
We also introduced a random variable $\vec{\tau}_\textrm{2nd}$ to fulfill the requirement that every observable node must be associated with a parentless random variable in order to apply \cref{eq:baynet_upd_cov,eq:baynet_upd_mean}. 
We can now extend the Bayesian network to accomodate the vector $\vec{y}_\textrm{2nd}$ with the second derivatives of $\vec{y}_\textrm{GP}$:
\begin{center}
\tikz{
    \node[obs](H3){$\vec{\sigma}_\textrm{exp}$};
    \node[latent, below=of H3, yshift=0.7cm, xshift=-0.5cm](H2){$\vec{\tau}$};
    \node[latent, left=of H3, xshift=0.0cm](H1){$\vec{y}_\textrm{GP}$};
    \node[obs, below=of H1, yshift=0.5cm](H4){$\vec{y}_\textrm{2nd}$};
    \node[latent, left=of H4, xshift=0.5cm](H5){$\vec{\tau}_\textrm{2nd}$};
    \edge{H1,H2}{H3}
    \edge{H1}{H4}
    \edge{H5}{H4}
}
\end{center}
As a reminder, nodes filled with gray color are assumed to be observed.
The additional node $\vec{\tau}_\textrm{2nd}$ introduces stochasticity which weakens the link between the observed vector $\vec{y}_\textrm{2nd}$ and the function values in $\vec{y}_\textrm{GP}$.
Without this additinal noise term, the observation of $\vec{y}_\textrm{2nd}$ would fully determine $\vec{y}_\textrm{GP}$ up to two integration constants, which could be for instance the function value of $y_1$ and the slope at the same energy.

\begin{figure}[tb]
    \centering
    \includesvg{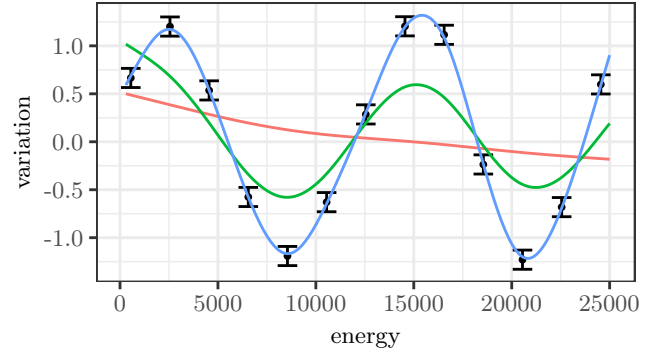}
    \caption{Impact of the uncertainty assigned to the pseudo-observations of the second derivative on the evaluated curve.
    The computational mesh contains 1000 points and the synthetic measurements follow a sine wave perturbed by random noise with a standard deviation of 0.2.
    The uncertainty used for the pseudo-observations at each of the thousand mesh points to obtain the evaluated blue, green and red curve is $10^{-6}$, $10^{-7}$ and $10^{-8}$, respectively.
    }
    \label{fig:sparsegp-reg-example}
\end{figure}

The corresponding compound vectors $\vec{y}$ and $\vec{z}$ need to be extended to
\begin{equation}
    \vec{y} = \begin{pmatrix}
        \vec{\sigma}_\textrm{exp} \\
        \vec{y}_\textrm{GP} \\
        \vec{y}_\textrm{2nd}
    \end{pmatrix}
    \;\;\;\textrm{and}\;\;\;
    \vec{z} = \begin{pmatrix}
        \vec{\tau} \\
        \vec{y}_\textrm{GP} \\
        \vec{y}_\textrm{2nd}
    \end{pmatrix}
\end{equation}
and the reference vectors to
\begin{equation}
    \vec{y}_\textrm{ref} = \begin{pmatrix}
        \vec{\sigma}_\textrm{exp}^\textrm{ref} \\
        \vec{y}_\textrm{GP}^\textrm{ref} \\
        \vec{y}_\textrm{2nd}^\textrm{ref}
    \end{pmatrix}
    \;\;\;\textrm{and}\;\;\;
    \vec{z}_\textrm{ref} = \begin{pmatrix}
        \vec{\tau}^\textrm{ref} \\
        \vec{y}_\textrm{GP}^\textrm{ref} \\
        \vec{\tau}_\textrm{2nd}^\textrm{ref}
    \end{pmatrix} \,.
\end{equation}
The overall mapping matrix is now given by
\begin{equation}
    \mat{S} = \begin{pmatrix}
        \mathbb{1} & \mat{J}_\textrm{GP} & \mat{0} \\
        \mat{0} & \mathbb{1} & \mat{0} \\
        \mat{0} & \mat{J}_\textrm{2nd} & \mathbb{1}
    \end{pmatrix} \,.
\end{equation}
Please note that also $\mat{J}_\textrm{2nd}$ is very sparse as it only contains three non-zero elements per row.
The associated prior covariance matrix is of the form
\begin{equation}
    \mat{U} = \begin{pmatrix}
        \mat{U}_{\vec{\tau}} & \mat{0} & \mat{0} \\
        \mat{0} & \mat{U}_\textrm{GP} & \mat{0} \\
        \mat{0} & \mat{0} & \mat{U}_\textrm{2nd}
    \end{pmatrix} \,.
\end{equation}
We can use diagonal matrices for all the submatrices $\mat{U}_{\vec{\tau}}, \mat{U}_\textrm{GP}, \mat{U}_\textrm{2nd}$.
Let us denote by $P$ and $Q$ the index sets associated with the subvectors $\vec{\sigma}_\textrm{exp}$ and $\vec{y}_\textrm{2nd}$, respectively.
The index set $D$ indicating the observed variables is therefore given by $D = P \cup Q$. 
Finally, the vector $v$ in \cref{eq:vvec_general} required to apply the GLS method in \cref{eq:baynet_upd_mean} is defined by
\begin{equation}
    \vec{v}_{\neg D} = \vec{u}_{\neg D},
    \vec{v}_{P} = \vec{r}, \textrm{and}\;
    \vec{v}_{Q} = \vec{0} \,.
\end{equation}
The vector $\vec{r}$ contains the observed values of the vector $\vec{\sigma}_\textrm{exp}$.
By imposing $\vec{y}_Q = \vec{0}$ we assume that the observed second derivatives at all energies $E_1,\dots,E_{M-2}$ are zero but these measurements were affected by a statistical error whose uncertainty is reflected in $\mat{U}_\textrm{2nd}$.
The use of pseudo-observations of the second derivative is a way to implicitly define a Gaussian process prior on $\vec{y}_\textrm{GP}$ that favors a certain degree of smoothness.
Effectively, this construction penalizes solutions according to the magnitude of the sum of squared second derivatives at the various mesh points.
The continuous version of the same penalty criterion is employed in the fitting of smoothing splines, e.g.,~\cite{greenNonparametricRegressionGeneralized1993}.
Alternatively, one may regard the construction as a kind of Tikhonov regularization or ridge regression, e.g.,~\cite{tikhonovSolutionsIllposedProblems1977}, in the space of second derivatives.
\Cref{fig:sparsegp-reg-example} gives an impression of how the uncertainties assigned to the pseudo-observations of the second derivatives impact the result of the Bayesian inference.

As each element in $\vec{y}_\textrm{2nd}$ is associated with the second derivative at a particular energy $E_i$, we can use different variances along the diagonal of $\mat{U}_\textrm{2nd}$ to change the degree of smoothness in different energy ranges.
In addition, if we use very large variances in the diagonal of $\mat{U}_\textrm{GP}$, we only impose our knowledge about the expected smoothness.
In principle, we can also augment the Bayesian network with a node with pseudo-observations of the first derivative (slope).
The construction is analogous to the one to incorporate pseudo-observations on the second derivative.

To conclude this section, we remark that observations of first derivatives (gradients) are also taken into account in the construction of surrogate models to emulate expensive computer models, e.g., in the engineering context. 
If the surrogate model is based on a Gaussian process, this approach is referred to as gradient-enhanced Kriging, e.g.,~\cite{debaarImprovementsGradientenhancedKriging2014}.
Whereas in gradient-enhanced Kriging real observations of the gradient of a computer model are used to better constrain the surrogate model, the uncertainties of the pseudo-observations of the second derivatives introduced in this section act as additional free parameters that permit to regulate the smoothness of feasible solutions.

\subsection{\label{subsec:imp-nonlinear-maps}Two important non-linear mappings}
The Bayesian network framework in the form presented is general and can incorporate any linear or non-linear mapping as long as the functions for propagation and for the computation of the Jacobian are available.
Two experimental aspects, which are relative normalization errors and errors in the energy calibration, are very relevant for many experiments.
As they cannot be properly addressed in the GLS method, we highlight the treatment of these experimental aspects in the Bayesian network framework.

In time-of-flight (TOF) experiments, the true energy $E$ of the particle beam is related to the assumed energy $E'$ in first approximation by
\begin{equation}
    \label{eq:encalibunc}
    E = \alpha + (1+\beta) E' \,,
\end{equation}
with $\alpha$ being a constant energy shift and $(1+\beta)$ being a scaling factor.
In the previous section, we established \cref{eq:pwlinint} to interpolate the values from a mesh with knot points at energies $E_1, E_2, \dots, E_M$ to obtain the value at an energy $E$,
\begin{equation}
    \label{eq:linint_enunc}
    g(E) =
    \sum_{i=1}^{M} f_i(E) y_{i} \,.
\end{equation}
If the experimental energy is perfectly known, the factor $f_i(E)$ to map the values on the mesh to the experimental energy is a constant and consequently the interpolation result is a linear function of the values $\vec{y}_i$ on the mesh.
However, by acknowledging that $E$ is also uncertain because we do not perfectly know the precise values of $\alpha$ and $\beta$, this relationship becomes non-linear.

In the visualization of the Bayesian network, we can choose which collection of variables should be summarized to separate nodes.
As the variables $\vec{y} = (y_1, y_2, \dots, y_M)$ referring to the true cross section are independent of specific experiments, it is reasonable to group them to a node.
The variables $\alpha$ and $\beta$ are specific for an experiment and therefore may be grouped together to $\vec{\gamma} = (\alpha, \beta)^T$ as one node.
These choices lead to the following visualization of the Bayesian network:
\begin{center}
\tikz{
    \node[latent](H3){$\vec{\sigma}_\textrm{exp}$};
    \node[latent, left=of H3, yshift=-0.7cm](H2){$\vec{\gamma}$};
    \node[latent, left=of H3, yshift=0.7cm](H1){$\vec{y}$};
    \edge{H1,H2}{H3}
    \draw[dashed] (H1) -- (H2);
}
\end{center}
We did not attach a parentless node $\vec{\tau}$ to $\vec{\sigma}_\textrm{exp}$ but its presence for nodes with parents is implicitly understood.

For cases where the non-linearity arises due to non-linear interactions between source variables, such as between $\vec{y}$ and $\vec{\gamma}$, we introduce the convention that those nodes are connected by a dashed line.

The non-linear propagation of the variables $\vec{y}$ and $\vec{\gamma}$ is already established by \cref{eq:encalibunc} in combination with \cref{eq:linint_enunc}.
Also the partial derivatives to construct the Jacobian matrix can be readily computed,
\begin{align}
    \frac{\partial g(E)}{\vec{y}_k} &= f_k(E)  \,, \\
    \frac{\partial g(E)}{\partial \alpha} &= \sum_{i=1}^{M} \frac{d f_i(E)}{d E} y_{i} \,, \\
    \frac{\partial g(E)}{\partial \beta} &= E \sum_{i=1}^{M} \frac{d f_i(E)}{d E} y_{i} \,.
\end{align}
The appearance of the input variables $y_i$ in the Jacobian matrix is the signature of a non-linear relationship.

Analogous to the examples in previous sections, we can define the column vector $\vec{y} = (\vec{\sigma}_\textrm{exp}^T, \vec{y}^T, \vec{\gamma}^T )^T$.
With the positions of the subvectors in $\vec{y}$ defined, the source index set $\mathcal{S}$ of the mapping refers to the positions of $\vec{y}$ and $\vec{\gamma}$.
The target index set $\mathcal{T}$ refers to the indices associated with $\vec{\sigma}_\textrm{exp}$.
Such a mapping can be used in Bayesian networks as a building block.

Another experimental aspect is the energy resolution of the measurement.
Finite resolutions can be modeled by convolutions, which like linear interpolation represents a linear relationship between the variables on the mesh and the value propagated to the experimental energy.
The combination of convolutions with energy calibration errors works analogous to the case of linear interpolation.

As a second example, we elaborate on the modeling of relative normalization errors.
Within the GLS method, relative normalization uncertainties must be converted to absolute ones by using the measured values as reference.
However, this approach leads to a systematic underestimation of the uncertainty of measured values below and a systematic overestimation of measured values above the true value, effectively biasing the results of the GLS method towards lower values.
The correct way to deal with a relative normalization error is to use the true cross section as the reference.
Using again the same construction as in \cref{eq:linint_enunc} to map from the true cross sections given on the computational mesh to the experimental energy, the relationship between an experimental measurement and the true cross section can be written as:
\begin{equation}
    \vec{\sigma}_\textrm{exp} = 
    g(\vec{E}_\textrm{exp}) + \eta \, g(\vec{E}_\textrm{exp}) \,,
\end{equation}
where $g(\vec{E}_\textrm{exp})$ represents the vector given by $(g(E_{\textrm{exp},1}), g(E_{\textrm{exp},2}), \dots)^T$ with $E_\textrm{exp},i$ being the experimental energies.
The variable $\eta$ is the relative normalization error.
The associated Bayesian network is given by
\begin{center}
\tikz{
    \node[latent](H3){$\vec{\sigma}_\textrm{exp}$};
    \node[latent, left=of H3, yshift=-0.7cm](H2){$\eta$};
    \node[latent, left=of H3, yshift=0.7cm](H1){$\vec{y}$};
    \edge{H1,H2}{H3}
    \draw[dashed] (H1) -- (H2);
}
\end{center}

Conceptually, we can think of $\vec{\sigma}_\textrm{exp}$ as the superposition of two mappings, which are the linear interpolation from the computational mesh to the experimental energies and the contribution of the relative normalization error. 
The mapping associated with the first term $g(\vec{E}_\textrm{exp})$ is linear and we do not discuss it further.
However, the mapping of the relative normalization error is non-linear.
The partial derivatives to construct the associated Jacobian matrix are
\begin{equation}
    \frac{\partial g(E)}{\vec{y}_k} = \eta \, f_k(E)
    \;\;\;\textrm{and}\;\;\;
    \frac{\partial g(E)}{\partial \eta} = g(\vec{E})  \,. 
\end{equation}

In both the case of the energy calibration error and the relative normalization error, the mapping used as input the true cross section vector, which is of course not known.
During the iterative scheme of the LM algorithm, the current best estimate of the true cross section vector is used as reference for the relative normalization error. 
As already described in~\cref{subsec:nonlinearfuns}, the LM algorithm is able to locate an assignment of values to the variables, e.g., to $\vec{\eta}$ and $\vec{y}$, that corresponds to the maximum of the posterior distribution taking into account the exact non-linear relationships.

\subsection{Few remarks on practical Bayesian network modeling}
Before we present the Bayesian network examples in the next section, which make use of the algorithms developed in the method part in orchestration, we want to summarize the possibilities in modeling and give an idea of how Bayesian network modeling is done in practice using the \textit{nucdataBaynet} package.

We aimed to provide a comprehensive account of the mathematics involved to deal with Bayesian networks with multivariate normal distributions as priors on variables and non-linear, deterministic and nested relationships between variables.
We also discussed a flexible sparse Gaussian process prior construction that scales well to large datasets and elaborated on two important non-linear mappings associated with experimental data, which are relative normalization errors and energy calibration errors.

However, from the user point of view, the benefit of Bayesian networks is that they are a helpful mental abstraction to find valid Bayesian models.
Furthermore, Bayesian networks can be quickly created on the computer by assembling common mappings, such as linear interpolation mappings, convolutions and non-linear transformations, like a puzzle.
Therefore, in the following, we want to convey a basic understanding of how Bayesian models are defined in practice using the \textit{nucdataBaynet} package from the user point of view.

The first step is to define a data table with the information about the variables present in the Bayesian network.
An example data table is shown in \cref{tbl:nodedef}.
Each row is associated with a variable.
The column IDX establishes the position of all independent variables in the vector $\vec{z}$ defined in \cref{eq:zdef}.
The column NODE establishes names for the nodes in the Bayesian network.
Several variables can be grouped to a single node, hence the multiple appearance of the same node label.
The column PRIOR contains the prior expectation in $\vec{u}$ and the column UNC the prior uncertainty associated with each variable in $\vec{z}$.
Please note that the column UNC is for convenience in the case that the prior covariance matrix $\mat{U}$ is diagonal and the diagonal elements are given by the squared values of the UNC column.
However, it is also possible to introduce covariance elements between variables, but then instead of the column UNC, a full (and better sparse) prior covariance matrix must be defined along the data table.
The column OBS contains the observed values of dependent nodes.
The special value NA (=not available) means that the dependent node was not observed.
Please be aware that the values in PRIOR and UNC refer to the values in $\vec{z}$ whereas the values in OBS refer to the observed values of dependent nodes $\vec{y}$.
The (linear) relationship between the variables in $\vec{z}$ and those in $\vec{y}$ was stated in \cref{eq:baynet_mapping}.
\begin{table}[ht]
\centering
\begin{tabular}{clrrrrcr}
  \hline
IDX & NODE & PRIOR & UNC & OBS & REAC & ENERGY & EXPID \\ 
  \hline
  1 & truexs  & 1000 & 500 & NA & (N,TOT) & 1.0 & NA \\ 
  2 & truexs  & 1000 & 500 & NA & (N,TOT) & 1.5 & NA \\ 
  3 & truexs  & 1000 & 500 & NA & (N,TOT) & 2.0 & NA \\ 
  4 & normerr &    0 & 100 & NA & (N,TOT) & NA  & 20733  \\ 
  5 & exp     &    0 & 50  & NA & (N,TOT) & 1.7 & 20733  \\
   \hline
\end{tabular}
\caption{Structure of a data table with the definition of the variables of the Bayesian network}
\label{tbl:nodedef}
\end{table}
The columns REAC, ENERGY and EXPID are provided as examples of columns that are helpful for the definition of the mappings between nodes, as we will see in a moment.
Depending on the evaluation situation, columns with other information may be more pertinent.

A mapping is defined by a list with several named variables to specify its characteristics.
The essential variables present in the definition of any mapping are \textit{maptype} to define the type of mapping (e.g., linear interpolation), \textit{mapname} to assign a unique name to the mapping, \textit{src\_idx} to define the source index set $\mathcal{S}$ and \textit{tar\_idx} to define the target index set $\mathcal{T}$ of the specific mapping.
Depending on the type of mapping, additional variables may be required, such as \textit{src\_x} and \textit{tar\_x} to define the (e.g., energy) mesh of the source variables and the mesh of the target variables in the case of a mapping that implements linear interpolation.
These lists are preferably not created by hand but by query operations on the node data table.
An example R code to define a linear interpolation mapping is presented in \cref{lst:example_mapping_def}.
The variable name \textit{dt} refers to the node data table, in this example given by \cref{tbl:nodedef}.
Query operations, such as \textit{NODE=="truexs"} are employed to select specific rows and to retrieve the values of a particular column, e.g., ENERGY.
The query syntax is very flexible and different criteria can be combined to retrieve the relevant information of the rows of interest in order to define a mapping.
\begin{lstlisting}[language=R,caption=Example R code to create a list with the defiition of a linear interpolation mapping,captionpos=b,
label=lst:example_mapping_def,float]
list(
  maptype = "linearinterpol_map",
  mapname = "truexs_to_experiment",
  src_idx = dt[NODE=="truexs", IDX],
  tar_idx = dt[NODE=="exp", IDX],
  src_x = dt[NODE=="truexs", ENERGY],
  tar_x = dt[NODE=="exp", ENERGY],
)
\end{lstlisting}
The collection of those lists with the mapping definitions determine the link structure of the Bayesian network and are bundled together to a so-called compound mapping, which is also specified by a similar list as the one given in \cref{lst:example_mapping_def} but containing as variable a list of all the individual mapping definitions.
The list with the compound mapping specification is used to instantiate a mapping object (or more precisely a closure in R, which can be regarded for all practical matters as an object).

The functions implementing the inference algorithms, such as the GLS method and the LM algorithm, are called with the compound mapping object and the relevant prior specifications given in the node data table.
Additional functions exist to evaluate parts of the posterior covariance matrix and to create samples from the (approximate) posterior distribution.
Functions to visualize Bayesian networks are also available.

\section{\label{sec:examples}Examples}

We demonstrate the use of Bayesian networks in relevant evaluation scenarios.
The examples emerged in the context of the neutron data standards project and the INDEN evaluation efforts of structural materials.
As for the GLS method, the quality of the results using Bayesian networks depends on the validity of the assumptions imposed.
The objective of these examples is to demonstrate the modeling possibilities outlined in the method part in practice.
To this end, some assumptions were taken for the sake of demonstration and not because they are believed to be a sound reflection of expert knowledge.
Therefore, we do not claim or recommend that the results in this section should replace established evaluations.
Nevertheless they serve as proof of the technical feasibility and make the benefits of Bayesian network modeling in the nuclear data context tangible.

\subsection{\label{subsec:example-usu}Energy-dependent USU error component}

Recently the concept of so-called Unrecognized Sources of Uncertainty (USU) has been extensively discussed in~\cite{capoteUnrecognizedSourcesUncertainties2020}.
It was suggested that even if the greatest care in the determination of experimental error sources is applied, there may be still error contributions remaining which the evaluators are unaware of.
If the possibility of such contributions is not taken into account, evaluated uncertainties may be too optimistic.
As an additional complication for some quantities, the magnitude of these unrecognized errors may be energy-dependent.

In this example, we want to demonstrate how the sparse GP construction introduced in~\cref{subsec:sparse-gp-construction} can be used within the Bayesian network framework to model energy-dependent USU error components in the evaluation procedure.

As an important disclaimer, please be aware that USU as defined in~\cite{capoteUnrecognizedSourcesUncertainties2020} refers to unrecognized errors missed by an evaluator after a diligent effort to comprehensively quantify all uncertainties due to error sources associated with an experiment. 
It may well be that a large part of the contribution we refer to as USU error component in this example can be reattributed to known error sources in a serious evaluation effort.
How we name an error component is therefore a matter of available knowledge and interpretation but the mathematical modelling can be the same.

We use measurement data of the $^{235}$U(n,f) cross section, which is a neutron data standard~\cite{carlsonInternationalEvaluationNeutron2009,carlsonEvaluationNeutronData2018} between 150\,eV and 200 MeV, and an important reaction to design and assess nuclear reactors.
We consider the five datasets listed in~\cref{tbl:u5nf-datasets}.
\begin{table}[ht]
\centering
\begin{tabular}{lrlrl}
  \hline
EXFOR & NUM & AUTHOR & YEAR & REF \\ 
  \hline
20483 & 205 & Blons et al & 1971 & \cite{blonsMeasurementAnalysisFission1971} \\ 
  20783 & 308 & Migneco et al & 1975 & \cite{mignecoIntermediateStructureKeV1975} \\ 
  20826 & 155 & Wagemans et al & 1976 & \cite{wagemansNeutronInducedFission1976} \\ 
  12877 & 482 & Weston et al & 1984 & \cite{westonSubthresholdFissionCross1984} \\ 
  23294 & 310 & Paradela et al & 2016 & \cite{paradelaHighAccuracy2352016} \\ 
   \hline
\end{tabular}
\caption{Experimental datasets used in the example evaluation of $^{235}$U(n,f) reaction between seven and 9 keV. The column EXFOR contains the EXFOR accession number, NUM the number of datapoints of the datasets in the energy range considered. } 
\label{tbl:u5nf-datasets}
\end{table}

\begin{figure}[htb]
    \centering
    \includesvg{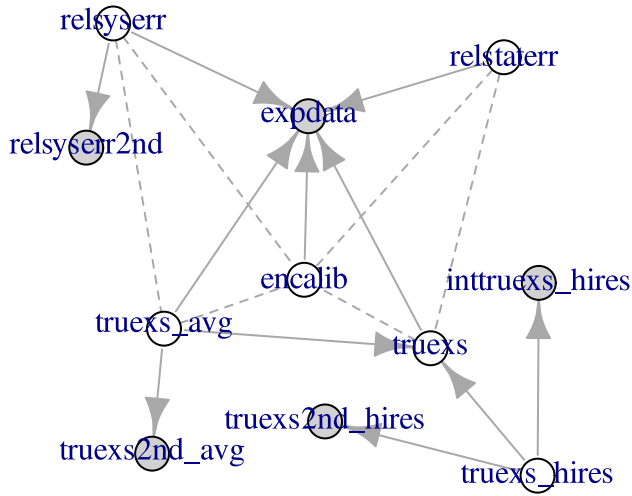}
    \caption{Bayesian network for the evaluation of the $^{235}$U(n,f) cross section}
    \label{fig:U5nf-baynet}
\end{figure}
We created the Bayesian network in \cref{fig:U5nf-baynet} to model the dependencies between the pieces of information.
The node \textit{truexs} represents the variables that contain the true cross sections on a densely defined computational mesh with a spacing of 1\,eV.
It is made up of the transformed sum of two components \textit{truexs\_avg} and \textit{truexs\_hires} which represent a smoothly varying cross section to capture an average trend and a high-resolution component to capture an overlaid resonance structure.
The transformation keeps the result of the sum as it is if the sum is positive and otherwise yields zero.
This transformation is known as Rectified Linear Unit (ReLU) in the machine learning field, e.g.,~\cite{fukushimaVisualFeatureExtraction1969}, and ensures in our case that all cross sections are non-negative.
The mapping from the mesh of \textit{truexs\_avg} and \textit{truexs\_hires} to the one of \textit{truexs} is established by linear interpolation. 
Prior knowledge about the smoothness of these components is incorporated by using a coarser mesh for \textit{truexs\_avg} (spacing 50\,eV) than for \textit{truexs\_hires} (spacing 1\,eV) and by introducing for both components pseudo-observations of the second derivative as explained in~\cref{subsec:sparse-gp-construction} elaborating on a sparse Gaussian process construction.
The nodes associated with the second derivatives of \textit{truexs\_avg} and \textit{truexs\_hires} are named \textit{truexs2nd\_avg} and \textit{truexs2nd\_hires}, respectively.
We use smaller prior uncertainties for the variables of \textit{truexs2nd\_avg} than for \textit{truexs2nd\_hires} to ensure that one component really captures an average trend and the other one the high-resolution behavior.
However, the posterior uncertainties of \textit{truexs\_avg} and \textit{truexs\_hires} would remain quite large without further prior constraints.
The reason is that the prior on the second derivative is indifferent to shifting the function up or down and only the sum of \textit{truexs\_avg} and \textit{truexs\_hires} relates to the experiments, hence the distribution of possible global shifts to these two components is only determined with an uncertainty given by the one barn prior uncertainty associated with \textit{truexs\_hires} as it is smaller than the ten barn prior uncertainty of \textit{truexs\_avg}.
To solve this issue, we introduce a node \textit{inttruexs\_hires} that represents binned averages of \textit{truexs\_hires} and associate this node with pseudo-observations with a zero value and small uncertainty.
With this additional constraint, the average of the posterior estimate of the high-resolution component is close to zero for large enough energy intervals covering several peaks and valleys.
Consequently the global shift of the \textit{truexs\_hires} component is fixed to be close to zero. 

The other nodes reflect the experimental information of the five datasets.
The experimental data points represented by \textit{expdata} are modeled as a sum consisting of the convoluted true cross section \textit{truexs}, a statistical error component \textit{relstaterr} and an energy-dependent systematic error component \textit{relsyserr}.
Without further knowledge about the experimental error(s) associated with the energy-dependent systematic error component, we can regard it as an USU error component.

The employed convolution mapping from \textit{truexs} to \textit{expdata} averages the true cross section within a window of certain size centered at the true experimental energy given by the linear transformation stated in~\cref{eq:encalibunc}.
The respective parameters of the energy transformation are summarized in the node \textit{encalib}.
The non-linearity in the convolution mapping due to the energy transformation is indicated by the dashed lines between those two nodes in \cref{fig:U5nf-baynet}.

The energy-dependent systematic error component is given relative to the true average cross section \textit{truexs\_avg} whereas the statistical error component relative to the high-resolution true cross section \textit{truexs\_hires}. 
The specification as relative errors make these mappings non-linear, which is indicated by the dashed lines between \textit{relstaterr} and \textit{truexs}, and \textit{relsyserr} and \textit{truexs\_avg}, respectively.
The mapping from \textit{relsyserr} to \textit{expdata} is given by the same convolution operation as used in the mapping from \textit{truexs} to \textit{expdata} (including the energy transformation), but also taking into account that the values in \textit{relsyserr} are given relative to \textit{truexs\_avg}.
Therefore, \textit{relsyserr} is connected by dashed lines to the \textit{truexs\_avg} and \textit{encalib} nodes.
We use the approach of pseudo-observation of second derivatives another time to impose a smoothness prior on \textit{relsyserr}.
The node associated with the second derivatives is named \textit{relsyserr2nd}.

To keep the representation of the Bayesian network compact, we aggregated the experimental components referring to individual datasets together.
Differently stated, each of the nodes \textit{encalib}, \textit{relsyserr}, \textit{relsyserr2nd}, \textit{relstaterr} and \textit{expdata} can be split up into five nodes referring to the various dataset.
The nodes associated with one dataset do not have any direct connections to the nodes of another dataset.
They are only indirectly linked over the nodes \textit{truexs} and \textit{truexs\_avg}.

\begin{table}[ht]
\centering
\begin{tabular}{l|ccrrr}
  \hline
NODE & PRIOR & UNC & EMIN & EMAX & NUM \\
  \hline
expdata &   0 & 0.01 & 7001 & 11998 & 1460 \\
  relstaterr &   0 & 0.03 & 7001 & 11998 & 1460 \\
  relsyserr (all) &   0 & 0.05 & 6851 & 12115 & 490 \\
  relsyserr (Paradela) &   0 &   0 & 6851 & 12115 & 490 \\
  relsyserr2nd &   0 & $1 \times 10^{-6}$ & 6901 & 12065 & 480 \\
  truexs &   0 &   0 & 6000 & 14000 & 8001 \\
  truexs\_avg &   0 & $1 \times 10^{4}$ & 6000 & 14000 & 161 \\
  truexs2nd\_avg &   0 & $1 \times 10^{-5}$ & 6050 & 13950 & 159 \\
  truexs\_hires &   0 & $1 \times 10^{3}$ & 6000 & 14000 & 8001 \\
  truexs2nd\_hires &   0 &   1 & 6001 & 13999 & 7999 \\
  inttruexs\_hires &   0 & 0.001 & 6101 & 13801 &  78 \\
  encalib (resolution) &   4 &   2 &  &  &  15 \\
  encalib (shift $\alpha$) &   0 & 0.05 &  &  &  15 \\
  encalib (scaling $\beta$) &   0 & 0.05 &  &  &  15 \\
   \hline
\end{tabular}
\caption{Prior estimates and uncertainties of the variables aggregated to the various nodes.
The number of variables associated with a node is given in column NUM.
The range of the associated energy mesh is given by columns EMIN and EMAX.
Priors on cross sections (truexs, truexs\_hires and truexs\_avg) are given in barn.
All energy related quantities (encalib (resolution), EMIN, EMAX) are given in eV.
}
\label{tbl:U5nf-priorspec}
\end{table}
Information about the nodes and the associated prior specifications are displayed in~\cref{tbl:U5nf-priorspec}.
No effort was undertaken to base these specifications on the experimental details of the measurements.
We assumed that each experimental point is affected by a relative statistical uncertainty of 3\% (relative to the true high-resolution cross section). 
With the exception of the measurement by Paradela and colleagues, we assumed a 5\% prior uncertainty for the relative energy-dependent systematic error (relative to the true average cross section). 
For the measurement of Paradela and colleagues, we specified that no energy-dependent systematic error is present.
This choice is only for the sake of demonstration and should not be taken as judgment about the quality of this or the other experiments.
The large prior uncertainties of \textit{truexs\_avg}, \textit{truexs\_hires} and \textit{truexs} are chosen to express our a priori indifference about the cross section values.
The prior uncertainties of \textit{truexs\_avg} are assumed larger than of \textit{truexs\_hires} to favor larger adjustments of the former component to ensure that it captures an average trend and the latter component additive fluctuations.
The prior uncertainties associated with the pseudo-observations of the second derivatives were manually fine-tuned to obtain visually pleasing fits and were guided by the following considerations.
The spacing of equidistant energy associated with \textit{truexs\_avg} is 50\,eV.
The square of the energy spacing appears in the denominators in~\cref{eq:finitediff2nd} so we work already with a baseline of $1/50^2= 1/2500 = 4 \times 10^{-4}$.
If we assume that changes within 50\,eV of the smooth trend are somewhere between $0.01$ and $0.1$ barn, which appears to be a plausible assumption considering~\cref{fig:U5nf-evaluation}, we end up with a prior uncertainty somewhere around $10^{-5}$.
To see this, we can choose, e.g., $y_i = y_{i+1} = 0$ and $y_{i+2}=0.01$ in~\cref{eq:finitediff2nd}.
Similar considerations lead to the prior uncertainty of one for the \textit{truexs2nd\_hires} node.
Together with an energy spacing of 1\,eV, this prior uncertainty implies the prior assumption that \textit{truexs\_hires} is expected to exhibit changes of about one barn in a one electronvolt energy interval.
The spacing of the energy mesh of the relative energy-dependent systematic error associated with each experiment is 50\,eV.
The prior uncertainty of $10^{-6}$ implies that we expect it to change less quickly as a function of energy than the average trend given by \textit{truexs\_avg}. 

Please note that all these assumptions are choices, which are ideally informed by physics knowledge but will be in pratice based to some extent on the subjective judgement of the modeler.
Data-driven approaches, such as proposed in~\cite{schnabelFittingAnalysisTechnique2018}, to determine good values for uncertainties may help in specific cases but can be expected to be less useful in complex Bayesian networks due to the large amount of degeneracy.

The Bayesian network comprises in total 28304 variables.
The associated prior matrix is very sparse with a proportion of $2.5\times 10^{-5}$ non-zero elements.
Only 2\% of the elements in the inverse posterior covariance matrix associated with the independent variables are non-zero. 

Due to the non-linear mappings between nodes, such as associated with the relative normalization error, several local maxima exist in the posterior distribution, which we discovered by performing several optimization runs.
Finally, to arrive at a good local maximum, we divided the optimization into several stages and only optimized a subset of the nodes in each stage.
We followed the general heuristic to first adjust global components, such as \textit{truexs\_avg} and subsequently include the optimization of local or highly fluctuating components.
Therefore, in the first stage, we adjusted only the true average cross section \textit{truexs\_avg} for a maximum of ten iterations.
\begin{figure}[!t]
    \centering
    \includesvg{img/example1/U5nf-evaluation.svg}
        \caption{Posterior estimate and credibility band of the true average cross section (\textit{truexs\_avg}).
	The brown highly fluctuating curve in the background represents the posterior estimate of the true cross section (\textit{truexs}), which is the sum of \textit{truexs\_avg} and \textit{truexs\_hires}. 
	}
    \label{fig:U5nf-evaluation}
    \vspace{0.5cm}
    \includesvg{img/example1/U5nf-relnormerr.svg}
        \caption{Posterior estimates and credibility bands of the energy-dependent systematic error components (\textit{relsyserr}) of two experimental datasets: Blons (71) in orange and Weston (84) in green. }
    \label{fig:U5nf-relnormerr}
\end{figure}

Starting from the obtained values for the variables, we continued with the joint optimization of \textit{truexs\_avg}, \textit{relnormerr} and \textit{relstaterr} for a maximum of 30 iterations.
Afterwards, we optimized jointly \textit{relnormerr}, \textit{relstaterr} and \textit{truexs\_hires} for a maximum of 30 iterations.
In the final stage, we included all variables in the optimization with a maximum number of 300 iterations.
These scheme was adopted after some trial and error.
The complete optimization procedure took a couple of minutes.

Finally, we provide impressions of some evaluated quantities.
The most likely true average cross section according to the posterior distribution is shown in~\cref{fig:U5nf-evaluation}.
The most likely relative energy-dependent normalization error for two experimental datasets is displayed in~\cref{fig:U5nf-relnormerr}.

\FloatBarrier

\subsection{\label{subsec:example-ironslow}No-model evaluation of cross sections with resonance structure}

One part of the efforts within the \textit{International Nuclear Data Evaluation Network} (INDEN) is focused on the evaluation of neutron-induced reactions of structural materials.
One such material is $^{56}$Fe, which is notoriously difficult to evaluate in the incident energy range between a few hundred keV and about five MeV with either R-matrix fits and nuclear models.  
R-matrix fits are difficult because there is a large number of resonances, which cannot be experimentally resolved.
Nuclear models cannot describe well this energy range because they rely on the assumption that the structure due to resonances is completely averaged out, which is however also not true.
A third option is a no-model fit, i.e., to use a mathematical function, such as a spline, instead as a fitting function. 
There are several aspects that make a no-model fit in this region difficult, which are:
\begin{itemize}
    \item The large number of mesh points to capture the resonant structure of the cross sections,
    \item The preservation of consistency between the elastic, inelastic and total cross section, and,
    \item The enforcement of the constraint that cross sections must be non-negative.
\end{itemize}
Here we explore a potential solution, which uses the sparse Gaussian process construction described in this paper as a fitting function for cross section within the Bayesian framework for a consistent evaluation of the elastic, inelastic and total cross section.
We have included the datasets listed in~\cref{tbl:fe56-slow-datasets}.
\begin{table}[ht]
\centering
\begin{tabular}{llrlrl}
  \hline
REAC & EXFOR & NUM & AUTHOR & YEAR & REF \\ 
  \hline
EL & 40532014 &   2 & Korzh et al & 1977 & \cite{korzhStudyCrossSections1977} \\ 
  INL & 11700002 &   1 & Barrows & 1965 & \cite{barrowsStudyNgReactions1965} \\ 
  INL & 10529004 & 378 & Perey et al & 1971 & \cite{pereyHighResolutionInelastic1971} \\ 
  INL & 32201002 &   4 & Korzh et al & 1994 & \cite{korzhStudyCrossSections1994} \\ 
  INL & 23134005 &  11 & Beyer et al & 2014 & \cite{beyerInelasticScatteringFast2014} \\ 
  TOT & 13764002 & 426 & Harvey & 1987 &  \\ 
  TOT & 22316003 & 2050 & Rohr et al & 1995 &  \\ 
   \hline
\end{tabular}
\caption{Experimental datasets used in the example evaluation of neutron-induced reactionf of $^{56}$Fe between one and two MeV. The column EXFOR contains the EXFOR accession number, NUM the number of datapoints of the datasets in the energy range considered. A missing reference means that no accessible publication is known to the authors.} 
\label{tbl:fe56-slow-datasets}
\end{table}

\begin{figure*}[htb]
    \centering
    \includesvg{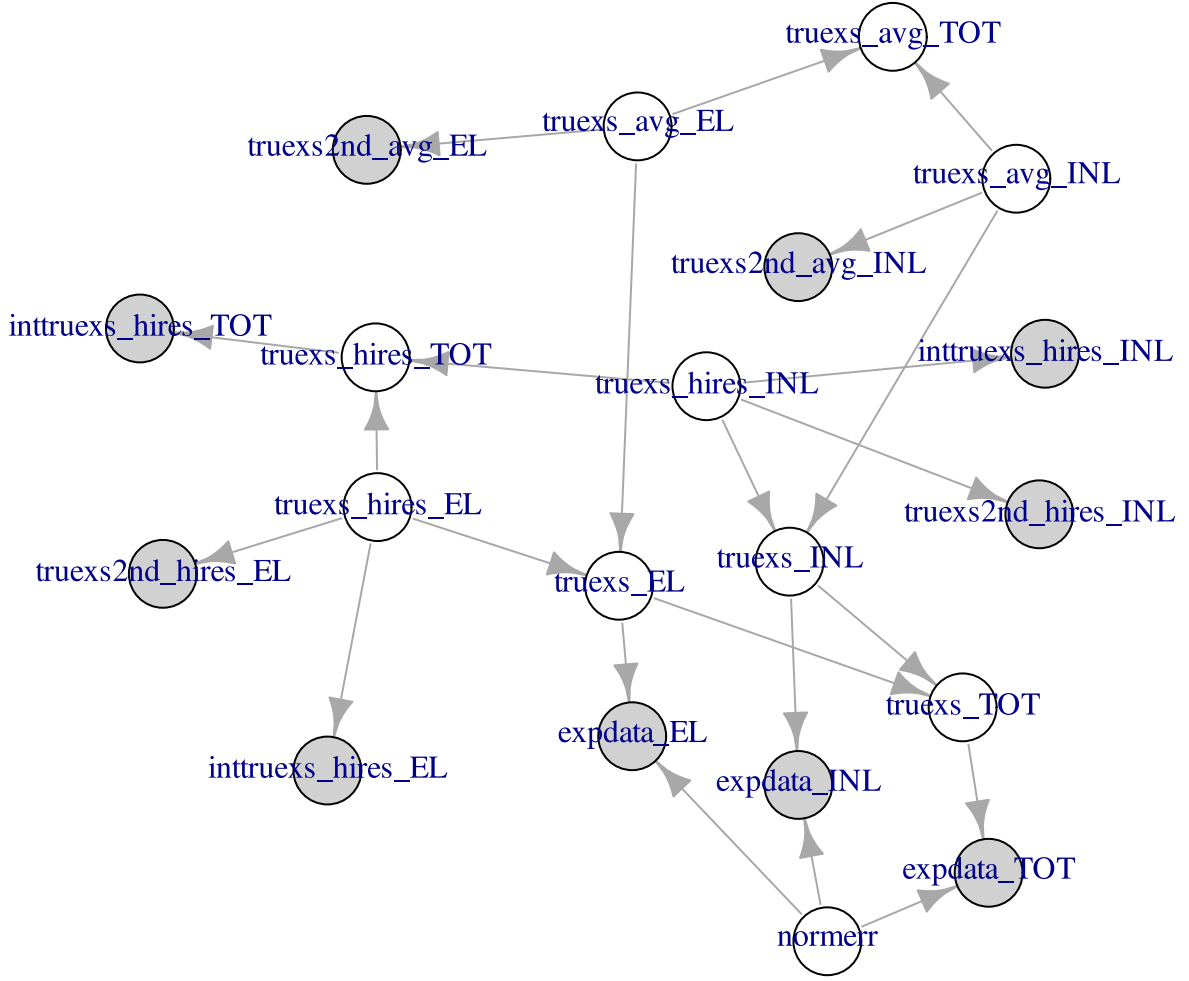}
    \caption{Bayesian network for an evaluation of neutron-induced reactions of $^{56}$Fe in the energy range between one and two MeV.
    The meaning of the nodes and the mappings between them are described in the text.}
    \label{fig:fe56low_baynet}
\end{figure*}

We have created the Bayesian network depicted in \cref{fig:fe56low_baynet}.
In the following description, we use \textit{REAC} in a node name to indicate that there is a node of this form for each reaction channel.
For example, the string \textit{truexs\_REAC} indicates that the nodes \textit{truexs\_EL}, \textit{truexs\_INL} and \textit{truexs\_TOT} are present in the network.

As an important clarification, we will use the term \textit{true cross section} to refer to the quickly fluctuating cross section curve followed by the experimental data. 
However, as single resonances cannot be resolved anymore in this energy range, what we label as true cross section is in reality an average trend of the true cross section over several resonances.
A more rigorous investigation of the pertinence of the Bayesian network approach to deal with unresolved resonances, the relationship to approaches based on probability tables, and the possibility of a coupling to processing codes for the resolved and unresolved resonance region, e.g., CALENDF~\cite{subletProbabilityTableBased2002}, remains as future work.
Quantitative analysis tools, such as the autocorrelation function to investigate the presence of structure~\cite{bertschFluctuations235Cross2018}, are expected to help in this regard.
We also note that we neglected the capture cross section in this example for the sake of simplicity.

The true cross section \textit{truexs\_REAC} of each reaction channel is modeled as a transformed sum of an average trend component \textit{truexs\_avg\_REAC} and a high-resolution component \textit{truexs\_hires\_REAC}.
Regarding the transformation, let $x$ be the value at energy $E$ of \textit{truexs\_avg\_REAC} and $y$ be the value at the same energy of \textit{truexs\_hires\_REAC}, the transformation is given by $\max(0, x+y)$.
This non-linear transformation, known as rectified linear unit (ReLU) in the machine learning field, ensures that all cross sections remain non-negative quantities. 
We enforce different degrees of smoothness of \textit{truexs\_avg\_REAC} and \textit{truexs\_hires\_REAC} by introducing pseudo-observations of their second derivatives represented by \textit{truexs2nd\_avg\_REAC} and \textit{truexs2nd\_hires\_REAC}, respectively.
As we use non-informative priors for \textit{truexs\_REAC} and the second derivative is invariant to shifts of the cross section, we need to ensure that \textit{truexs\_hires\_REAC} is in average zero so that \textit{truexs\_avg\_REAC} captures the average cross section. 
As in the last example, we achieve this by introducing pseudo-observations \textit{inttruexs\_hires\_REAC} of the high-resolution component, which represent averages of the high-resolution component in several overlapping intervals.
These pseudo-observations are taken to be zero with a low uncertainty compared to the magnitude of the resonance-like fluctuations.
The window size was manually fixed by inspecting the width of the resonance-like  structures.
Each experimental dataset was associated with an absolute normalization error represented by \textit{normerr}.
A summary of the individual nodes including their prior specifications and number of variables is given in \cref{tbl:fe56low-node-summary}.
\begin{table}[b]
\centering
\begin{tabular}{l|ccrrr}
  \hline
NODE & PRIOR & UNC & EMIN & EMAX & NUM \\ 
  \hline
expdata\_INL &   0 &  70 & 1.00 & 2.00 & 394 \\ 
  expdata\_TOT &   0 & $3 \times 10^{2}$ & 1.00 & 2.00 & 2476 \\ 
  expdata\_EL &   0 & $2 \times 10^{2}$ & 1.50 & 2.00 &   2 \\ 
  truexs\_INL &   0 &   0 & 0.75 & 2.25 & 1501 \\ 
  truexs\_TOT &   0 &   0 & 0.75 & 2.25 & 1501 \\ 
  truexs\_EL &   0 &   0 & 0.75 & 2.25 & 1501 \\ 
  normerr &   0 & $1 \times 10^{2}$ &  &  &   7 \\ 
  truexs\_hires\_INL &   0 & $1 \times 10^{4}$ & 0.75 & 2.25 & 1501 \\ 
  truexs\_hires\_TOT &   0 &   0 & 0.75 & 2.25 & 1501 \\ 
  truexs\_hires\_EL &   0 & $1 \times 10^{4}$ & 0.75 & 2.25 & 1501 \\ 
  inttruexs\_hires\_INL &   0 &  50 & 1.10 & 1.90 &   9 \\ 
  inttruexs\_hires\_TOT &   0 &  50 & 1.10 & 1.90 &   9 \\ 
  inttruexs\_hires\_EL &   0 &  50 & 1.10 & 1.90 &   9 \\ 
  truexs2nd\_hires\_INL &   0 & $1 \times 10^{8}$ & 0.75 & 2.25 & 1499 \\ 
  truexs2nd\_hires\_EL &   0 & $1 \times 10^{8}$ & 0.75 & 2.25 & 1499 \\ 
  truexs\_avg\_INL &   0 & $1 \times 10^{8}$ & 0.75 & 2.25 &  31 \\ 
  truexs\_avg\_TOT &   0 &   0 & 0.75 & 2.25 &  31 \\ 
  truexs\_avg\_EL &   0 & $1 \times 10^{8}$ & 0.75 & 2.25 &  31 \\ 
  truexs2nd\_avg\_INL &   0 & $1 \times 10^{4}$ & 0.80 & 2.20 &  29 \\ 
  truexs2nd\_avg\_EL &   0 & $1 \times 10^{4}$ & 0.80 & 2.20 &  29 \\ 
   \hline
\end{tabular}
    \caption{Information of the nodes in the Bayesian network to evaluate $^{56}$Fe in the energy range between one and two MeV.
    The column NUM shows the number of variabes aggregated in the node, PRIOR and UNC show the prior estimate and uncertainty, respectively, and are the same for each variable belonging to a node.
    There are no prior correlations between the variables of one node (and by the construction of the Bayesian network none between variables belonging to different nodes).
    If UNC is zero, it means the associated node is a deterministic function of the values associated with other nodes.
    EMIN and EMAX show the energy range covered by the variables of a node.
    These two columns do not apply to the \textit{normerr} node.
    The Bayesian network contains 15061 variables in total.
    }
    \label{tbl:fe56low-node-summary}
\end{table}

\begin{figure}[b]
    \centering
    \includegraphics{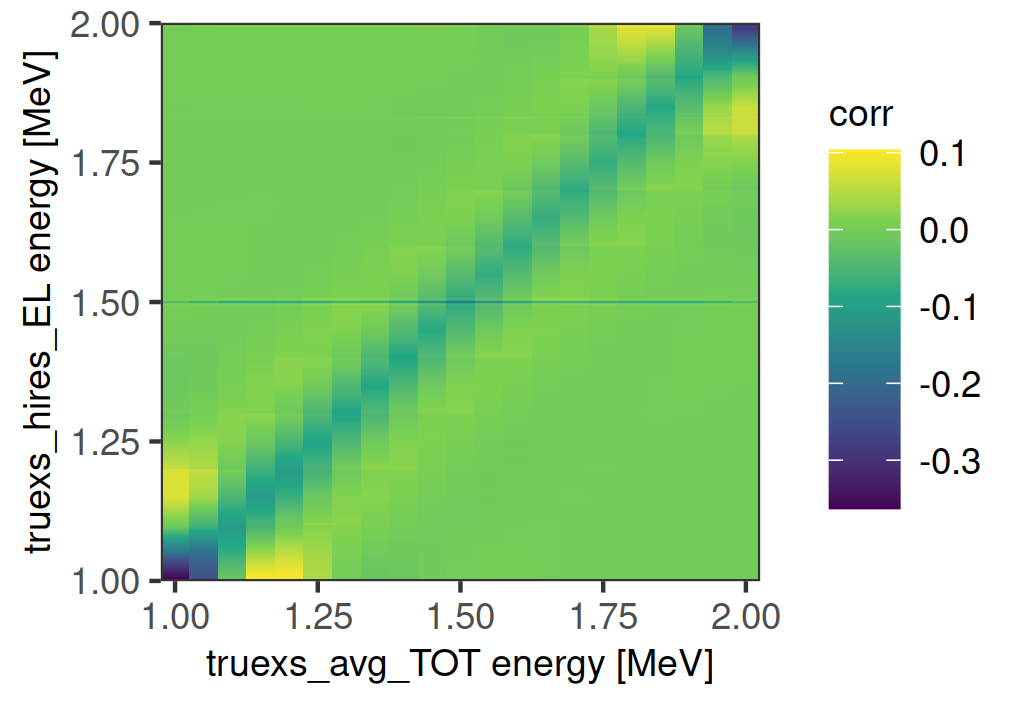}
    \caption{Correlation matrix between the variables associated with the node \textit{truexs\_hires\_EL} and those associated with the node \textit{truexs\_avg\_TOT}.
    The horizontal stripe of weak negative correlation at 1.50\,MeV of the high-resolution component is due to the experimental data point at 1.5\,MeV in the elastic channel.}
    \label{fig:fe56low-corr-example}
\end{figure}

The customized LM algorithm finds the most likely values of the variables according to the posterior distribution in a few iterations, and the complete optimization finished in about ten seconds.
In this example, all relationships are linear and the GLS method can therefore exactly locate the posterior maximum.
The LM algorithm takes a few more iterations because it starts out with a non-zero damping term which leads to more careful steps compared to the GLS method.

\begin{figure*}[!htb]
    \centering
    \includesvg{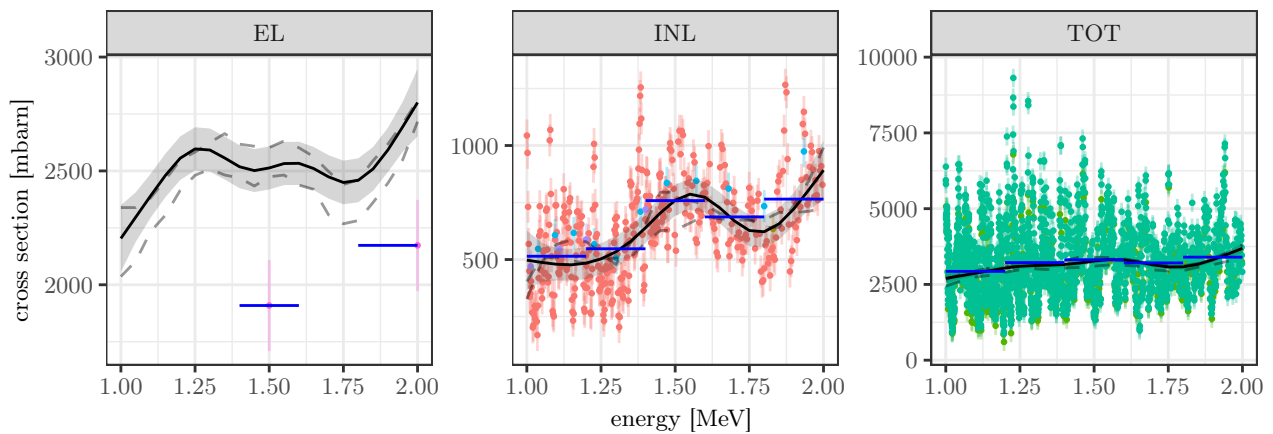}
    \caption{Most likely average trends (\textit{truexs\_avg\_REAC}) of the elastic, inelastic and total cross section curves between one and two MeV for neutrons on $^{56}$Fe according to the posterior distribution.
    The one-sigma credibility interval is also displayed.
    The horizontal blue segments represent the binned averages of the experimental data with the bin size of 0.2\,MeV indicated by the lengths of the segments.
    The dashed lines are samples from the approximate posterior distribution.}
    \label{fig:fe56low-truexs-avg}
\end{figure*}

\begin{figure*}[!htb]
    \centering
    \includesvg{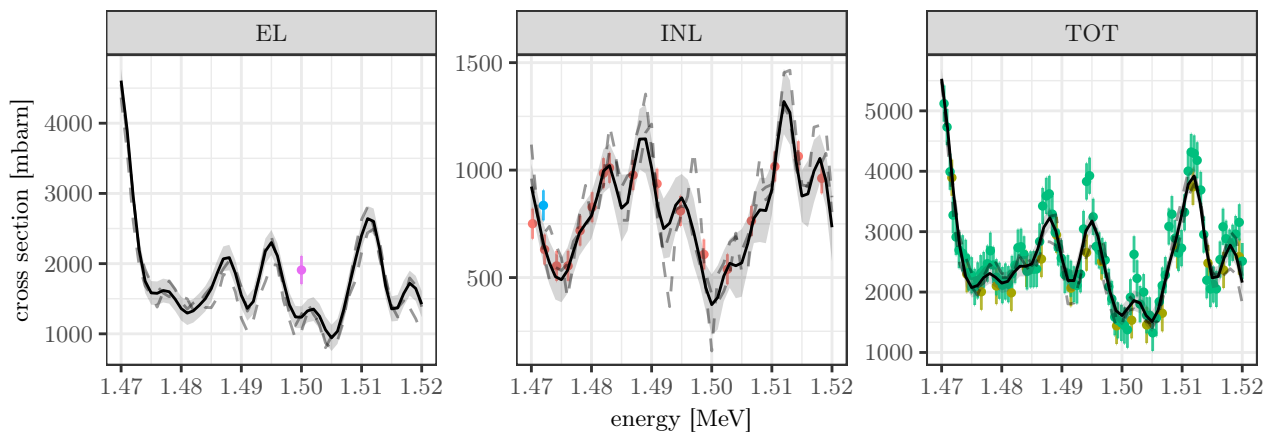}
    \caption{Most likely cross section (\textit{truexs\_REAC}) of the elastic, inelastic and total cross section curves for neutrons on $^{56}$Fe in a selected energy range.
    The one-sigma posterior credibility band is also shown and the dashed lines represent samples from the approximate posterior distribution.
    An energy resolution of 0.3\,keV has been assumed for all experimental data points in the Bayesian inference.}
    \label{fig:fe56low-truexs}
\end{figure*}

The posterior estimates of the average and the high-resolution cross section component are shown in \cref{fig:fe56low-truexs-avg} and \cref{fig:fe56low-truexs}, respectively.
The average components track well the averages of the binned cross section indicated by the blue horizontal segments.
The exception is the elastic cross section channel with only two data points, which are at first sight inconsistent with the measurements in the inelastic and total channel considering the sum rule.
The evaluated curve is far above the two experimental data points.
However, by inspecting the true elastic cross section (equaling sum of average and high-resolution component) in \cref{fig:fe56low-truexs} and taking into account the assumed energy resolution of 3\,keV for all experiments, the data point at 1.5\,MeV seems to be consistent with the evaluated curve. 
As an important remark, the Bayesian network framework enables the use of different energy resolutions for different experiments but in this example we assumed that all experiments have the same energy resolution for the sake of simplicity.

Overall, we see that both the average trend and the resonant structure in the cross sections are well reproduced by the tentative evaluation.
Additional fine-tuning of the prior uncertainties of the second derivatives at various energy locations can improve the coherence between the evaluation and the experimental data at the peaks and valleys.

The presence of the posterior uncertainty bands in the plots highlights the possibility to calculate selected blocks of the posterior covariance matrix described in~\cref{subsec:postcov-approx}.
The time needed to compute the posterior covariance block associated with the \textit{truexs\_avg\_REAC} was only a couple of seconds.
As a final demonstration of the possibility to selectively compute elements of the posterior covariance matrix, \cref{fig:fe56low-corr-example} shows the correlations between the variables of \textit{truexs\_avg\_TOT} and \textit{truexs\_hires\_EL}.
Weak negative correlations between those components only occur at the same energy, which is expected due to the short-range prior correlation within the high-resolution component. 
The horizontal stripe of weak negative correlation at 1.5 MeV of the high-resolution component is induced by the data point in the elastic channel at the same energy.

Even though there are a total of 15061 variables constituing the Bayesian network, there are only 3071 independent variables, which are associated with the nodes \textit{normerr}, \textit{truexs\_avg\_EL}, \textit{truexs\_hires\_EL}, \textit{truexs\_avg\_INL} and \textit{truexs\_hires\_INL}.
Only $12\%$ of the elements in the inverse posterior covariance matrix of dimension $3071\times 3071$ are non-zero.

\subsection{\label{subsec:example-ironfast}Model based evaluation with consistent model defects}

Nuclear model codes, such as CCONE~\cite{iwamotoCCONECodeSystem2016}, CoH$_3$~\cite{kawanoCoH3CoupledChannelsHauserFeshbach2021}, EMPIRE~\cite{hermanEMPIRENuclearReaction2007,hermanEMPIRE3MaltaModularSystem2013}, GNASH~\cite{youngGNASHPreequilibriumStatistical1977}, OPTMAN~\cite{soukhovitskiProgramOPTMANVersion2013} and TALYS~\cite{koning_modern_2012,koningTENDLCompleteNuclear2019}, can be used for an evaluation in the fast energy region.
However, due the complexity of nuclear processes and the limitations of nuclear models, discrepancies between the experimental data and the model predictions may remain even after the model parameters have been adjusted using the data.
Two approaches relying on Gaussian processes have been explored in the past to address the issue of imperfect nuclear models in the fast energy range:
\begin{enumerate}
    \item The use of Gaussian processes as priors for energy-dependent model parameters to inject more flexibility into the model~\cite{helgessonTreatingModelDefects2018,schnabelConceptionSoftwareImplementation2021}, and,
    \item The combination of model predictions and model defects modeled by Gaussian processes to explain the experiments, e.g., ~\cite{pigni_uncertainty_2003,leeb_consistent_2008,schnabelDifferentialCrossSections2016,helgessonFittingDefectNonlinear2017b}.
\end{enumerate}
Whereas the first approach has already been applied in a tentative full scale evaluation of neutron-induced reactions of $^{56}$Fe~\cite{schnabelConceptionSoftwareImplementation2021}, the second approach has only been explored in schematic examples.
Obstacles to the application of the second approach are preserving the consistency between the various reaction channels, such as the sum of the exclusive reaction channels equaling the total cross section, and enforcing that all cross sections are non-negative according to the posterior distribution.
In this example, we show the application of a Bayesian network to perform an evaluation following the second approach.

We employ the sparse Gaussian process construction explained in~\cref{subsec:sparse-gp-construction} within the Bayesian network framework to perform an evaluation of neutron-induced reactions of $^{56}$Fe between 5 MeV and 30 MeV with sum rule and positivity constraints preserved.
We remark that a sum rule Gaussian process construction has already been suggested and studied in a toy example in~\cite{schnabel_large_2015} and an evaluation based on GPs under positivity constraint has been presented in~\cite{iwamotoGenerationNuclearData2020a}.
Also the fitting of splines using linear regression can be regarded as a special case of Gaussian process regression and therefore~\cite{kawanoSimultaneousEvaluationFission2000} can also be considered as an example of GP regression under a positivity constraint.
However, to the best of our knowledge, these two constraints have not been modeled together in a nuclear data evaluation so far. 
Here we elaborate on a tentative full scale evaluation enforcing both constraints at the same time.
The experimental datasets used in this example evaluation are listed in~\cref{apx:3rdexample-expdata}.

\begin{figure*}[ht]
    \centering
    \includesvg{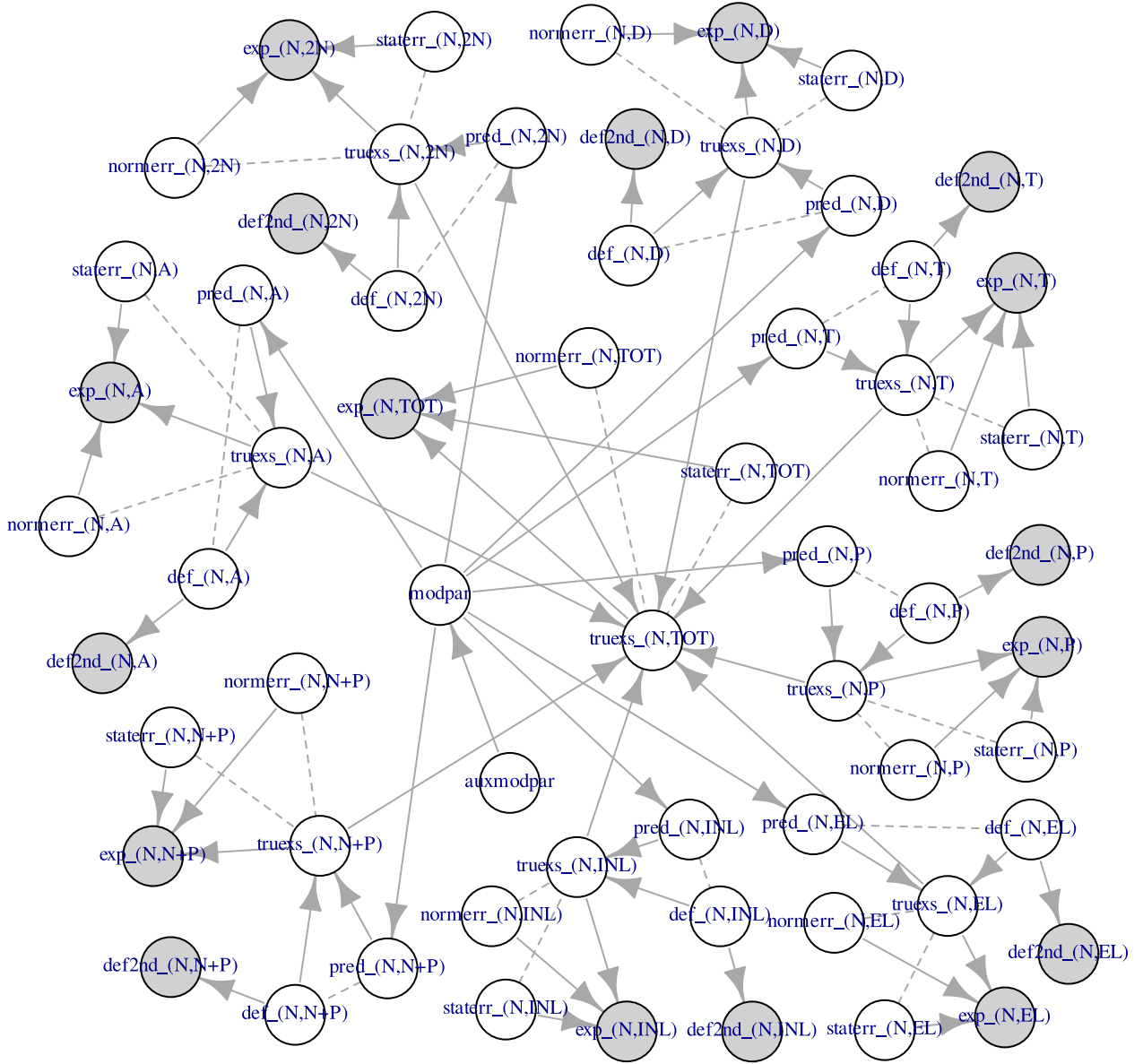}
    \caption{Bayesian network for an evaluation of neutron-induced reactions of $^{56}$Fe in the energy range between 5 and 30 MeV using TALYS as a nuclear model code and model defects defined by Gaussian processes.
    The meaning of the nodes and the mappings between them are explained in the text.}
    \label{fig:fe56fast-baynet}
\end{figure*}
\begin{figure*}[htb]
    \centering
    \includesvg{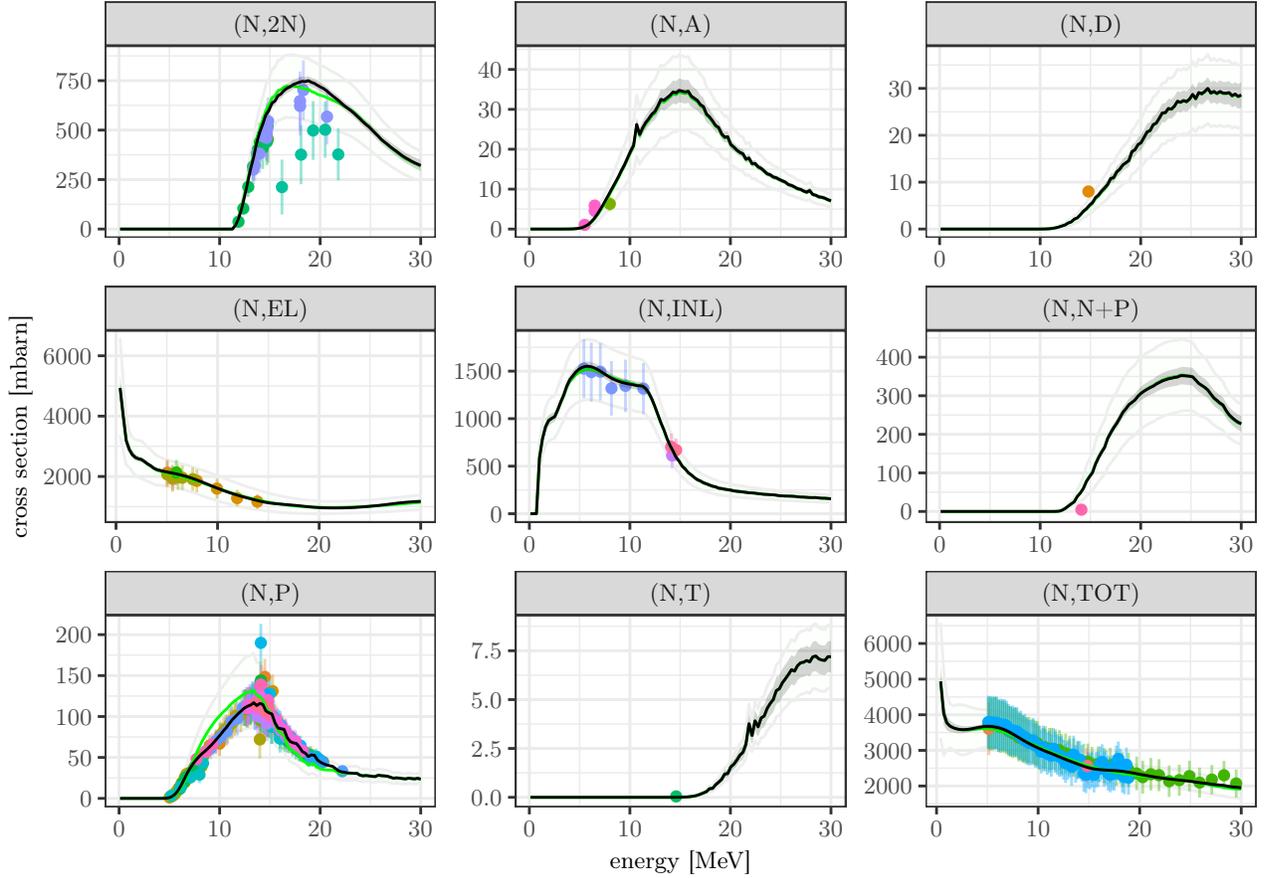}
    \caption{Most likely true cross section (\textit{truexs\_REAC}, black) of the neutron-induced reaction channels of $^{56}$Fe between 5 and 30 MeV and associated one-sigma credibility band according to the posterior distribution.
    The prior curves (green) and associated uncertainty band (pale green) are also displayed.}
    \label{fig:fe56fast-truexs-eval}
\end{figure*}

We created the Bayesian network depicted in~\cref{fig:fe56fast-baynet} to perform a consistent evaluation of neutron-induced reactions of $^{56}$Fe using 91 experimental datasets with a total number of 2072 datapoints between 5 and 30\,MeV in the following nine reaction channels: (N,EL), (N,A), (N,T), (N,D), (N,P), (N,INL), (N,N+P), (N,2N) and (N,TOT).  
Regarding the total cross section, we aggregated the data in bins of 0.2 MeV and use the bin averages as the experimental data. 
This measure was taken to make the plots less busy and avoid the modeling of a potential high-resolution component as done in the previous examples for the sake of simplicity.
After this aggregation, the total number of data points is reduced to 596.
The experimental data along with the evaluated cross sections based on the Bayesian network are shown in~\cref{fig:fe56fast-truexs-eval}.

Following we describe the Bayesian network in more detail.
If we use \textit{REAC} in a node name, it implies that a node of this type exists for each reaction channel except the total cross section.
For instance, \textit{truexs\_REAC} makes reference to \textit{truexs\_(N,A)}, \textit{truexs\_(N,P)} and all the other channels but not \textit{truexs\_(N,TOT)}.

The node \textit{auxmodpar} contains the multipliers to be applied to the default model parameter values of the nuclear model code TALYS, as they are specified by the \textit{rvadjust} and similar keywords in a TALYS input file.
If we use the term model parameters from now on, we refer to these multiplication factors.
The prior estimate of each multiplication factor is one, meaning that the default value employed by TALYS is a priori considered the best guess, and the associated prior uncertainty is taken as $0.05$.
These parameters are propagated to \textit{modpar} by a (deterministic) non-linear transformation that limits their range to $0.9$ and $1.1$ in order to avoid excessive adjustments and counteract potential overfitting to available experimental data at the cost of predictive power. 
Any value of a parameter below $0.9$ in \textit{auxmodpar} is mapped to $0.9$ in \textit{modpar} and any value above $1.1$ to $1.1$. 
The model parameters in \textit{modpar} yield the predictions of the nuclear model in the \textit{pred\_REAC} nodes.
In this example, we use a linear approximation of the nuclear model using the default parameters as reference vector to simplify the example.
The feasibility of incorporating the exact non-linear deterministic model at scale into the Bayesian inference procedure has already been demonstrated in~\cite{schnabelConceptionSoftwareImplementation2021}.

As TALYS is a deterministic code, the links from \textit{modpar} to \textit{pred\_REAC} are deterministic as well.
The predictions are an additive contribution to the true reaction cross sections in \textit{truexs\_REAC}.
The other contribution is given by the model defects \textit{def\_REAC} associated with each reaction channel.
We also apply a non-linear transformation to the sum of \textit{pred\_REAC} and \textit{def\_REAC} to ensure that the cross sections in \textit{truexs\_REAC} are non-negative.
If the result of the sum at any energy is negative, it is mapped to zero.

The model defects are given relative to the model predictions, which introduces a non-linear interaction between \textit{def\_REAC} and \textit{pred\_REAC} indicated by the dashed connections.
The prior estimate of these model defects at all energies is zero.
The prior uncertainty is 20\% for energies above the reaction specific reaction threshold and zero below.
To ensure a certain smoothness, we introduced the nodes \textit{def2nd\_REAC} representing the second derivatives of \textit{def\_REAC} and assumed that a zero value was observed at all energy locations with small uncertainty.
This construction, as explained in~\cref{subsec:sparse-gp-construction}, enforces a certain degree of smoothness of the \textit{def\_REAC} nodes and guards against kinks caused for instance by inconsistent experimental datasets.
Because the variables of the node \textit{def\_REAC} are relative to \textit{pred\_REAC}, also the observations of smoothness in \textit{def2nd\_REAC} are defined in terms of relative changes. 
This feature appears helpful to us as the expected absolute variations of different cross sections can be quite different depending on their magnitude.

The sum of the true reaction cross sections \textit{truexs\_REAC} yields the total cross section \textit{truexs\_(N,TOT)}.
Please note that in this example, to keep the display of the Bayesian network in this paper manageable, we did not account for some channels, such as $(N,N+A)$, $(N,3N)$ and $(N,2N+P)$ which yield contributions of a few millibarn at 15\,MeV due to $(N,N+A)$, several dozen millibarns at 20\,MeV and about hundred millibarn at 25\,MeV, which introduces a bias of respective size distributed over the reaction channels in this example evaluation.

At this point in the explanation, we have covered the construction of the true cross sections as a function of a nuclear physics model and model defect components with the constraints that the individual reaction channels should yield the total cross section and all cross sections are non-negative quantities.
The remaining part of the explanations covers the modelization of the experimental data and their relationship to the true cross sections.
From now on, the appearance of \textit{REAC} in a node name implies that a node of this type is present for \textit{all} reaction channels, hence also including the total cross section $(N,TOT)$.
Furthermore, we aggregated the nodes of different experimental datasets in each channel together to one node for a manageable display of the Bayesian network.
A node, such as \textit{staterr\_(N,A)} in the display of the Bayesian network can be expanded into a node for each individual dataset without connections between those nodes. 

Each experimental dataset \textit{exp\_REAC} is modeled as a sum of three components, which are the true reaction cross section \textit{truexs\_REAC}, a relative normalization error \textit{normerr\_REAC} and a relative statistical error \textit{staterr\_REAC}.
Both error components are given relative to the true reaction cross section, and the non-linear interactions between the source nodes are indicated by dashed lines from \textit{normerr\_REAC} to \textit{truexs\_REAC} and from \textit{staterr\_REAC} to \textit{truexs\_REAC}. 
The prior uncertainty of all relative statistical errors is assumed to be 1\% and the prior uncertainty of the normalization errors of each experimental dataset is $10\%$.
This description of the nodes associated with the properties of the experiments completes the explanation of the Bayesian network topology depicted in~\cref{fig:fe56fast-truexs-eval}, which comprises a total of 4895 variables associated with the nodes.

Tentative optimization runs using the customized Levenberg-Marquardt algorithm revealed several local maxima in the posterior distribution.
We obtained the highest value of the posterior distribution, which may represent the global maximum, by dividing the optimization procedure in two stages.
In the first stage we allowed only the adjustment of the variables in the nodes \textit{def\_REAC}, \textit{normerr\_REAC} and \textit{staterr\_REAC}.
In the second stage, starting from the optimized values of the first stage, we allowed the adjustment of all independent variables, which includes the variables of the first stage and in addition \textit{auxmodpar}.
The optimization procedure finished in a couple of minutes.

\begin{figure}[t]
    \centering
    \includesvg{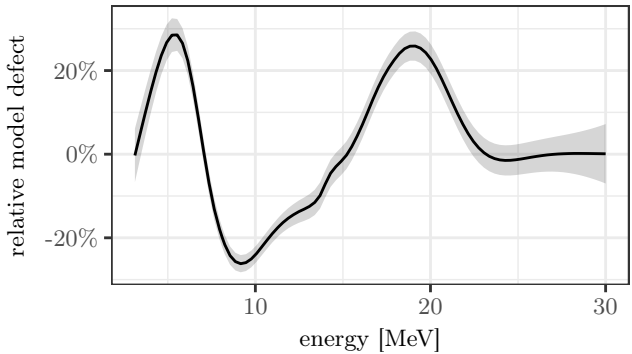}
    \caption{Posterior estimate and associated one-sigma credibility band of the model defect in the (N,P) channel of $^{56}$Fe.
    The model defect is given relative to the posterior model prediction, i.e., the prediction based on the most likely model parameter values according to the posterior distribution.}
    \label{fig:fe56fast-defect-example}
\end{figure}

The most likely cross sections according to the posterior distribution, which we refer to as evaluated cross sections, are depicted in~\cref{fig:fe56fast-truexs-eval}.
In most cases, the evaluated cross sections coincide with the best prior estimate given by the model prediction using the default values of the model parameters.
Only the evaluated $(N,P)$ cross section is significantly below the prior estimate and this difference is associated with the evaluated model defect component shown in~\cref{fig:fe56fast-defect-example}.
We double checked that this reduction of the cross section in $(N,P)$ due to the model defect component is done consistently so that the sum of exclusive cross sections still equals the total cross section.

\begin{figure}[t]
    \centering
    \includesvg{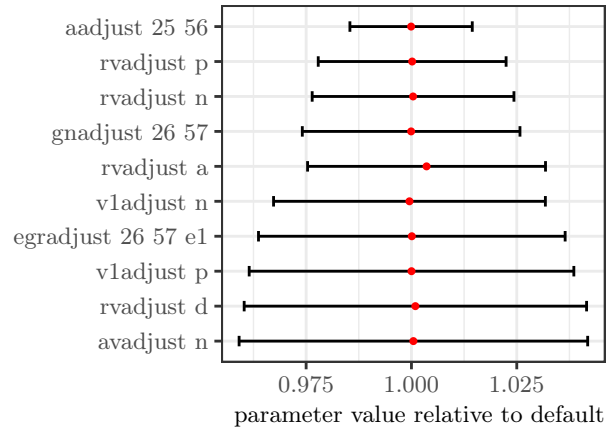}
    \caption{
        Posterior estimate and associated one-sigma credibility interval of the most constrained TALYS model parameters.
        }
    \label{fig:fe56fast-modelpars-posterior}
\end{figure}

The most likely model parameters and their posterior credible interval are depicted in~\cref{fig:fe56fast-modelpars-posterior}. 
The posterior estimates of all model parameters are very close to the prior ones, only the posterior uncertainty of some parameters is significantly smaller compared to the prior uncertainty of $0.05$.
The additional flexibility introduced by the model defect components reduces the need to adjust model parameters to explain the experimental data.
A model defect with a prior uncertainty of 20\% renders the adjustment of model parameters unnecessary as most experimental data are in a 20\% corridor around the model prediction with default parameter values. 

From a physics point of view, it is interesting to know which model parameters are constrained most by the data.
The TALYS input keyword \textit{aadjust 25 56} is a multiplication factor to the TALYS default value of the level density parameter $a$ for $^{56}$Mn appearing in the Fermi gas model.
The \textit{adjust} keywords starting with \textit{rv}, \textit{av} and \textit{v1} represent multiplication factors applied to the default values of the optical model parameters for a specific projectile indicated at the end of the keyword string.
The value of the radius $r_V$ appearing in the Woods-Saxon factor of the real and imaginary component of the volume-central potential is adjusted by \textit{rvadjust}.
The value $v_1$ appears as a factor in the real component of the volume-central potential and is adjusted by \textit{avadjust}.
The diffuseness parameter $a_V$ in the Woods-Saxon factor appearing in the real and imaginary component of the volume-central potential is adjusted by \textit{avadjust}.
The keyword \textit{gnadjust 26 57} adjusts for $^{57}$Fe the single-particle neutron level density parameter $g_\nu$ that appears in the particle-hole state density expression of B\v et\'ak and Dobe\v s~\cite{dobesUnknown1983} used in the pre-equilibrium excition model.
The keyword \textit{egradjust 25 57 e1} is a normalization factor for the energy of the giant dipole resonance.
More information on these input keywords and how they impact the nuclear models can be found in the TALYS user manual.

The plots in~\cref{fig:fe56fast-truexs-eval} contain several outlying datasets. 
These datasets are also revealed as outliers by inspecting the posterior estimates and uncertainties of the relative normalization errors.
For example, for the dataset in the (N,2N) channel being located significantly below the peak (EXFOR identification 20854015, see~\cref{apx:3rdexample-expdata}), the posterior estimate of the relative normalization error is -44\% and the posterior uncertainty of this value is 0.9.
This result is incompatible with the 10\% prior uncertainty of the normalization error.
The possibility to obtain posterior estimates and uncertainties of all nodes in the Bayesian network is therefore an algorithmic way to identify outliers.

Finally, we remark that in tentative optimization runs that converged to local minima associated with a smaller value of the posterior density function, also the posterior estimates of the model parameters significantly differed from the prior ones.

\FloatBarrier

\section{Summary and outlook}

Starting from the Generalized Least Squares method, we presented the equations to perform inference in Bayesian networks with multivariate normal priors on the variables associated with each node and linear, non-linear, deterministic and nested relationships between the variables belonging to different nodes.
Furthermore, we elaborated on a sparse Gaussian process construction that blends well with the Bayesian network interpretation and is suitable for scenarios with a large number of variables.
One key ingredient of this construction is to use a discretized version of a function defined on a mesh and to use linear interpolation between mesh points.
The second ingredient is to use a diagonal prior covariance matrix for the function values at the mesh points and to regulate the degree of smoothness by pseudo-observations of the second derivative with the associated covariance matrix also being diagonal.
An advantage of this GP construction, besides increased computational and storage efficiency, is the possibility to define prior knowledge about the magnitude and smoothness of the unknown function in a fine-grained manner individually for different energy regions.  
We also reviewed two important non-linear mappings to model energy calibration errors and relative experimental errors, which can be rigorously taken into account in the outlined Bayesian network framework by using a customized Levenberg-Marquardt algorithm~\cite{helgessonFittingDefectNonlinear2017b,schnabelConceptionSoftwareImplementation2021} to exactly locate the posterior maximum even in the presence of non-linear relationships.
It was also explained how samples can be obtained from an approximate posterior distribution and how selected blocks of an approximated posterior covariance matrix can be computed, even in the case of a large number of variables by exploiting the sparseness of the inverse posterior covariance matrix.
The approximate posterior distribution is close to the true posterior distribution if the non-linearities are in good approximation linear in the parameter domain associated with significant values of the posterior density function.
If this assumption is strongly violated, Monte Carlo sampling schemes should be preferred to extract samples or blocks of the posterior covariance matrix from the posterior distribution.

We provided three proof-of-concept examples to perform nuclear data evaluations with Bayesian networks in combination with the outlined sparse GP construction, which are of practical relevance.
The first example demonstrated the evaluation of the $^{235}$U(n,f) cross section including a relative energy-dependent systematic error.
For example, such an energy-dependent error contribution can be used to model errors associated with Unrecognized Sources of Uncertainty (USU) in the neutron data standards project~\cite{carlsonInternationalEvaluationNeutron2009,carlsonEvaluationNeutronData2018}.
The second example was a no-model evaluation of neutron-induced reactions of the important structural material $^{56}$Fe in the energy range between one and two MeV, where it is difficult to obtain satisfactory evaluations with either R-matrix and nuclear model fits due to the many overlapping resonances causing rapid fluctuations of the cross sections.
The third example demonstrated a model-based evaluation of neutron-induced reaction of $^{56}$Fe between 5 and 30\,MeV including model defect components to account for discrepancies between the model predictions and the experimental data while preserving sum rule and positivity constraints.

The presentation of the examples was focused on the modeler point of view who can build a Bayesian network by first defining a data table with the variables and then link these variables to each other by using various mappings, such as a linear interpolation mapping, convolutions, and non-linear transformations as building blocks.
The diverse aspects of the examples, such as the presence or absence of a physics model or fluctuations, hinted at the general applicability of the Bayesian network interpretation to many different evaluation scenarios.

The examples also proved that the customized Levenberg-Marquardt algorithm is able to find good solutions in a reasonable amount of time.
The optimization process never took loger than a couple of minutes on a personal computer.
However, a point that needs improvement is the implementation of heuristics to ensure that a good local (and ideally the global) maximum of the posterior distribution is found.
Common non-linear mappings between variables, such as relative normalization error, usually yield multiple local maxima in the posterior distribution. 
In the examples, we dealt with this issue by performing the optimization in several stages and optimizing only a subset of the variables in each stage.
The determination of the final optimization strategy was the result of trial and error.

We believe that the modelization of an evaluation scenario as a Bayesian networks can help to make nuclear data evaluation more transparent and less error-prone.
The possibility to get posterior estimates and uncertainties of all quantities involved, such as the various systematic error components, helps to quickly identify problematic modeling assumptions by checking the compatibility of the posterior estimates with the prior estimates and uncertainties.
At some point in the future, templates of measurement uncertainties~\cite{neudeckerTemplateEstimatingUncertainties2018,neudeckerApplyingTemplateExpected2020} may be used for the specification of default priors on the experimental errors associated with different measurement types, and also to check the compatibility of the posterior estimates with sensible ranges given by a template.
The Bayesian network framework is also compatible with the procedures described in~\cite{schnabelFittingAnalysisTechnique2018,siefmanDataAssimilationPostirradiation2020a,siefmanDevelopmentApplicationMarginal2021,schnabelConceptionSoftwareImplementation2021} for the automated determination of missing or misspecified uncertainties.
As the relations between experiments in nuclear databases, such as EXFOR~\cite{otuka_towards_2014}, can also be represented as a graph~\cite{brownVisualizingConnectionsEXFOR2014,hirdtIdentifyingUnderstudiedNuclear2016}, with links established by common features, Bayesian networks may be used for the automatic correction of experimental datasets and also outlier detection there.
These procedures are complementary to machine learning approaches, such as presented in~\cite{whewellEvaluating239PuCross2020,neudeckerEnhancingNuclearData2020,dwivediTreesIslandsMachine2019}, to help humans make sense of data and enhance evaluations.


\bibliography{manuscript}

\onecolumngrid
\vspace{10ex}
\twocolumngrid

\appendix

\section{\label{app:GLS-LM-derivation}Derivation of GLS method and LM algorithm}

\def\muvec{\vec{\mu}}
\def\muobs{\vec{\mu}_\textrm{obs}}
\def\mutrue{\vec{\mu}_\textrm{true}}
\def\muhid{\vec{\mu}_\textrm{hid}}
\def\xobs{\vec{x}_\textrm{obs}}

\def\sigmamat{\mathbf{\Sigma}}
\def\covobs{\mathbf{\Sigma}_\textrm{obs}}
\def\covhid{\mathbf{\Sigma}_\textrm{hid}}
\def\covobshid{\mathbf{\Sigma}_\textrm{obs,hid}}

\def\obsvec{\vec{\sigma}_\textrm{exp}}
\def\covexp{\mathbf{\Sigma}_\textrm{exp}}

\def\modrefvec{\vec{\sigma}_\textrm{ref}}
\def\modparvec{\vec{p}}
\def\modparrefvec{\vec{p}_\textrm{ref}}
\def\syserrvec{\vec{\varepsilon}_\textrm{sys}}
\def\moderrvec{\vec{\varepsilon}_\textrm{mod}}
\def\staterrvec{\vec{\varepsilon}_\textrm{stat}}

\def\syserrmap{\mathbf{S}_\textrm{sys}}
\def\modparmap{\mathbf{S}_\textrm{mod}}
\def\totmap{\mathbf{S}}

\def\totunobs{\vec{u}}
\def\covtotunobs{\mathbf{U}}
\def\totunobsref{\vec{u}_\textrm{ref}}

\def\covstaterr{\mathbf{D}}

\def\totrefvec{\vec{\sigma}_0}

The GLS method is the basis for inference in Bayesian networks of continuous variables with a multivariate normal prior distribution and linear relationships between variables.
The customized LM algorithm~\cite{helgessonFittingDefectNonlinear2017b} is an iterative optimization algorithm that extends the GLS method to enable the determination of the most likely values of the variables also in the case of non-linear relationships between them.
Here we present a derivation of the GLS method and the LM algorithm that follows closely the derivation given in the appendix of~\cite{schnabelConceptionSoftwareImplementation2021}. 

Assume that we have a vector $\modparvec$ with variables of interest that are related to observed quantities $\obsvec$ by the sum of a vector-valued function $\mathcal{M}(\modparvec)$ and a noise term.
We denote the prior distribution on the vector $\modparvec$ by $\pi(\modparvec)$ and the likelihood connecting $\modparvec$ to the observations by $\ell(\obsvec \,|\, \modparvec)$.
According to Bayes theorem, the posterior distribution is proportional to the product of prior and likelihood,
\begin{equation}
    \pi(\modparvec \,|\, \obsvec) = \frac{1}{\pi(\obsvec)} \ell(\obsvec \,|\, \modparvec) \pi(\modparvec) \,.
\end{equation}
For both the GLS method and the customized LM algorithm, multivariate normal distributions are imposed for the likelihood and the prior distribution,
\begin{align}
    \ell(\obsvec \,|\, \modparvec) &= \mathcal{N}(\obsvec \,|\, \mathcal{M}(\vec{p}); \covexp) \,, \\
    \pi(\modparvec) &= \mathcal{N}(\modparvec \,|\, \vec{p}_0; \mat{P}) \,.
\end{align}
The functional form of the posterior density function (pdf) of a multivariate normal distribution is given by
\begin{multline}
    \mathcal{N}(\vec{x} \,|\, \vec{\mu}, \mat{\Sigma}) =
    \frac{1}{\sqrt{(2\pi)^N \det \mat{\Sigma}}}
    \\
    \exp\left(
    -\frac{1}{2}
    (\vec{x}-\vec{\mu})^T
    \mat{\Sigma}^{-1}
    (\vec{x}-\vec{\mu})
    \right)
\end{multline}
characterized by a mean vector $\vec{\mu}$ and a covariance matrix $\mat{\Sigma}$.
In the following, we make use of the natural logarithm of the multivariate normal pdf,
\begin{multline}
    \ln \mathcal{N}(\vec{x}\,|\,\vec{\mu},\mat{\Sigma}) =
    -\frac{N}{2}\ln (2\pi) - \frac{1}{2} \ln \det \mat{\Sigma} \\
    -\frac{1}{2} 
    (\vec{x}-\vec{\mu})^T
    \mat{\Sigma}^{-1}
    (\vec{x}-\vec{\mu}) = \\
    -\frac{1}{2} \vec{x}^T \mat{\Sigma}^{-1} \vec{x}
    -(\vec{x}-\vec{\mu})^T \mat{\Sigma}^T \vec{x} + \mathcal{C} 
    \label{eq:logmvnpdf}
\end{multline}
where we absorbed terms independent of $\vec{x}$ into $\mathcal{C}$ and made use of the fact that the second term is equivalent to its transpose due to the covariance matrix being symmetric.

For the GLS method, we assume that the function $\mathcal{M}(\modparvec)$ is linear and can therefore be written in the form
\begin{equation}
    \mathcal{M}_\textrm{lin}(\vec{p}) = \modparrefvec +
    \mat{J} \left( \modparvec - \modparrefvec \right) \,.
\end{equation}
The introduction of $\delta\modparvec = \modparvec - \modparrefvec$, $\vec{d} = \obsvec - \modrefvec$ allows us to write
\begin{multline}
    \label{eq:proplogpost}
    (-2) \ln \pi(\modparvec \,|\, \obsvec) =
    \left(
        \vec{d} - \mat{J} \, \delta\modparvec
    \right)^T
    \covexp^{-1}
    \left(
        \vec{d} - \mat{J} \, \delta\modparvec
    \right) 
    \\
    +
    \left(
        \delta\modparvec + \modparrefvec - \vec{p}_0
    \right)^T
    \mat{P}^{-1}
    \left(
        \delta\modparvec + \modparrefvec - \vec{p}_0
    \right)
    + \mathcal{C}
\end{multline}
with the constant $\mathcal{C}$ absorbing everything which is independent of $\delta\modparvec$.
The constant is of no significance as we are going to take derivatives with respect to elements in $\delta\modparvec$.

Now we isolate terms containing $\delta\modparvec$ and regroup the expression according to their order,
\begin{multline}
    \frac{1}{2}
    \delta\modparvec^T
    \left(
        \mat{J}^T \covexp^{-1} \mat{J} +
        \mat{P}^{-1}
    \right)
    \delta\modparvec \\
    -
    \delta\vec{p}^T
    \left(
    \mat{J}^T \covexp^{-1} \vec{d}
    + \mat{P}^{-1} (\vec{p}_0 - \modparrefvec)
    \right) + \mathcal{C}
    \label{eq:LMobjective}
\end{multline}
where all terms independent of $\delta\modparvec$ are absorved in $\mathcal{C}$.
The comparison of this expression with~\cref{eq:logmvnpdf} reveals that also the posterior pdf is a multivariate normal distribution. 
We can identify the inverse of the posterior covariance matrix $\mat{P}'$ being given by
\begin{equation}
    \left( \mat{P}' \right)^{-1} = 
    \mat{J}^T \covexp^{-1} \mat{J} +
    \mat{P}^{-1} \,.
    \label{eq:invpostcov}
\end{equation}

To determine the vector $\delta\modparvec$ that minimizes~\cref{eq:LMobjective}, we calculate the gradient of \cref{eq:LMobjective} and require it to vanish:
\begin{multline}
    \left(
        \mat{J}^T \covexp^{-1} \mat{J} +
        \mat{P}^{-1}
    \right)
    \delta\modparvec \\
    -
    \left(
    \mat{J}^T \covexp^{-1} \vec{d}
    + \mat{P}^{-1} (\vec{p}_0 - \modparrefvec)
    \right) = \vec{0} \,.
    \label{eq:gradLMobjective}
\end{multline}
Rearranging yields
\begin{equation}
    \left(
        \mat{J}^T \covexp^{-1} \mat{J} +
        \mat{P}^{-1}
    \right)
    \delta\modparvec
    =
    \left(
    \mat{J}^T \covexp^{-1} \vec{d}
    + \mat{P}^{-1} (\vec{p}_0 - \modparrefvec)
    \right) \,.
    \label{eq:GLSformula-notfinal}
\end{equation}

Using the identification in~\cref{eq:invpostcov} and rearranging~\cref{eq:GLSformula-notfinal} another time yields the GLS formula to compute the posterior center vector,
\begin{equation}
    \modparvec
    =
    \modparrefvec +
    \mat{P}' 
    \left(
    \mat{J}^T \covexp^{-1} \vec{d}
    + \mat{P}^{-1} (\vec{p}_0 - \modparrefvec)
    \right) \,.
    \label{eq:GLSupdate}
\end{equation}

Now we discuss the derivation of the LM algorithm.
For non-linear relationships $\mathcal{M}(\modparvec)$, the Jacobian $\mat{J}$ changes depending on the chosen expansion point $\modparrefvec$.
It may be still used to determine a small step in vicinity of the expansion point to climb up the posterior density function but cannot be trusted globally anymore as in the GLS method.  
Inspecting~\cref{eq:GLSformula-notfinal}, we see that the right-hand side is proportional to the gradient of \cref{eq:LMobjective} evaluated at the reference parameter vector, i.e., $\delta\modparvec=0$.
Therefore it is also proportional to the gradient of the logarithmized posterior distribution $\ln \pi(\modparvec\,|\,\obsvec)$ evaluated at $\modparvec=\modparrefvec$.
Taking this aspect into account, we can augment~\cref{eq:invpostcov} by a damping term,
\begin{equation}
    \left( \mat{P}' \right)^{-1} = 
    \mat{J}^T \covexp^{-1} \mat{J} +
    \mat{P}^{-1} + \lambda \mathcal{I} \,,
\end{equation}
with $\lambda$ being a non-negative number and $\mathcal{I}$ the identity matrix or a diagonal matrix.
An increase of $\lambda$ makes the matrix $(\mat{P}')^{-1}$ more diagonal and consequently also its inverse.    
In addition, the magnitude of the diagonal elements in the matrix $\mat{P}'$ decreases.
Consequently, for an increasing value of $\lambda$, the update prescription in~\cref{eq:GLSupdate} transforms gradually into the gradient ascent method.
On the other side, for $\lambda=0$, we recover the GLS update.

The LM algorithm adaptively changes the value of $\lambda$ from one iteration to the next depending on how well the expected improvement according to the linear approximation of $\mathcal{M}(\modparvec)$ matches the real improvement using the exact non-linear function.
A strategy for the adjustment of $\lambda$ was explained in~\cref{subsec:nonlinearfuns}.

\onecolumngrid
\clearpage

\section{\label{apx:3rdexample-expdata}Experimental data used in the $^{56}$Fe example evaluation}

\begin{table*}[ht]
\centering
\begin{tabular}{lllrllllrl}
  \hline
REAC & EXFOR & AUTHOR & YEAR & REF & REAC & EXFOR & AUTHOR & YEAR & REF \\ 
  \hline
(N,2N) & 20091004 & Wenusch and Vonach & 1962 & \cite{wenusch2nCrosssectionMeasurements1962} & (N,P) & 10031005 & Barrall et al & 1969 & \cite{barrallCrossSectionsReactions1969} \\ 
  (N,2N) & 20721021 & Molla and Qaim & 1977 & \cite{mollaSystematicStudyReactions1977} & (N,P) & 10289002 & Dyer and Hamilton & 1972 & \cite{dyer56Fe58FeCross1972} \\ 
  (N,2N) & 20854015 & Corcalciuc & 1978 & \cite{corcalciucStudyNeutronInduced1978} & (N,P) & 10309004 & Singh & 1972 & \cite{singhNeutronReactionCross1972} \\ 
  (N,2N) & 20416003 & Frehaut et al & 1980 & \cite{frehautStatus2nCross1980} & (N,P) & 20798003 & Robertson et al & 1973 & \cite{robertson56Fe56Mn27Al1973} \\ 
  (N,2N) & 13132002 & Bowers and Greenwood & 1989 & \cite{bowersAnalysisLonglivedIsotopes1989} & (N,P) & 12956012 & Spangler et al & 1975 & \cite{spangler14MeVCrossSection1975} \\ 
  (N,2N) & 23171003 & Wallner et al & 2011 & \cite{wallnerProductionLonglivedRadionuclides2011} & (N,P) & 21049005 & Mostafa & 1976 & \cite{mostafaMeasurementsRelativeNeutron1976} \\ 
  (N,A) & 12812012 & Saraf et al & 1991 & \cite{sarafCrossSectionsSpectra1991} & (N,P) & 10835002 & Sothras & 1977 & \cite{sothrasStudySystematics2n1977} \\ 
  (N,A) & 32737002 & Wang et al & 2015 & \cite{wangCrossSectionsFe2015} & (N,P) & 20721094 & Molla and Qaim & 1977 & \cite{mollaSystematicStudyReactions1977} \\ 
  (N,A) & 32737003 & Wang et al & 2015 & \cite{wangCrossSections562015} & (N,P) & 20993002 & Kudo & 1977 & \cite{kudoAbsoluteMeasurement56Fe1977} \\ 
  (N,D) & 10827031 & Grimes et al & 1979 & \cite{grimesChargedparticleEmissionReactions1979} & (N,P) & 41313002 & Ramendik et al & 1977 & \cite{ramendikDetermination56Fe56Mn1977} \\ 
  (N,EL) & 11708003 & Kinney & 1968 & \cite{kinneyNeutronElasticInelastic1968} & (N,P) & 20772003 & Ryves et al & 1978 & \cite{ryvesCrossSectionMeasurements1978} \\ 
  (N,EL) & 10037004 & Boschung et al & 1971 & \cite{boschungScatteringFastNeutrons1971} & (N,P) & 30483002 & Chi-Chou et al & 1978 & \cite{chi-chouCrossSectionMeasurement1978} \\ 
  (N,EL) & 10958012 & El-Khadi et al & 1982 & \cite{el-kadiElasticInelasticScattering1982} & (N,P) & 30483003 & Chi-Chou et al & 1978 & \cite{chi-chouCrossSectionMeasurement1978} \\ 
  (N,EL) & 14462004 & Ramirez et al & 2017 & \cite{ramirezNeutronScatteringCross2017} & (N,P) & 30676002 & Sharma et al & 1978 & \cite{sharmaAbsoluteMeasurementsFe561978} \\ 
  (N,INL) & 41316002 & Kozyr and Prokopets & 1978 & \cite{kozyrRadiativeTransitionsUnbound1978} & (N,P) & 40485002 & Nemilov and Tofimov & 1978 & \cite{nemilovCrosssectionsReactionsNi581978} \\ 
  (N,INL) & 30656021 & Xiamin et al & 1982 & \cite{xiaminMeasurementsInducedGamma1982} & (N,P) & 30562019 & Ngoc et al & 1980 & \cite{ngocInvestigations2nReactions1980} \\ 
  (N,INL) & 41156006 & Simakov et al & 1992 & \cite{simakov14MeVFacilityResearch1992} & (N,P) & 21868002 & Kudo & 1982 & \cite{kudo56Fe56MnCross1982} \\ 
  (N,INL) & 23134005 & Beyer et al & 2014 & \cite{beyerInelasticScatteringFast2014a} & (N,P) & 30644006 & Viennot et al & 1982 & \cite{viennotExcitationFunctionsReactions1982} \\ 
  (N,N+P) & 41118013 & Klochkova & 1992 & \cite{klochkovaInvestigationReactionsAl271992} & (N,P) & 30802002 & Ngoc et al & 1983 & \cite{ngocNeutronActivationCross1983} \\ 
  (N,P) & 11274031 & Paul and Clarke & 1953 & \cite{paulCrossSectionMeasurements1953} & (N,P) & 21923003 & Kudo & 1984 &  \\ 
  (N,P) & 11703002 & Mcclure and Kent & 1955 & \cite{mcclureInelasticScattering141955} & (N,P) & 30707013 & Gupta et al & 1985 & \cite{guptaPreequilibriumEmissionEffect1985} \\ 
  (N,P) & 20280004 & Yasumi & 1957 & \cite{yasumiNuclearReactionsInduced1957} & (N,P) & 30807008 & Garlea et al & 1985 & \cite{garleaNeutronCrossSections1985} \\ 
  (N,P) & 21487008 & Allan & 1957 & \cite{allanProtonsInteraction141957} & (N,P) & 12969013 & Meadows et al & 1987 & \cite{meadowsMeasurement14MeV1987} \\ 
  (N,P) & 11715003 & Terrell and Holm & 1958 & \cite{terrellExcitationFunctionFe1958} & (N,P) & 30755003 & Muyao et al & 1987 & \cite{muyaoShellEffectCross1987} \\ 
  (N,P) & 11715004 & Terrell and Holm & 1958 & \cite{terrellExcitationFunctionFe1958} & (N,P) & 22089042 & Ikeda et al & 1988 & \cite{ikedaActivationCrossSection1988} \\ 
  (N,P) & 11464006 & Thompson and Ferguson & 1959 & \cite{kernCrossSectionsReactions1959} & (N,P) & 22093011 & Kimura and Kobayashi & 1990 & \cite{kimuraCalibratedFissionFusion1990} \\ 
  (N,P) & 21419006 & Depraz et al & 1960 & \cite{deprazMesureSectionsEfficaces1960} & (N,P) & 12812010 & Saraf et al & 1991 & \cite{sarafCrossSectionsSpectra1991} \\ 
  (N,P) & 11718005 & Chittenden et al & 1961 & \cite{chittendenNewIsotopeManganese1961} & (N,P) & 22338048 & Ercan et al & 1991 & \cite{ercan14MeVNeutron1991} \\ 
  (N,P) & 21352002 & Pollehn and Neuert & 1961 & \cite{pollehnBestimmungWirkungsquerschnittenEiniger1961} & (N,P) & 30978022 & Viennot et al & 1991 & \cite{viennotCrossSectionMeasurementsNp1991} \\ 
  (N,P) & 21352007 & Pollehn and Neuert & 1961 & \cite{pollehnBestimmungWirkungsquerschnittenEiniger1961} & (N,P) & 31479002 & Fuga & 1991 & \cite{fugaStudyExcitationFunction1991} \\ 
  (N,P) & 11494009 & Gabbard and Kern & 1962 & \cite{gabbardCrossSectionsCharged1962} & (N,P) & 31479003 & Fuga & 1991 & \cite{fugaStudyExcitationFunction1991} \\ 
  (N,P) & 21339005 & Bormann et al & 1962 & \cite{bormannUeberWirkungsquerschnitteEiniger1962} & (N,P) & 31524008 & Belgaid et al & 1992 & \cite{belgaidMeasurement14MeV1992} \\ 
  (N,P) & 11696007 & Cross et al & 1963 & \cite{crossActivationCrossSections1963} & (N,P) & 41118012 & Klochkova et al & 1992 & \cite{klochkovaInvestigationReactionsAl271992} \\ 
  (N,P) & 11701002 & Santry and Butler & 1964 & \cite{santryExcitationCurvesReactions1964} & (N,P) & 22312004 & Ikeda et al & 1993 & \cite{ikedaAbsoluteMeasurementsActivation1993} \\ 
  (N,P) & 11701003 & Santry and Butler & 1964 & \cite{santryExcitationCurvesReactions1964} & (N,P) & 30993003 & Zongyu et al & 1993 & \cite{zongyuAbsoluteMeasurementCross1993} \\ 
  (N,P) & 11701004 & Santry and Butler & 1964 & \cite{santryExcitationCurvesReactions1964} & (N,P) & 41240012 & Filatenkov et al & 1999 & \cite{filatenkovSystematicMeasurementActivation1999} \\ 
  (N,P) & 20888004 & Bonazzola & 1964 & \cite{bonazzolaMeasurementActivationCross1964} & (N,P) & 22414017 & Fessler et al & 2000 & \cite{fesslerNeutronActivationCrossSection2000} \\ 
  (N,P) & 20377002 & Liskien and Paulsen & 1965 & \cite{liskienCrosssectionMeasurementThreshold1965} & (N,P) & 22497004 & Coszach et al & 2000 & \cite{coszachNeutroninducedReactionsContributing2000} \\ 
  (N,P) & 20887015 & Bormann et al & 1965 & \cite{bormannExcitationFunctionsNeutron1965} & (N,P) & 22976017 & Mannhart and Schmidt & 2007 & \cite{mannhartMeasurementNeutronActivation2007} \\ 
  (N,P) & 20387004 & Liskien and Paulsen & 1966 & \cite{liskienCrossSectionsCu631966} & (N,P) & 33045003 & Mulik et al & 2013 & \cite{mulikMeasurement56Fe56Mn2013} \\ 
  (N,P) & 21372003 & Hemingway et al & 1966 & \cite{hemingwayDeterminationCrossSections1966} & (N,P) & 41614019 & Filatenkov & 2016 & \cite{filatenkovNeutronActivationCross2016} \\ 
  (N,P) & 10417007 & Grundl & 1967 & \cite{grundlStudyFissionNeutronSpectra1967} & (N,T) & 20669004 & Qaim and Stocklin & 1976 & \cite{qaimInvestigationReactions141976} \\ 
  (N,P) & 20815014 & Vonach et al & 1968 & \cite{vonachPrecisionMeasurementsExcitation1968} & (N,TOT) & 10037005 & Boschung et al & 1971 & \cite{boschungScatteringFastNeutrons1971} \\ 
  (N,P) & 20890004 & Cuzzocrea et al & 1968 & \cite{cuzzocreaExcitationFunctionsNeutroninduced1968} & (N,TOT) & 41325003 & Tutubalin et al & 1973 & \cite{tutubalinTotalNeutronCross1973} \\ 
  (N,P) & 10022010 & Barrall et al & 1969 & \cite{barrallHighEnergyNeutron1969} & (N,TOT) & 13764002 & Harvey & 1987 &  \\ 
   \hline
\end{tabular}
\caption{Experimental datasets used in the example evaluation of neutron-induced reactionf of $^{56}$Fe between five and thirty MeV. The column EXFOR contains the EXFOR accession number. A missing reference means that no accessible publication is known to the authors.} 
\label{tbl:fe56-fast-datasets}
\end{table*}


\end{document}